\documentclass[reprint, amsmath,amssymb, aps, floatfix, showkeywords]{revtex4-2}

\usepackage{graphicx}
\usepackage{dcolumn}
\usepackage{bm}
\usepackage[colorlinks = true,linkcolor = blue,urlcolor  = blue,citecolor = blue,anchorcolor = blue]{hyperref}
\usepackage{txfonts}

\usepackage{microtype}
\usepackage{multirow}

\begin{document}
	
	\title{Neutron Stars and Gravitational Waves: The Key Role of Nuclear Equation of State}
	
	\author{P.S. Koliogiannis}
	\email{pkoliogi@physics.auth.gr}
	\author{A. Kanakis Pegios}
	\author{Ch.C. Moustakidis}
	\affiliation{Department of Theoretical Physics, Aristotle University of Thessaloniki, 54124 Thessaloniki, Greece}
	
	\date{\today}

\begin{abstract}
	Neutron stars are the densest known objects in the universe and an ideal laboratory for the strange physics of super-condensed matter. Theoretical studies in connection with recent observational data of isolated neutron stars, as well as binary neutron stars systems, offer an excellent opportunity to provide robust solutions on the dense nuclear problem. In the present work, we review  recent studies concerning  the applications of various theoretical nuclear models on a few recent observations of binary neutron stars or neutron-star--black-hole systems. In particular, using a simple and well-established model, we parametrize the stiffness of the equation of state with the help of the speed of sound. Moreover, in comparison to the recent observations of two events by LIGO/VIRGO collaboration, GW170817 and GW190425, we suggest possible robust constraints. We also concentrate our theoretical study on the resent observation of a compact object with mass~$\sim 2.59_{-0.09}^{+0.08}~M_{\odot}$ (GW190814 event), as a component of a system where the main companion was a black hole with mass $\sim 23~M_{\odot}$. There is scientific debate  concerning the identification of the low mass component, as it falls into the neutron-star--black-hole mass gap. This is an important issue since  understanding the nature of GW190814 event will offer rich information concerning  the upper limit of the speed of sound in dense matter and the possible phase transition into other degrees of freedom.  We systematically study the tidal deformability of a possible high-mass candidate existing as an individual star or as a component in a binary neutron star system. Finally, we provide some applications of equations of state of hot, dense nuclear matter in hot neutron stars (nonrotating and rapidly rotating with the Kepler frequency neutron stars), protoneutron stars, and binary neutron star merger remnants.
\end{abstract}


\maketitle

\section{Introduction}
One of the currently unsolved open problems  in nuclear physics is the properties of dense nuclear matter. In particular, compact objects, such as white dwarfs and especially neutron stars, offer the opportunity to study the behavior of nuclear matter at high densities~\cite{Shapiro-1983,Glendenning-2000,Haensel-2007,Weinberg-1972}. Neutron stars are a very promising tool for studying the properties of dense nuclear matter, such as the speed of sound and its possible upper~bound.

The main assumption for the speed of sound is that it cannot exceed the speed of light because of the causality. However, this is not determinant, as Zel'dovich~\cite{Zeldovich-62,Zeldovich-71} showed the importance of defining a rigorous limit of speed of sound upon the equation of state (EoS). To~be more specific, in~the electromagnetic interaction, the main assumption is that $v_s \leq c/\sqrt{3}$ is  generally  low in nature. Moreover, by~considering the interaction of baryons through a vector field, he noticed that the upper limit of the speed of sound is the causality, $v_s=c$. Therefore, the~only restriction imposed by general principles is that $v_s\leq c$~\cite{Zeldovich-62,Zeldovich-71}. On~the other hand, Hartle noticed that the causality is not enough to constrain the high-density part of the EoS~\cite{Hartle-1978}, while Weinberg pointed out that the speed of sound is much less than the speed of light for a cold nonrelativistic ﬂuid~\cite{Weinberg-1972}. In~addition, for~nonrelativistic and/or weakly coupled theories, the bound $v_s=c/\sqrt{3}$, according to Bedaque and Steiner seems to be valid, while in conformal theories, the upper bound is saturated~\cite{Bedaque-2015}. According to these authors, the~existence of a $2~M_{\odot}$ neutron star
, in~combination with the knowledge of the EoS of hadronic matter at the low density region, is not consistent with the limit $c/\sqrt{3}$. We notice that various recent studies have been conducted regarding the speed of sound and the tidal deformability of neutron stars~\cite{Moustakidis-2017,Reed-2020,Oeveren-2017,Ma-2019}.

One of the goals of this study is to apply a method that directly relates the observed tidal deformability, derived from binary neutron star mergers, to~the maximum mass of neutron stars, aiming to obtain constraints on the upper bound of the speed of sound. The~main idea is the fact that while the measured upper limit of the effective tidal deformability favors softer EoSs,  recent measurements of high neutron star masses favor stiffer EoSs. As~a basis in our study, we used a model in which we parametrized the EoS through the various bounds of the speed of sound (stiffness). Hence, the~EoS is a functional of the transition density and the speed of sound bound. In~our approach, we used the observation of  two recent events, GW170817~\cite{Abbott-2017} and GW190425~\cite{Abbott_2020}, as~well as the current observed maximum neutron star masses ($1.908 \pm  0.016 M_{\odot}$~\cite{Arzoumanian_2018}, $2.01 \pm 0.04 \ M_{\odot}$~\cite{Antoniadis-2013}, $2.14^{+0.10}_{-0.09}\ M_{\odot}$~\cite{Cromartie-2019}, and $ 2.27^{+0.17}_{-0.15}\ M_{\odot}$~\cite{Linares-2018}). The~need for (a) a soft EoS for the low density region (to be in accordance with the observed upper limit of the effective tidal deformability) and (b) a stiff EoS for the high density region (to provide the high neutron star masses) leads to robust constraints on the EoS. In~addition, this method allows making postulations about the kind of future measurements that would be more informative and help to improve our~knowledge.

Furthermore, we highlight the very recent observation of the GW190814 event, where a gravitational wave has been detected from the merger of a 22.2--24.3 $M_{\odot}$ black hole with a non-identified compact object with mass 2.5--2.67\;$M_{\odot}$~\cite{Abbott_2020_a,Tan-2020}. Although~the authors of the mentioned references suggest that is unlikely for the second component's mass to belong to a neutron star, they do leave \emph{open the window} that the improved knowledge of the neutron star EoS and further observations of the astrophysical population of compact objects could alter this~assessment.

It is worth  pointing out that the observation of the GW190814 event has some additional general benefits, apart from the measurement of 2.6 $M_{\odot}$ of the second partner~\cite{Abbott_2020_a}. Firstly, this binary system has the most unequal mass ratio yet measured with gravitational waves close to the value of $0.112$. Secondly, the~dimensionless spin of the primary black hole is constrained to $\leq$0.07, where various tests of general relativity confirm this value, as~well as its predictions of higher multiple emissions at high confidence intervals. Moreover, the~GW190814 event poses a challenge for the understanding of the population of merging compact binaries. It was found, after~systematic analysis, that the merger rate density of the GW190814-like binary system was $7_{-6}^{+16}$ Gpc$^{-3}$yr$^{-1}$~\cite{Abbott_2020_a}. More relevant to the present study, the~observation of the GW190814 event led to the following conclusion: due to the source's asymmetric masses, the~lack of detection of an electromagnetic counterpart and of clear signature of tides or the spin-induced quadrupole effect in the waveform of the gravitational waves, we are not able to distinguish between a black-hole--black-hole and black-hole--neutron-star system~\cite{Abbott_2020_a}. In~this case, one must count only  the comparison between the mass of the second partner with the estimation of maximum neutron star mass~\cite{Datta-2020}. This is one of the main subjects which has been revised in the  present work. It should be emphasized that the measurements of neutron star mass can also inform us about a bound on the maximum gravitational mass independently of the assumptions of the specific EoS. For~example, Alsing~et~al.~\cite{Alsing-2018}, fitting the known population of neutron stars in binaries to double-Gaussian mass distribution, obtained the empirical constraint that $M_{\rm max} \leq 2.6 \ M_{\odot}$ (with $90 \%$ confidence interval ).

Farr and Chatziionannou updated knowledge from previous studies, including recent measurements~\cite{Farr-2020}. Their study constrains the maximum mass $M_{\rm max}=2.25_{-0.26}^{+0.81} \ M_{\odot}$, leading to the conclusion that the posterior probability (for the mass of the second partner $m_2 \leq M_{\rm max}$) is around only $29 \%$. However, the~prediction of $M_{\rm max}$ is sensitive to the selection mass rules of neutron stars (not only on binary systems but also isolated) as well as to the discovery of new events, and this consequently remains an open problem. Finally, the~conclusion of the recent GW190814 event in comparison with previous ones (for example the GW170817 event~\cite{Abbott-2018}) may shed light on the problem of the $M_{\rm max}$. For~example, the~spectral EoSs, which are conditioned by the GW170817 event, are once more elaborated to include the possibility that  the prediction of $M_{\rm max}$  is at least equal to $m_2$. This approach leads to significant constraints on the radius and tidal deformability of a neutron star with mass of $1.4 \ M_{\odot}$ ($R_{1.4}=12.9_{-0.7}^{+0.8}$ km and $\Lambda_{1.4}=616_{-158}^{+273}$, respectively, \cite{Abbott_2020_a}).

The matter of the finite temperature and its effect on the nuclear EoS, as~well as astrophysical applications, has been extensively studied by Bethe~et~al.~\cite{BETHE1979487}, Brown~et~al.~\cite{BROWN1982481}, Lamb~et~al.~\cite{PhysRevLett.41.1623}, Lattimer and Ravenhall~\cite{As.J.223.314}, and Lattimer~\cite{AnnRNPS.31.337}. Lattimer and Swesty~\cite{LATTIMER1991331}, as~well as Shen~et~al.~\cite{SHEN1998435},  constructed the most used hot neutron star EoSs. The~first one is based on the liquid drop-type model, and the second one is based on the relativistic mean field model. Afterwards, Shen~et~al.~\cite{SHEN1998435} broadened their study in order to study supernovae, binary neutron star mergers, and~black hole formations by developing EoSs for various temperatures and proton fractions~\cite{PhysRevC.83.065808}. In~addition, the~density and temperature dependence of the nuclear symmetry free energy using microscopic two- and three-body nuclear potentials constructed from Chiral effective field theory have been studied in a series of works, including the one of Wellenhofer~et~al.~\cite{PhysRevC.92.015801}. Furthermore, hot neutron star and supernova properties have been studied by Constantinou~et~al.~\cite{PhysRevC.89.065802,PhysRevC.92.025801}, where a suitable hot EoS is produced. Finally, the~interplay between the temperature and the neutron star matter was probed by Sammarruca~et~al.~\cite{doi:10.1142/S0217732320501564}, by~considering the framework of chiral effective field~theory.

EoSs at finite temperature constructed within the Brueckner--Hartree--Fock approach and the properties of hot $\beta$-stable nuclear matter were studied in a series of papers~\mbox{\cite{AA.451.1.2010,AA.518.A17.2010, BALDO2016203,10.1093/mnras/sty147,PhysRevC.100.054335,PhysRevC.103.024307,PhysRevD.102.043006,PhysRevC.101.065801,10.1093/mnras/staa1879}}. In~addition, a~model for cold nucleonic EoSs, extended to include temperature and proton fractions for simulations of astrophysical phenomena, was constructed by Raithel~et~al.~\cite{Raithel_2019}. Pons~et~al.~\cite{Pons_1999}, as~well as Prakash~et~al.~\cite{Prakash_2000}, focused on describing the thermal and chemical evolution of protoneutron stars by~considering neutrino opacities consistently calculated with the EoS. Finally, neutron stars, along with the hot EoS of dense matter, are presented in a recent review of Lattimer and Prakash~\cite{LATTIMER2016127}.

The processes that occupy the stages of the merger and postmerger phases of a binary neutron star system have been extensively studied in recent years. However, matters that concern the remnant evolution are still under consideration or~even unsolved. In~particular, the~remnant evolution contains (a) the collapse time, (b) the threshold mass, (c)~the possible phase transition in the interior of the star, and~(d) the disk ejecta and neutrino emission (for an extended discussion and applications, see Perego~et~al.~\cite{Perego2019}). It has to be noted here that the possibility of a phase transition will affect the signal of the emitted gravitational wave. Relevant previous work is also be available in Bauswein~et~al.~\cite{PhysRevD.82.084043}, Kaplan~et~al.~\cite{Kaplan_2014}, Tsokaros~et~al.~\cite{PhysRevLett.124.071101}, Yasin~et~al.~\cite{PhysRevLett.124.092701}, Radice~et~al.~\cite{doi:10.1146/annurev-nucl-013120-114541}, Sarin~et~al.~\cite{PhysRevD.101.063021}, Soma and Bandyopadhyay~\cite{Soma_2020}, and~Sen~\cite{Sen_2020}.

In the present work, we review  some applications of  the thermal effects on neutron star properties.  In~particular, we apply a momentum-dependent effective interaction (MDI) model, where thermal effects can be studied simultaneously on the kinetic part of the energy and also on the interaction one. In~addition, the~extension of the proposed model can lead to EoSs with varying stiffness with respect to the parameterized symmetry energy. In~fact, Gale~et~al.~\cite{PhysRevC.35.1666} presented a model aimed at the influence of MDI on the momentum flow of heavy-ion collisions. However, the~model has been successfully applied in studying the properties of cold and hot nuclear and neutron star matter (for an extensive review of the model, see References~\cite{PRAKASH19971,Libook,LI2008113}).

Moreover, we review specific properties of neutron stars (including mainly the mass and radius, moment of inertia, Kerr parameter, etc.) with~respect to the EoS, both at nonrotating and rapidly rotating (considering the mass-shedding limit) configurations. The~above properties are well applied in studying hot neutron stars, protoneutron stars, and the remnants of binary neutron star~systems.

To summarize, in~the present work, we review the applications of various theoretical nuclear models on a few recent observations of binary neutron stars or neutron-star--black-hole systems, including mainly the  GW170817,  GW190425, and GW190814 events. Our results have been published recently in sections, in~the following journals~\cite{Koliogiannis-2020,Koliogiannis-2021,Kanakis-2020,Kanakis-2021}.

It has to be noted that the present study is mainly dedicated to the case of the emission of the gravitation waves due to the merger of a binary neutron star system. However, there are other mechanisms by which gravitational waves are emitted by a neutron star and thus, have additional ways of studying its internal structure (for a recent review, see Reference~\cite{universe5110217}). These mechanisms, which are considered as continuous gravitational waves sources, include, for example, (a) the case of radiation of gravitational waves by a rigidly rotating aligned triaxial ellipsoid (radiation of purely quadrupolar waves), (b) the emission of gravitational waves due to asymmetry in the magnetic field distribution in the interior of the neutron star, and~(c) the radiation of gravitational waves from the rapidly rotating neutron stars. In~this case, neutron stars may suffer a number of different instabilities with a general feature in common: they can be directly associated with unstable modes of oscillation (for exable g-modes, f-modes, w-modes, r-modes; for a review see Reference~\cite{Glampedakis2018}). The~more notable mechanism is the r-mode oscillations. In~these oscillations, the restoring force is the Coriolis force. The~r-mode mechanics have been proposed as an explanation for the observed relatively low spin frequencies of young neutron stars, as~well as of accreting neutron stars in low-mass X-ray binaries. This instability only occurs when the gravitational radiation driving timescale is shorter compared to the ones of the various dissipation mechanisms, which occur in the neutron star matter~\cite{doi:10.1142/S0218271801001062}. The~free procession may cause deformation of the neutron star, leading to a better understanding of some neutron star matter properties, including breaking strain, viscosity, rigidity, and elasticity~\cite{universe5110217}. We expect that in the future, the~development of the sensitivity of LIGO and Virgo detectors in cooperation with and new instruments will help to significantly improve  our knowledge of neutron star interiors with the detection of the emitted gravitational waves~\cite{universe5110217}.

Another way to study the properties of dense nuclear matter that exists inside neutron stars can be done with the help of the statistical study of the observed properties of neutron stars. To~be more specific, we refer to the observational data concerning both isolated neutron stars and those that exist in binary systems. In~the first case, there are extensive studies where observational data are applied to estimations with the help of statistical studies, for~example, the maximum possible mass of a neutron star, but~mainly the radius of neutron stars with masses of about $1.4~M_{\odot}$. In~each case, valuable information and ideas can be extracted and utilized in terms of knowledge of the EoS of neutron star matter, in~order to evaluate the reliability of the existing EoS (see the review article~\cite{doi:10.1146/annurev-astro-081915-023322} and references therein). In~the case of binary systems, the~analysis of the emitted gravitational waves from the fusion of a binary system of neutron stars, where a large amount of information is received by studying the amplitude and phase of gravitational waves, is utilized. These studies mainly focus on the measurement (and utilization of the measurement) of tidal deformability. In~any case, extensive and systematic statistical estimation of data can lead to valuable knowledge of the structure and composition of neutron stars (for a recent review see~\cite{10.1007/s10714-020-02754-3} and references therein).

The article  is organized  as follows: in Section~\ref{sec:2}, we present the theory concerning the EoS and the structure of cold neutron stars. In~Section~\ref{sec:3}, we present the construction of the hot EoSs (both isothermal and isoentropic) and briefly discuss the stability equations of hot rapidly rotating neutron stars. The~results and the discussion are provided in Section~\ref{sec:4}, while Section~\ref{sec:5} includes the most noteworthy concluding remarks of the present~review.

\section{Cold Neutron~Stars} \label{sec:2}
\subsection{The Momentum-Dependent Interaction Nuclear~Model}
The description of the interior structure of neutron stars demands the use of a nuclear  model suitable to describe the properties of dense nuclear matter. In~the present work, the~EoS of nuclear matter is studied using the MDI model. In~this model, the~energy per baryon is given by the formulae~\cite{PRAKASH19971,Moustakidis-15}
\begin{widetext}
	\begin{eqnarray}
		E(n,I)&=&\frac{3}{10}E_F^0u^{2/3}\left[(1+I)^{5/3}+(1-I)^{5/3}\right]+
		\frac{1}{3}A\left[\frac{3}{2}-X_{0}I^2\right]u
		+
		\frac{\frac{2}{3}B\left[\frac{3}{2}-X_{3}I^2\right]u^{\sigma}}
		{1+\frac{2}{3}B'\left[\frac{3}{2}-X_{3}I^2\right]u^{\sigma-1}}
		\nonumber \\
		&+&\frac{3}{2}\sum_{i=1,2}\left[C_i+\frac{C_i-8Z_i}{5}I\right]\left(\frac{\Lambda_i}{k_F^0}\right)^3
		\left(\frac{\left((1+I)u\right)^{1/3}}{\frac{\Lambda_i}{k_F^0}}-
		\tan^{-1} \frac{\left((1+
			I)u\right)^{1/3}}{\frac{\Lambda_i}{k_F^0}}\right)\nonumber \\
		&+&
		\frac{3}{2}\sum_{i=1,2}\left[C_i-\frac{C_i-8Z_i}{5}I\right]\left(\frac{\Lambda_i}{k_F^0}\right)^3
		\left(\frac{\left((1-I)u\right)^{1/3}}{\frac{\Lambda_i}{k_F^0}}-
		\tan^{-1}
		\frac{\left((1-I)u\right)^{1/3}}{\frac{\Lambda_i}{k_F^0}}\right),
		\label{eq:mdi_model}
		\label{e-T0}
	\end{eqnarray}
\end{widetext}
where $u$ is the baryon density normalized with respect to the saturation density $(n_{s}=0.16~{\rm fm^{-3}})$, $I=(n_{n}-n_{p})/n$ is the asymmetry parameter, $X_{0} = x_{0} + 1/2$, and $X_{3} = x_{3} + 1/2$. The~parameters $A$, $B$, $\sigma$, $C_i$, and~$B'$ appear in the description of symmetric nuclear matter (SNM) and are determined so that the relation $E(n_{s},0) = -16~{\rm MeV}$ holds. $\Lambda_{1}$ and $\Lambda_{2}$ are finite range parameters equal to $1.5k_{F}^{0}$ and $3k_{F}^{0}$, respectively, where $k_{F}^{0}$ is the Fermi momentum at the saturation density. The~remaining parameters, $x_0$, $x_3$, $Z_i$, appear in the description of asymmetric nuclear matter (ANM) and, with~a suitable parametrization, are used in order to obtain different forms for the density dependence of symmetry energy, as~well as the value of the slope parameter L and the value of the symmetry energy $E_{\rm sym}$ at the saturation density, defined as~\cite{Koliogiannis-2020}
\begin{equation}
	L=3n_{s}\frac{dE_{\rm sym}(n)}{dn}\bigg\vert_{n_{s}} \quad \text{and} \quad E_{\rm sym}(n)=\frac{1}{2\!}\frac{\partial^{2} E(n,I)} {\partial I^{2}}\bigg\vert_{I=0},
\end{equation}
and consequently different parametrizations of the EoS~stiffness.

The specific choice of the MDI model is based on the following: (a) it combines both density and momentum dependent interaction among the nucleons, (b) it is suitable for studying neutron star matter at zero and finite temperature (due to the momentum term), (c) it reproduces with high accuracy the properties of SNM at the saturation density, including isovector quantities, (d) it reproduces the microscopic properties of the Chiral model for pure neutron matter (PNM) and the results of \emph{state-of-the-art} calculations of Akmal~et~al.~\cite{Akmal-1998} with suitable parametrizations, (e) it predicts higher maximum neutron star mass  than the observed ones~\cite{Antoniadis-2013,Cromartie-2019,Linares-2018}, and~(f) it maintains the causal behavior of the EoS even at densities higher than the ones that correspond to the maximum mass~configuration.

\subsection{Speed of Sound~Formalism}
An EoS can be parametrized in order to reproduce specific values of the speed of sound in the interior of the neutron star. This parametrization is possible following the formula available from References~\cite{Margaritis-2020,Rhoades-1974,Kalogera-1996,Koranda-1997,Chamel-2013a,Alsing-2018,Podkowka-2018,Xia-2021}:
\begin{eqnarray}
	P({\cal E})&=&\left\{
	\begin{array}{ll}
		P_{\rm crust}({\cal E}), \quad  {\cal E} \leq {\cal E}_{\rm c-edge}  & \\
		\\
		P_{\rm NM}({\cal E}), \quad  {\cal E}_{\rm c-edge} \leq {\cal E} \leq {\cal E}_{\rm tr}  & \\
		\\
		\left(\frac{v_{\rm s}}{c}  \right)^2\left({\cal E}-{\cal E}_{\rm tr}  \right)+
		P_{\rm NM}({\cal E}_{\rm tr}), \quad  {\cal E}_{\rm tr} \leq {\cal E} , & \
	\end{array}
	\right.
	\label{eq:eos_sos}
\end{eqnarray}
where $P$ and $\mathcal{E}$ denote the pressure and the energy density, respectively, and~the corresponding subscript ``tr'' in energy density denotes the energy density at the transition~density.

However, following the approach in Equation~\eqref{eq:eos_sos}, only the continuity in the EoS is ensured. The~artificial character of Equation~\eqref{eq:eos_sos} does not take into account the continuity in the speed of sound. Therefore, in~order to ensure the continuity and a smooth transition, we followed the method presented in Reference~\cite{Tews-2018}. We proceeded with the matching of the EoSs with the transition density by considering that, above~this value, the~speed of sound is parametrized as follows (for more details, see Reference~\cite{Tews-2018}):
\begin{equation}
	\frac{v_{\rm s}}{c}=\left(a-c_1\exp\left[-\frac{(n-c_2)^2}{w^2} \right]\right)^{1/2}, \quad a\in [1/3,1]
	\label{speed-matc-1}
\end{equation}
where the parameters $c_1$ and $c_2$ are fit to the speed of sound and its derivative at $n_{{\rm tr}}$ and~also to the demands $v_{\rm s}(n_{{\rm tr}})=[c/\sqrt{3},c]$~\cite{Margaritis-2020} according to the value of $\alpha$. The~remaining parameter $w$ controls the width of the curve, which in our case is equal to $10^{-3}~{\rm fm^{-3}}$, in order to preserve the neutron star properties. Using Equation~\eqref{speed-matc-1}, the~EoS for $n \geq n_{{\rm tr}}$ can be constructed with the help of the following ~\cite{Tews-2018}:
\begin{equation}
	{\cal E}_{i+1} = {\cal E}_i+\Delta {\cal E}, \quad P_{i+1} = P_i+\left(\frac{v_s}{c}(n_i)\right)^2\Delta {\cal E},
	\label{eq:5}
\end{equation}
\begin{equation}
	\Delta {\cal E} = \Delta n\left(\frac{{\cal E}_i+P_i}{n_i} \right),
	\label{eq:6}
\end{equation} 
\begin{equation}
	\Delta n = n_{i+1}-n_i.
	\label{eq:7}
\end{equation}

\subsection{Construction of the~EoS}
The construction of the EoS for the interior of neutron stars is based on the MDI model and data provided by Akmal~et~al.~\cite{Akmal-1998}. More specifically, we utilized the data for the A18+UIX (hereafter APR-1) EoS from Akmal~et~al.~\cite{Akmal-1998} for the energy per particle of SNM and PNM in the density range $[0.04,0.96]~{\rm fm^{-3}}$. Due to the complexity of the microscopic data, we divided the density region into three sections---(a) low-density region $[0.04,0.2]~{\rm fm^{-3}}$, (b) medium-density region $[0.2,0.56]~{\rm fm^{-3}}$, and~(c) high-density region $[0.56,0.96]~{\rm fm^{-3}}$---in~order to acquire the best fitting using Equation~\eqref{eq:mdi_model}. From~this process emerged the coefficients for the SNM and ANM, and~eventually the EoS, hereafter~MDI-APR.

In the case of the speed-of-sound-parametrized EoSs, the~construction of the EoSs follows the procedure: (a) in region ${\cal E} \leq {\cal E}_{\rm c-edge}$, we used the equation of Feynman~et~al.~\cite{Feynman-1949} and also of Baym~et~al.~\cite{Baym-1971} for the crust and low densities of neutron star; (b) in the intermediate region, ${\cal E}_{\rm c-edge} \leq {\cal E} \leq {\cal E}_{\rm tr}$, we employed the MDI-APR EoS; and~(c) for ${\cal E}_{\rm tr}\geq \mathcal{E}$ region, the~EoS is maximally stiff with the speed of sound, defined as  $v_s=c\sqrt{\left(\partial P / \partial {\cal E}\right)}_{\rm S}$ (where $S$ is the entropy) fixed in the present work in the range $[c/\sqrt{3},c]$. The~lowest allowed value of the speed of sound, that is $(v_{s}/c)^{2}=1/3$, is introduced in order to be consistent with the possibility of a phase transition in quark matter. In~this case, the~theoretical predictions lead to this value as an upper bound of the speed of sound. The~implementation of speed of sound values between the limited ones will lead to results well constrained by the two mentioned limits. Although~the energy densities below the ${\cal E}_{\rm c-edge}$ have negligible effects on the maximum mass configuration, we used them in calculations for the accurate estimation of the tidal~deformability.

In this study, two cases, based on the transition density, $n_{{\rm tr}} = p n_{\rm s}$ and~the speed of sound, are employed, in~particular, (a) the ones where $p$ takes the values $[1.5,2,3,4,5]$ while the speed of sound is parametrized in the two limiting cases, $(v_{s}/c)^{2}=1/3$ and $(v_{s}/c)^{2}=1$, and (b) the ones where $p$ takes the values $[1.5,2]$ while the speed of sound is parametrized in the range $(v_{s}/c)^{2}=[1/3,1]$.

\begin{figure*}
	\includegraphics[width=\textwidth]{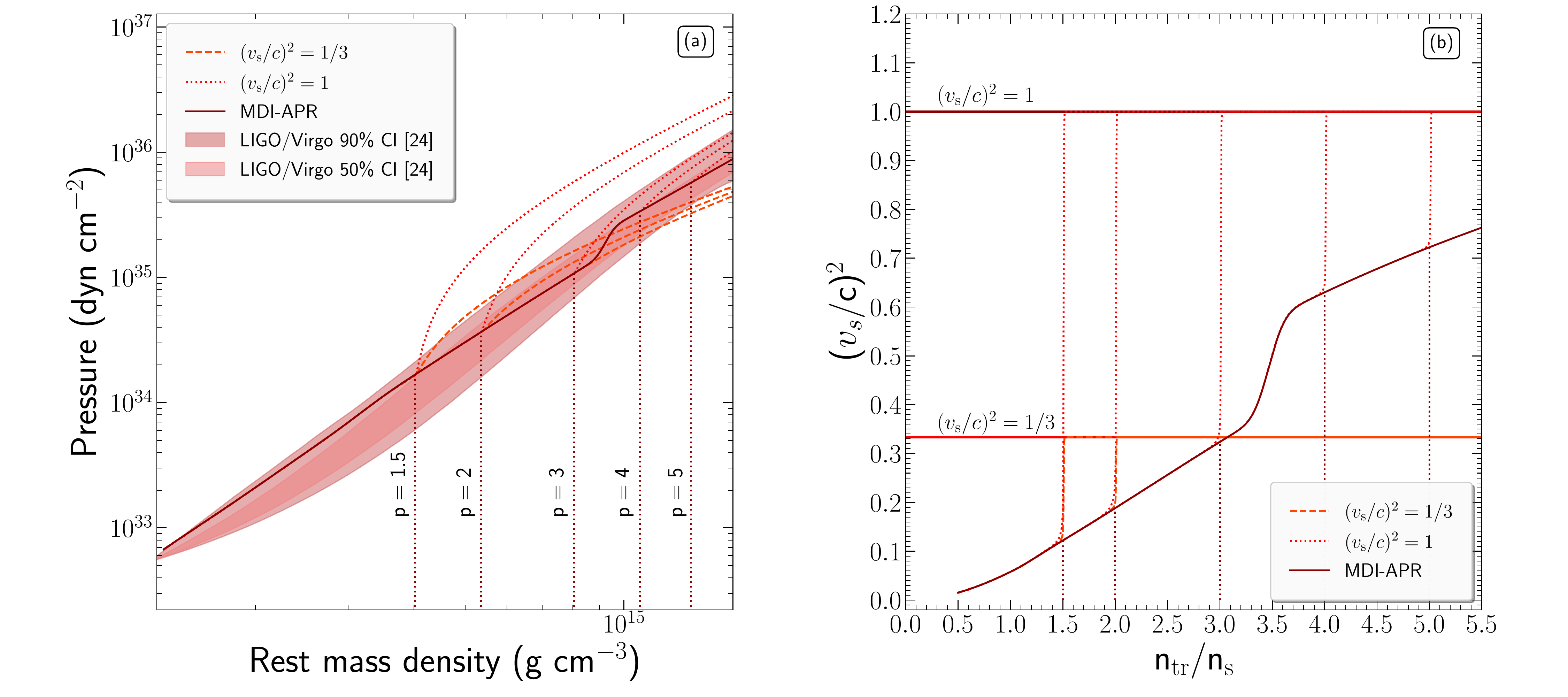}
	
	\includegraphics[width=\textwidth]{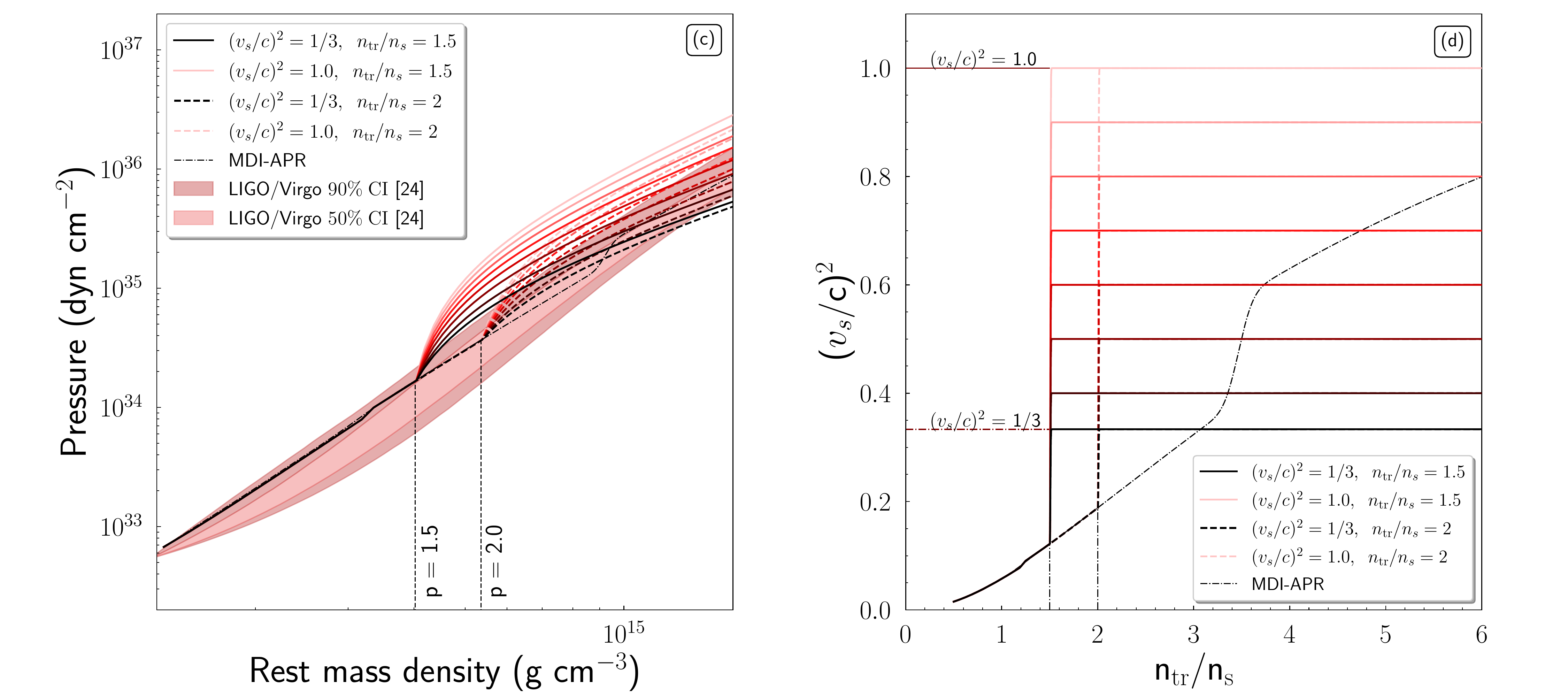}
	\caption{(a,c) Dependence of the pressure on the rest mass density and (b,d) dependence of the square sound speed in units of light speed on the transition density. (a,b) The speed of sound is fixed at the two boundary cases, $(v_{s}/c)^{2}=1/3$ and $(v_{s}/c)^{2}=1$, and~the value $p$ takes the arguments $[1.5,2,3,4,5]$. (c,d) The value $p$ takes the arguments $[1.5,2]$, and~the speed of sound is parametrized in the range $(v_{s}/c)^{2}=[1/3,1]$ (the lower values of the speed of sound correspond to the darker colored curves). In~all Figures, the~vertical dotted lines indicate the transition cases, while the shaded regions note the credibility interval derived from Reference~\cite{Abbott-2018}.}
	\label{fig:pressure_density_sos}
\end{figure*}

For reasons of completeness, the~treatment with both discontinuity and continuity in the speed of sound is presented in Table V of Reference~\cite{Margaritis-2020}. The~main point was that the two approaches converge and consequently the effects of the discontinuity are~negligible.

In Figure~\ref{fig:pressure_density_sos}, we present the pressure as a function of the rest mass density ($\rho_{\rm rest} = n_{b}m_{n}$) and the square speed of sound in units of speed of light as a function of the transition density for the EoSs constructed in cases (a) and (b). In~addition, we display the credibility intervals proposed by Reference~\cite{Abbott-2018} from LIGO/Virgo collaboration for the GW170817 event. It is clear from these figures that the pure MDI-APR EoS is well-defined in the proposed limits of LIGO/Virgo collaboration and also fulfills the speed of light limit at high~densities. 

\subsection{Structure~Equations}
\subsubsection{Nonrotating Neutron~Stars}
For a static spherical symmetric system, which is the case of a nonrotating neutron star,  the~metric can be written as follows~\cite{Shapiro-1983,Glendenning-2000}:
\begin{equation}
	ds^2=e^{\nu(r)}dt^2-e^{\lambda(r)}dr^2-r^2\left(d\theta^2+\sin^2\theta d\phi^2\right).
	\label{GRE-1}
\end{equation}
The  density distribution and the local pressure is related to the  metric functions $\lambda(r)$ and $\nu(r)$  according to the relations~\cite{Shapiro-1983,Glendenning-2000}
\begin{equation}
	\frac{8\pi G}{c^2}\rho(r)=\frac{1}{r^2}\left(1-e^{-\lambda(r)}\right)+ e^{-\lambda(r)}\frac{\lambda'(r)}{r},
	\label{GRE-2}
\end{equation}
\begin{equation}
	\frac{8\pi G}{c^4}P(r)=-\frac{1}{r^2}\left(1-e^{-\lambda(r)}\right)+ e^{-\lambda(r)}\frac{\nu'(r)}{r},
	\label{GRE-3}
\end{equation}
where derivatives with respect to the radius are denoted by $'$. The combination of \mbox{Equations~(\ref{GRE-2})} and (\ref{GRE-3}) leads to the well known Tolman--Oppenheimer--Volkoff (TOV) equations~\mbox{\cite{Shapiro-1983,Glendenning-2000}}:
\begin{eqnarray}
	\frac{dP(r)}{dr}&=&-\frac{G\rho(r) M(r)}{r^2}\left(1+\frac{P(r)}{\rho(r) c^2}\right)\left(1+\frac{4\pi P(r) r^3}{M(r)c^2}\right) \\\nonumber && \left(1-\frac{2GM(r)}{c^2r}\right)^{-1},
	\label{TOV-1}
\end{eqnarray}
\begin{equation}
	\frac{dM(r)}{dr}=4\pi r^2\rho(r).
	\label{TOV-2}
\end{equation}

It is difficult to obtain exact solutions to TOV equations in closed analytical form, and they are solved numerically with an equation of state specified~\cite{Moustakidis-2017a}. Actually, there are hundreds of analytical solutions of TOV equations, but there are three that satisfy the criteria that the pressure and energy density vanish on the surface of the star, and they also both decrease monotonically with increasing radius. These three solutions are the Tolman VII, the~ Buchdahl, and~the Nariai IV. Actually, the~Tolman VII and the Buchdahl have already been analyzed and employed in Reference~\cite{Lattimer-01}. It is worth pointing out that in the present work we use the Tolman VII solution, which contains two parameters, that is, the central density $\rho_c$ and the compactness parameter $\beta=GM/Rc^2$. All the mentioned solutions have been presented and analyzed in detail in References~\cite{Postnikov-2010,Lattimer-01,Lattimer-05,Lattimer-2000}.

One of the most significant sources for the detectors of terrestrial gravitational waves is the gravitational waves from inspiraling binary neutron star systems before their merger~\cite{Postnikov-2010,Baiotti-2019,Flanagan-08,Hinderer-08,Damour-09,Hinderer-10,Fattoyev-13,Lackey-015,Takatsy-2020}. The~component masses of these binary systems can be measured. Additionally, during~the last orbits before the merger, the~tidal effects that are present can also be measured~\cite{Flanagan-08}.

The dimensionless parameter that describes the response of a neutron star to the induced tidal field is called tidal Love number $k_2$. This parameter depends on the neutron star structure (i.e., the mass of the neutron star, and~the EoS). Specifically, the~tidal Love number $k_2$ is a proportional parameter between the induced quadrupole moment $Q_{ij}$ and the applied tidal field $E_{ij}$~\cite{Flanagan-08,Thorne-1998} given below:
\begin{equation}
	Q_{ij}=-\frac{2}{3}k_2\frac{R^5}{G}E_{ij}\equiv- \lambda E_{ij},
	\label{Love-1}
\end{equation}
where $R$ is the neutron star's radius and $\lambda=2R^5k_2/3G$ is a key-role quantity, which is called tidal deformability. The~tidal Love number $k_2$ is given by~\cite{Flanagan-08,Hinderer-08}
\begin{widetext}
\begin{eqnarray}
	k_2 &=& \frac{8\beta^5}{5}\left(1-2\beta\right)^2\left[2-y_R+(y_R-1)2\beta \right] \times
	\left[\frac{}{} 2\beta \left(6  -3y_R+3\beta (5y_R-8)\right) + 4\beta^3  \left(13-11y_R+\beta(3y_R-2)+2\beta^2(1+y_R)\right) \right. \nonumber \\
	&+& \left. 3\left(1-2\beta \right)^2\left[2-y_R+2\beta(y_R-1)\right] {\rm ln}\left(1-2\beta\right)\right]^{-1}.
	\label{k2-def}
\end{eqnarray}
\end{widetext}
The quantity $y_R$ is determined by solving the following differential equation
\begin{equation}
	r\frac{dy(r)}{dr}+y^2(r)+y(r)F(r)+r^2Q(r)=0, 
	\label{D-y-1}
\end{equation}
with the initial condition $ y(0)=2$~\cite{Hinderer-10}. $F(r)$ and $Q(r)$ are functionals of ${\cal E}(r)$, $P(r)$, and $M(r)$, defined as~\cite{Postnikov-2010,Hinderer-10}
\begin{equation}
	F(r)=\left[ 1- \frac{4\pi r^2 G}{c^4}\left({\cal E} (r)-P(r) \right)\right]\left(1-\frac{2M(r)G}{rc^2}  \right)^{-1},
	\label{Fr-1}
\end{equation}
and
\begin{widetext}
\begin{eqnarray}
		r^2Q(r)&=&\frac{4\pi r^2 G}{c^4} \left[5{\cal E} (r)+9P(r)+\frac{{\cal E} (r)+P(r)}{\partial P(r)/\partial{\cal E} (r)}\right]
		\times\left(1-\frac{2M(r)G}{rc^2}  \right)^{-1} - 6\left(1-\frac{2M(r)G}{rc^2}  \right)^{-1} \nonumber \\
		&-&\frac{4M^2(r)G^2}{r^2c^4}\left(1+\frac{4\pi r^3 P(r)}{M(r)c^2}   \right)^2\left(1-\frac{2M(r)G}{rc^2}  \right)^{-2}.
		\label{Qr-1}
\end{eqnarray}
\end{widetext}

The Equation~(\ref{D-y-1}) must be integrated self-consistently with the TOV equations using the boundary conditions $y(0)=2$, $P(0)=P_c$ and $M(0)=0$~\cite{Postnikov-2010,Hinderer-08}. The~numerical solution of these equations provides the mass $M$, the~radius $R$ of the neutron star, and~the value of $y_R=y(R)$. The~latter parameter along with the quantity $\beta$ are the basic ingredients of the tidal Love number $k_2$.

Moving on to the parameters of a binary neutron star system, a~well-measured quantity by the gravitational waves detectors is the chirp mass {\it $\mathcal{M}_c$} of the system~\cite{Abbott-2017,Abbott-3}:
\begin{equation}
	\mathcal{M}_c=\frac{(m_1m_2)^{3/5}}{(m_1+m_2)^{1/5}}=m_1\frac{q^{3/5}}{(1+q)^{1/5}},
	\label{chirpmass}
\end{equation}
where $m_1$ and $m_{2}$ are the masses of the heavier and lighter components. Therefore, the~binary mass ratio $q=m_2/m_1$ is within the range $0\leq q\leq1$.

Another quantity that can be constrained from the analysis of the gravitational wave signal and is of great interest, is the effective tidal deformability~\cite{Abbott-2017,Abbott-3}
\begin{equation}
	\tilde{\Lambda}=\frac{16}{13}\frac{(12q+1)\Lambda_1+(12+q)q^4\Lambda_2}{(1+q)^5},
	\label{L-tild-1}
\end{equation}
where~$\Lambda_i$ is the dimensionless tidal deformability, defined as~\cite{Abbott-2017,Abbott-3}
\begin{equation}
	\Lambda_i=\frac{2}{3}k_2\left(\frac{R_i c^2}{M_i G}  \right)^5\equiv\frac{2}{3}k_2 \beta_i^{-5}  , \quad i=1,2.
	\label{Lamb-1}
\end{equation}
By observing Equations~(\ref{k2-def}) and (\ref{Lamb-1}), one can find that  $\Lambda_i$  depends both on the star's compactness and the value of $y(R)$. More specifically, $\Lambda_i$  depends directly on the stiffness of the EoS through the compactness $\beta$ and indirectly through the speed of sound which appears in Equation~(\ref{Qr-1}). In~addition, the~applied EoS also affects  the behavior of $\Lambda$ regarding the neutron star's mass $M$ and radius $R$.

\subsubsection{Rotating Neutron~Stars}
In a fully general relativistic framework, the~rotating neutron stars are studied with the use of the stationary axisymmetric spacetime metric, which is given by~\cite{friedman_stergioulas_2013}
\begin{equation}
	ds^{2} = -e^{2\nu} dt^{2} + e^{2\psi}\left(d\phi - \omega dt\right)^{2} + e^{2\mu} \left(dr^{2} + r^{2}d\theta^{2}\right)
\end{equation}
where the metric functions $\nu$, $\psi$, $\omega$, and $\mu$ depend only on the coordinates $r$ and $\theta$. In~order to describe a rapidly rotating neutron star, in~addition to the above metric, we need the matter inside the neutron star described as a perfect fluid. By~neglecting sources of non-isotropic stresses, as~well as viscous ones and heat transport, then the matter inside the neutron star can be fully described by the stress-energy tensor~\cite{friedman_stergioulas_2013},
\begin{equation}
	T^{\alpha \beta} = \left(\mathcal{E} + P\right) u^{\alpha} u^{\beta} + P g^{\alpha \beta}
\end{equation}
where $u^{\alpha}$ is the fluid's 4-velocity and~$\mathcal{E}$ and $P$ are the energy density and~pressure.

For the stability of cold rotating neutron stars, the~turning-point criterion is being used. It has to be noted that this is only a sufficient and not a necessary condition. Actually, the~neutral stability line is positioned to the left of the turning-point line in $\left(M,\rho_{c}\right)$ space, which implies that the star will collapse before reaching the turning-point line~\cite{10.1111/j.1745-3933.2011.01085.x,10.1093/mnrasl/slx178}.

The numerical integration of the equilibrium equations was conducted under the RNS code~\cite{Stergioulas-1996} by Stergioulas and Friedman~\cite{Stergioulas-1995}, which is actually based on the method developed by Komatsu, Eriguchi, and Hachisu~\cite{Komatsu-1989} and modifications introduced by Cook, Shapiro, and Teukolsky~\cite{Cook-2-1994}.

\section{Hot Neutron~Stars} \label{sec:3}
\subsection{Thermodynamical Description of Hot Neutron Star~Matter}
The description of nuclear matter at finite temperature includes the Helmholtz free energy, where the differentials of the total free energy, as~well as the total internal energy (assuming that the baryons are contained in volume $V$) are given by~\cite{Goodstein-85,Fetter-03}
\begin{eqnarray}
	dF_{\rm tot} &=& -S_{\rm tot}dT-PdV+\sum_i \mu_i dN_i, \label{eq_s3:Ftot} \\
	dE_{\rm tot} &=& TdS_{\rm tot}-PdV+\sum_i \mu_i dN_i,
	\label{eq_s3:Etot}
\end{eqnarray}
where $S_{\rm tot}$ is the total entropy of baryons, $\mu_i$ is the chemical potential of each species, and~$N_i$ is the number of particles of each species, respectively. Furthermore, the~free energy per particle can be expressed as
\begin{eqnarray}
	F(n,T,I)&=&E(n,T,I)-TS(n,T,I) \nonumber \\ &=& \frac{1}{n} \left[\mathcal{E}(n,T,I) - Ts(n,T,I)\right],
	\label{eq_s3:fe}
\end{eqnarray}
where $E=\mathcal{E}/n$ and $S=s/n$ are the internal energy and entropy per particle, respectively. It has to be noted here that for $T=0~{\rm MeV}$, Equation~\eqref{eq_s3:fe} leads to the equality between free and internal~energy.

In addition, the~entropy density $s$ has the same functional form as a noninteracting gas system, calculated through the form
\begin{equation}
	s_{\tau}(n,T,I)=-g\int \frac{d^3k}{(2\pi)^3}\left[f_{\tau} \ln
	f_{\tau}+(1-f_{\tau}) \ln(1-f_{\tau})\right],
	\label{eq_s3:s}
\end{equation}
with spin degeneracy $g=[1,2]$. The~first case corresponds to protons, neutrons, electrons, and~muons, and~the second case corresponds to neutrinos. Finally, the~pressure and chemical potentials, which depend on the above quantities, are described as::
\begin{equation}
	P=-\frac{\partial E_{\rm tot}}{\partial V}\Bigg\vert_{S,N_i}=n^2\frac{\partial \left(\mathcal{E}/n\right)}{\partial n}\Bigg\vert_{S,N_i},
	\label{eq_s3:p_and_m_1}
\end{equation}
\begin{equation}
	\mu_{i}=\frac{\partial E_{\rm tot}}{\partial N_i}\Bigg\vert_{S,V,N_{j\neq i}}=\frac{\partial
		\mathcal{E}}{\partial n_i}\Bigg\vert_{S,V,n_{j\neq i}}.
	\label{eq_s3:p_and_m_2}
\end{equation}

\subsection{Bulk Thermodynamic~Quantities}
The pressure and chemical potentials, which are essential for the thermodynamical description of nuclear matter, can also be connected to the key quantity, that is, the~free energy, as
\begin{equation}
	P=-\frac{\partial F_{\rm tot}}{\partial V}\Bigg\vert_{T,N_i}=n^2\frac{\partial \left(f/n\right)}{\partial
		n}\Bigg\vert_{T,N_i},
	\label{eq_s3:p_and_m_isothermal_1}
\end{equation}
\begin{equation}
	\mu_{i}=\frac{\partial F_{\rm tot}}{\partial N_i}\Bigg\vert_{T,V,N_{j\neq i}}=\frac{\partial f}{\partial n_i}\Bigg\vert_{T,V,n_{j\neq i}},
	\label{eq_s3:p_and_m_isothermal_2}
\end{equation}
where $f$ denotes the free energy density. It is worth mentioning that the pressure $P$ can also be determined by~\cite{Goodstein-85,Fetter-03},
\begin{equation}
	P=Ts-\mathcal{E}+\sum_{i}\mu_in_i.
	\label{eq_s3:p_isothrmal}
\end{equation}
In this case, the~calculation of the entropy per particle is possible by differentiating the free energy density $f$ with respect to the temperature,
\begin{equation}
	S(n,T)=- \frac{\partial \left(f/n\right)}{\partial T}\Bigg\vert_{V,N_i}=-\frac{\partial F}{\partial T}\Bigg\vert_{n}.
	\label{eq_s3:S_isothermal}
\end{equation}
By applying Equation~\eqref{eq_s3:p_and_m_isothermal_2}, the~chemical potentials take the form~\cite{Prakash-94,AA.451.1.2010,Burgio2007}
\begin{eqnarray}
	\mu_n&=&F+u\frac{\partial F}{\partial u}\Bigg\vert_{Y_p,T}-Y_p\frac{\partial F}{\partial Y_p}\Bigg\vert_{n,T}, \\
	\mu_p&=&\mu_n+\frac{\partial F}{\partial Y_p}\Bigg\vert_{n,T}, \\
	\hat{\mu}&=&\mu_n-\mu_p=-\frac{\partial F}{\partial Y_p}\Bigg\vert_{n,T}.
	\label{eq_s3:m_isothermal}
\end{eqnarray}
The free energy $F(n,T,I)$ and the internal energy $E(n,T,I)$, as~well as the entropy, which depends on the latter quantities, can have the following quadratic dependence from the asymmetry parameter~\cite{AA.451.1.2010,Burgio2007,XU2007348,PhysRevC.78.054323,PhysRevC.79.045806}:
\begin{eqnarray}
	F(n,T,I)&=&F(n,T,I=0)+I^2F_{\rm sym}(n,T),
	\label{eq_s3:F_parabolic} \\
	E(n,T,I)&=&E(n,T,I=0)+I^2E_{\rm sym}(n,T),
	\label{eq_s3:E_parabolic} \\ 
	S(n,T,I)&=&S(n,T,I=0)+I^2S_{\rm sym}(n,T),
	\label{eq_s3:S_parabolic}
\end{eqnarray}
as the parabolic approximation (PA) suggests, where
\begin{eqnarray}
	F_{\rm sym}(n,T)&=& F(n,T,I=1)-F(n,T,I=0),
	\label{eq_s3:F_sym_parabolic} \\
	E_{\rm sym}(n,T)&=& E(n,T,I=1)-E(n,T,I=0),
	\label{eq_s3:E_sym_parabolic} \\
	S_{\rm sym}(n,T)&=&S(n,T,I=1)-S(n,T,I=0) \\
	&=& \frac{1}{T}(E_{\rm sym}(n,T)-F_{\rm sym}(n,T)).
	\label{eq_s3:S_sym_parabolic}
\end{eqnarray}
While the above approximation is valid for the specific model, in~general, it is mandatory to check it. For~completeness, the~PA is satisfied in both the internal energy and the free energy, as References~\cite{PhysRevC.78.054323,PhysRevC.79.045806,Burgio2007,AA.451.1.2010,XU2007348} state. However, there are studies~\cite{PhysRevC.93.035806} in which the PA contains uncertainties. In sum, the~PA and its validity strongly depend on the specific character of the selected nuclear~model.

Equation~\eqref{eq_s3:m_isothermal}, which is mandatory for this approximation, can be acquired by substituting Equation~\eqref{eq_s3:F_parabolic} as
\begin{equation}
	\hat{\mu}=\mu_n-\mu_p=4(1-2Y_p)F_{\rm sym}(n,T).
	\label{eq_s3:m_hat}
\end{equation}
This equation is similar to that obtained for cold catalyzed nuclear matter by replacing $E_{\rm sym}(n)$ with $F_{\rm sym}(n,T)$.

\subsection{Lepton's~Contribution}
Stable nuclear matter and a chemical equilibrium state are explicitly connected at high densities for all reactions. Specifically, electron capture and $\beta$ decay take place simultaneously
\begin{equation}
	p +e^{-}\longrightarrow n+ \nu_e \quad \text{and}  \quad n \longrightarrow p+e^{-}+\bar{\nu}_e.
	\label{eq_s3:beta_decay}
\end{equation}
In consequence, a~change in the electron fraction $Y_{e}$ is in order. Considering that the generated neutrinos have left the system, a~significant effect on the EoS is presented by changing the values of proton fraction $Y_{p}$~\cite{10.1143/ptp/92.4.779,10.1143/PTP.95.901}. In~particular, the~absence of neutrinos implies that
\begin{equation}
	\hat{\mu}=\mu_n-\mu_p=\mu_e.
	\label{eq_s3:cp}
\end{equation}
It is has to be mentioned that, in~principle, the~matter contains neutrons, protons, electrons, muons, photos, and~antiparticles, which are in a thermal equilibrium state. However, in~the present study, we consider only the contribution of neutron, protons, and~electrons, as~the contribution of the remaining particles is negligible~\cite{10.1143/ptp/92.4.779}. As~a consequence, the~following relation holds:
\begin{equation}
	\mu_n=\mu_p+\mu_e.
	\label{eq_s3:cps}
\end{equation}
The energy density and pressure of leptons are calculated through the following formulae:
\begin{equation}
	\mathcal{E}_{l}(n_l,T)=\frac{g}{(2\pi)^3}\int \frac{d^{3}k~\sqrt{\hbar^2 k^2 c^2+m_l^2c^4}}{1+\exp\left[\frac{\sqrt{\hbar^2k^2c^2+m_l^2c^4}-\mu_l}{T}\right]},
	\label{eq_s3:e_leptons}
\end{equation}
\begin{eqnarray}
	P_l(n_l,T)&=&\frac{1}{3}\frac{g(\hbar c)^2}{(2\pi)^3}\int \frac{1}{\sqrt{\hbar^2 k^2 c^2+m_l^2c^4}}\\ 
	&\times& \frac{d^{3}k~k^2}{1+\exp\left[\frac{\sqrt{\hbar^2k^2c^2+m_l^2c^4}-\mu_l}{T}\right]}.
	\label{eq_s3:p_leptons}
\end{eqnarray}

The chemical potential of electrons is available through the Equations~\eqref{eq_s3:m_hat} and~\eqref{eq_s3:cps}, as~%
\begin{equation}
	\mu_{e}=\mu_n-\mu_p=4I(n,T)F_{\rm sym}(n,T),
	\label{eq_s3:cp_leptons}
\end{equation}
which is crucial for the calculation of the proton fraction as a function of both the baryon density and the temperature. The~construction of the EoS of hot nuclear matter in the $\beta$-equilibrium state is provided through the calculation of the total energy density $\mathcal{E}_{\rm t}$, as~well as the total pressure $P_{\rm t}$. The~total energy density (with all terms) is given by
\begin{equation}
	\mathcal{E}_{\rm t}(n,T,I)=\mathcal{E}_b(n,T,I)+\sum_{l}\mathcal{E}_l(n,T,I) +\sum_{\bar{l}}\mathcal{E}_{\bar{l}}(n,T,I)+\mathcal{E}_{\gamma}(n,T),
	\label{eq_s3:e_total}
\end{equation}
where $\mathcal{E}_b(n,T,I)$ is the contribution of baryons, $\mathcal{E}_l(n,T,I)$, $\mathcal{E}_{\bar{l}}(n,T,I)$ are the contributions of particles and antiparticles of leptons, and~$\mathcal{E}_{\gamma}(n,T)$ is the contribution of photons. The~total pressure (with all terms) is
\begin{equation}
	P_{\rm t}(n,T,I)=P_b(n,T,I)+\sum_{l}P_l(n,T,I) +\sum_{\bar{l}}P_{\bar{l}}(n,T,I)+P_{\gamma}(T),
	\label{eq_s3:p_total}
\end{equation}
where $P_b(n,T,I)$ is the contribution of baryons,
\begin{equation}
	P_b(n,T,I)= T\sum_{\tau=p,n}s_{\tau}(n,T,I) +\sum_{\tau=n,p}n_{\tau}\mu_{\tau}(n,T,I)-\mathcal{E}_b(n,T,I),
	\label{eq_s3:p_baryon}
\end{equation}
while $P_l(n,T,I)$, $P_{\bar{l}}(n,T,I)$ are the contributions of particles and antiparticles of leptons, and~$P_{\gamma}(T)$ is the contribution of~photons.

\subsection{Isothermal~Configuration}
In the isothermal temperature profile, by~considering that for each value of temperature, the~value of the proton fraction is a well-known function of the baryon density, the~total energy density is evaluated as
\begin{equation}
	\mathcal{E}_{\rm t}(n,T,Y_{p})=\mathcal{E}_b(n,T,Y_{p})+\mathcal{E}_e(n,T,Y_{p}),
	\label{eq_s3:e_total_1}
\end{equation}
where
\begin{equation}
	\mathcal{E}_b(n,T,Y_{p})=nF_{\rm PA} + nTS_{\rm PA},
	\label{eq_s3:e_baryon_1}
\end{equation}
$\mathcal{E}_e(n,T,Y_{p})$ is given by Equation~\eqref{eq_s3:e_leptons}, replacing the leptons with electrons and  $\mu_{e}$ from Equation~\eqref{eq_s3:cp_leptons}, and~$F_{\rm PA}$ and $S_{\rm PA}$ are given by Equations~\eqref{eq_s3:F_parabolic} and~\eqref{eq_s3:S_parabolic}, respectively.
Moreover, the~total pressure is evaluated as
\begin{equation}
	P_{\rm t}(n,T,Y_{p})=P_b(n,T,Y_{p})+P_e(n,T,Y_{p}),
	\label{eq_s3:p_total_1}
\end{equation}
where
\begin{equation}
	P_b(n,T,Y_{p})=n^2\frac{\partial F_{\rm PA}(n,T,Y_{p})}{\partial n}\Bigg\vert_{T,n_i},
	\label{eq_s3:p_baryon_1}
\end{equation}
and  $P_e(n,T,Y_{p})$ is given by Equation~\eqref{eq_s3:p_leptons}, replacing the leptons with the electrons and $\mu_{e}$ from Equation~\eqref{eq_s3:cp_leptons}.

Thus, the~Equations~\eqref{eq_s3:e_total_1} and~\eqref{eq_s3:p_total_1} for the energy density and the pressure, respectively, correspond to the ingredients for the construction of isothermal EoSs of hot nuclear matter in $\beta$ equilibrium~state.

\subsection{Isentropic Configuration and Neutrino~Trapping}
In an isentropic configuration, we assume that the entropy per baryon and lepton fraction are constant in the interior of the neutron star (protoneutron star). Specifically, according to Equation~\eqref{eq_s3:beta_decay}, the~neutrinos are trapped in the interior of the star and the proton fraction is significantly increased. The~relevant chemical equilibrium can be expressed in terms of the chemical potentials for the four species,
\begin{equation}
	\mu_n+\mu_{\nu_e}=\mu_p+\mu_e.
	\label{chem-equil-PN-1}
\end{equation}  
In addition, the~charge neutrality demands the equality between proton and electron fraction, while the total fraction of leptons is equal to $Y_l=Y_e+Y_{\nu_e}$. Henceforth, the~chemical equilibrium is expressed as
\begin{equation}
	\mu_e-\mu_{\nu_e}=\mu_n-\mu_p=4(1-2Y_p)F_{\rm sym}(n,T).
	\label{chem-equil-PN-2}
\end{equation}
In this case too, the~relevant system of equations can provide us with the density and temperature dependence of proton and neutrino fractions, and~their chemical potentials, by~assuming a constant entropy per baryon. Nonetheless, we applied the approximation of $Y_p\simeq 2/3 Y_l+0.05$ ($3 \%$ accuracy) introduced by Takatsuka~et~al.~\cite{10.1143/ptp/92.4.779}, in~order to avoid computational complications. The~ingredients for the construction of isentropic EoSs are given by the Equations~\eqref{eq_s3:e_total} and ~\eqref{eq_s3:p_total}.

\subsection{Construction of the Hot~EoSs}
The construction of the EoS for the interior of neutron stars at finite temperature and entropy per baryon is based on the MDI model and data provided by Akmal~et~al.~\cite{Akmal-1998}. In~particular, we utilized the APR-1 EoS data from Akmal~et~al.~\cite{Akmal-1998} for the energy per particle of SNM and PNM in the density range $[0.04,0.96]~{\rm fm^{-3}}$. The~process leads to the evaluation of the coefficients for the symmetric and asymmetric nuclear matter, and~finally, to~the construction of the EoS, hereafter MDI+APR1.

In the case of the isothermal temperature profile, we have constructed 10 EoSs in the temperature range $[1,60]~{\rm MeV}$, while in the isentropic case, we have constructed nine EoSs in entropy per baryon and lepton fraction ranges $[1,3]~k_{B}$ and $[0.2,0.4]$, respectively. For~the crust region of the finite temperature cases and the low-density region $(n_{b}\leq 0.08~{\rm fm^{-3}})$, as~well as the finite entropies per baryon and lepton fractions, the~EoSs of Lattimer and Swesty~\cite{LATTIMER1991331} and the specific model corresponding to the incomprehensibility modulus at the saturation density of SNM $K_{s}=220~{\rm MeV}$ are used \linebreak (\url{https://www.stellarcollapse.org}, accessed on 4 March 2020).

\subsection{Rapidly Rotating Hot Neutron~Stars}
The stability of hot neutron stars is acquired via a specific version of the secular instability criterion of Friedman~et~al.~\cite{ApJ.325.722}, which follows Theorem I of Sorkin~\cite{ApJ.257.847}. In~a continuous sequence of equilibria at a fixed baryon number $N_{\rm bar}$ and total entropy of the neutron star $S_{\rm t}^{\rm ns}$, the~extremal point of the stability loss is found when~\cite{ApJ.321.822}
\begin{equation}
	\frac{\partial J}{\partial n_{b}^{c}}\Bigg\vert_{N_{\rm bar},S_{\rm t}^{\rm ns}} = 0,
\end{equation}
\textls[-15]{where $J$ and $n_{b}^{c}$ are the angular momentum and central baryon density of the star, respectively.}

Furthermore, in~a sequence, the turning-point appears in the case where three out of four following derivatives vanish,
\begin{equation}
	\frac{\partial M_{\rm gr}}{\partial n_{b}^{c}}, \quad \frac{\partial M_{b}}{\partial n_{b}^{c}}, \quad \frac{\partial J}{\partial n_{b}^{c}}, \quad \text{and} \quad \frac{\partial S_{\rm t}^{\rm ns}}{\partial n_{b}^{c}},
\end{equation}
with $M_{\rm gr}$ and $M_{\rm b}$ denoting the gravitational and baryon mass \cite{Kaplan_2014,PhysRevC.96.045806}. In~addition, the~turning-point theorem shows that at this point, the~fourth derivative also vanishes, and~the sequence has transitioned from stable to~unstable.

It has to be mentioned that the criterion for secularly stable/unstable configurations is essential only for constant entropy per baryon or temperature~\cite{PhysRevC.96.045806}. In~this review, the~entropy per baryon and the temperature in~each case are constant throughout the neutron star. Therefore, the~remaining criteria vanish at the maximum mass configuration (the last stable point). Furthermore, we considered that in the rotating configuration, the~maximum mass and the maximum angular velocity coincide, which generally is not the case~\cite{friedman_stergioulas_2013}. However, the~existing difference is very small, and~it could not be detected within the precision of our calculations~\cite{ApJ.321.822}.

The numerical integration of the equilibrium equations was conducted under the publicly available numerical code \emph{nrotstar} from the C++ Lorene/Nrotstar library~\cite{lorene}.

\section{Results and~Discussion} \label{sec:4}
\subsection{Speed of Sound and Tidal~Deformability}
In our study, we used two cases for the value of speed of sound, the~lower bound of $(v_s/c)^2=1/3$ and the upper one of $(v_s/c)^2=1$, and~four transition densities \linebreak $n_{{\rm tr}}=\{1,1.5,2,3\} n_s$~\cite{Kanakis-2020}.

\begin{figure}
	\centering
	\includegraphics[width=0.48\textwidth]{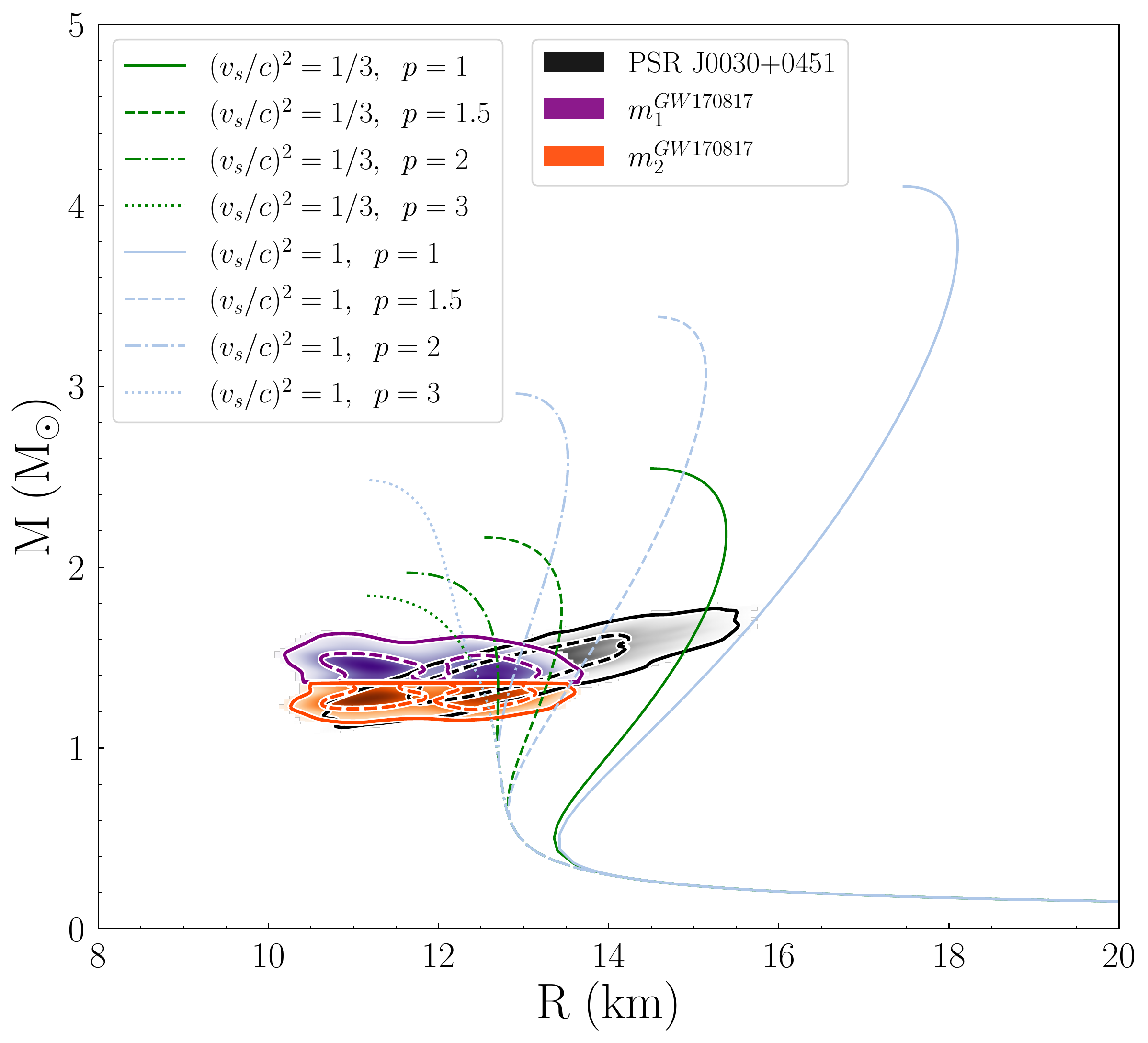}
	\caption{Mass vs. radius for an isolated neutron star and~for the two cases of speed of sound bounds. The~blue (green)
		lines correspond to the upper (lower) bound. The~black
		diagonal shaded region corresponds to NICER’s observation
		(data taken from Reference~\cite{Miller-2019}), while the purple upper
		(orange lower) shaded region corresponds to the higher
		(smaller) component of GW170817 event (data taken from
		Reference~\cite{Abbott-2018}). The~solid (dashed) contour lines describe the
		90\% (50\%) confidence interval.}
	\label{mass-radius2}
\end{figure}

In Figure~\ref{mass-radius2}, we display the corresponding mass-radius (M-R) diagram, which we obtained from the numerical solution to the TOV system of equations. The~green colored lines correspond to the $(v_s/c)^2=1/3$ limit, while the blue ones correspond to the \mbox{$(v_s/c)^2=1$} limit. One can observe that each transition density leads to bifurcations in the M-R diagram. Between~the same kind of linestyle, the~lower and upper bounds, $(v_s/c)^2=1/3$  and $(v_s/c)^2=1$ of speed of sound correspond to lower  and higher masses, respectively. In~general, the~higher the transition density, the~softer the EoS, with~the lower bound of $(v_s/c)^2=1/3$ leading to a more soft EoS compared to the $(v_s/c)^2=1$ case. In~addition, the~estimation of the GW170817 event and the NICER's data are also displayed~\cite{Abbott-2018,Miller-2019}. We notice that while there is an accordance between the two observations, the~GW170817 event (from the gravitational-waves perspective) is more informative for our study than the NICER's detection, as~it restricts the cases leading to the exclusion of EoS at least with transition density $n_{{\rm tr}}=n_s$, for~both bounds of speed of~sound.

Our study takes into consideration the observation of binary neutron stars mergers from the gravitational waves detectors. Therefore, we used the measured upper limit of the effective tidal deformability $\tilde{\Lambda}$ provided by the events GW170817 and GW190425~\cite{Abbott-2018,Abbott-3,Abbott_2020}. The~chirp masses for the two events are $\mathcal{M}_c=1.186\;M_\odot$~\cite{Abbott-2017} and  $\mathcal{M}_c=1.44\;M_\odot$~\cite{Abbott_2020}, respectively. The~component masses vary in the ranges  $m_1\in(1.36,1.60)\;M_\odot$ and  $m_2\in(1.16,1.36)\;M_\odot$~\cite{Abbott-3} (GW170817) and $m_1\in(1.654,1.894)\;M_\odot$ and $m_2\in(1.45,1.654)\;M_\odot$ (GW190425). We notice that we modified the range of the component masses (especially in the second event) to have an equal mass boundary, i.e.,~$q\leq1$.

\begin{figure*}
	\includegraphics[width=0.5\textwidth]{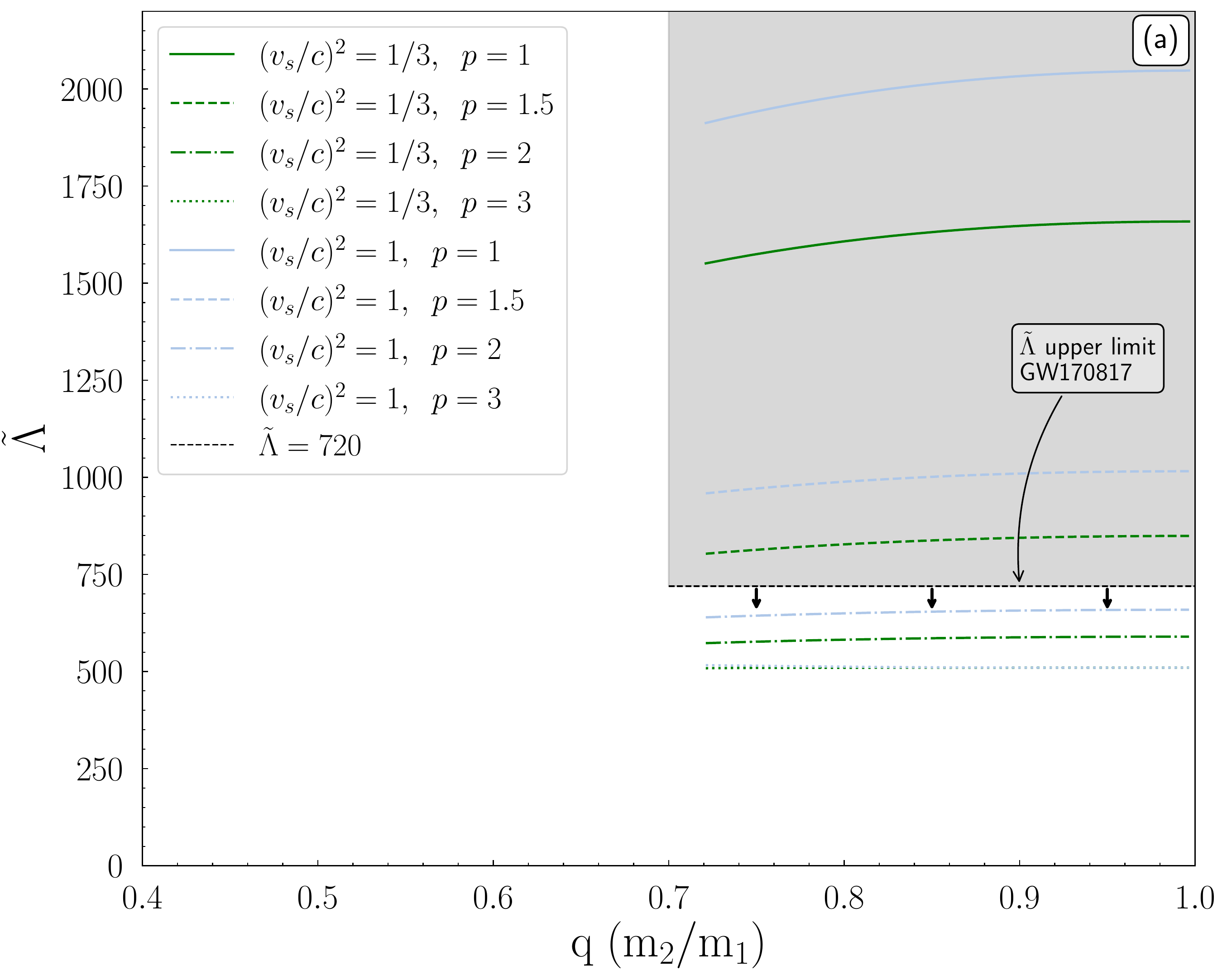}~
	\includegraphics[width=0.5\textwidth]{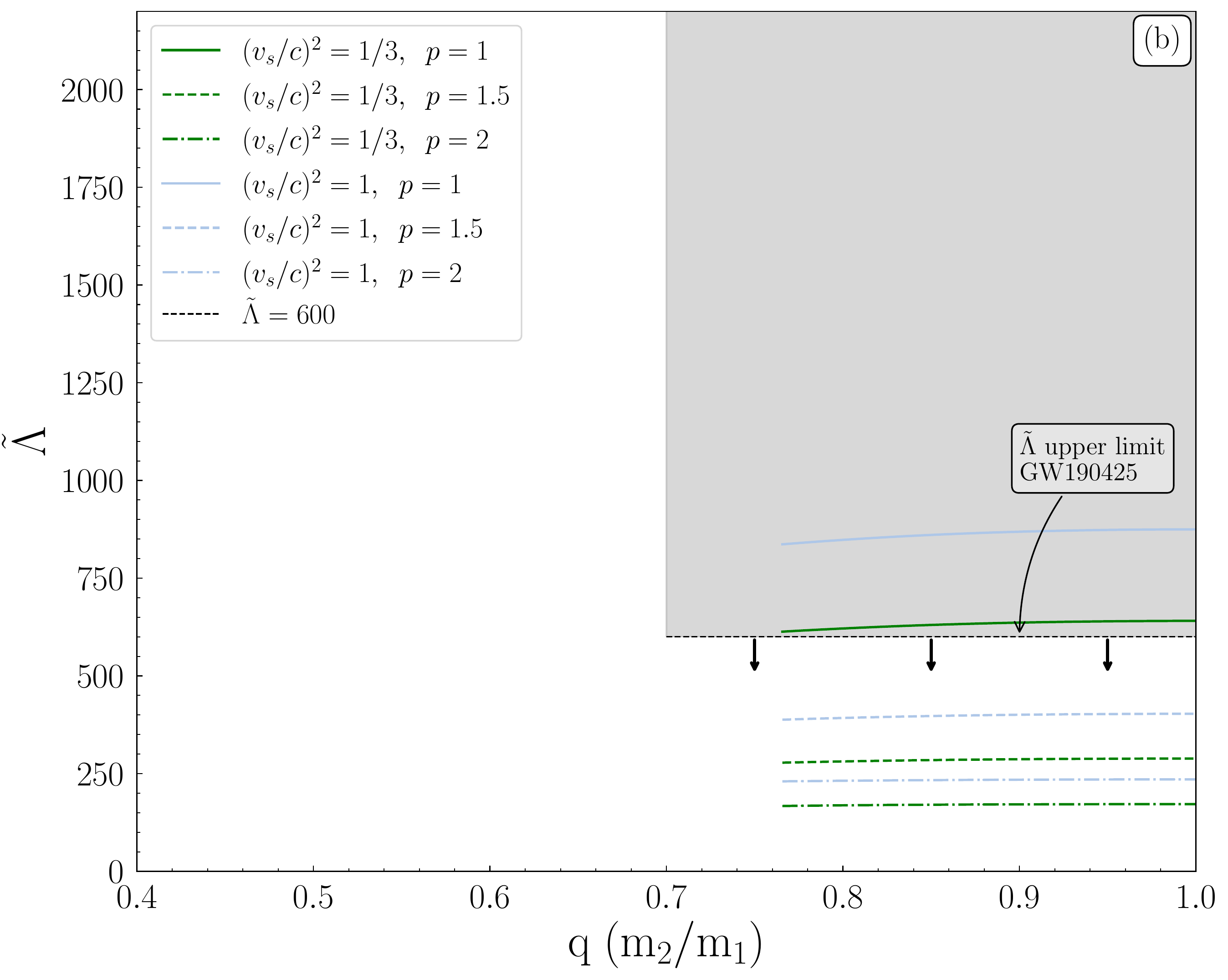}
	\caption{The effective tidal deformability $\tilde{\Lambda}$ as a function of the binary mass ratio $q$ for the event (a) GW170817 and (b)~GW190425. The~measured upper limits for $\tilde{\Lambda}$ are also indicated, with~the grey shaded region corresponding to the excluded area. The~green (blue) curves correspond to the $(v_s/c)^2=1/3$ ($(v_s/c)^2=1$) case.}
	\label{Ltildeq1}
\end{figure*}

\begin{figure*}
	\includegraphics[width=0.5\textwidth]{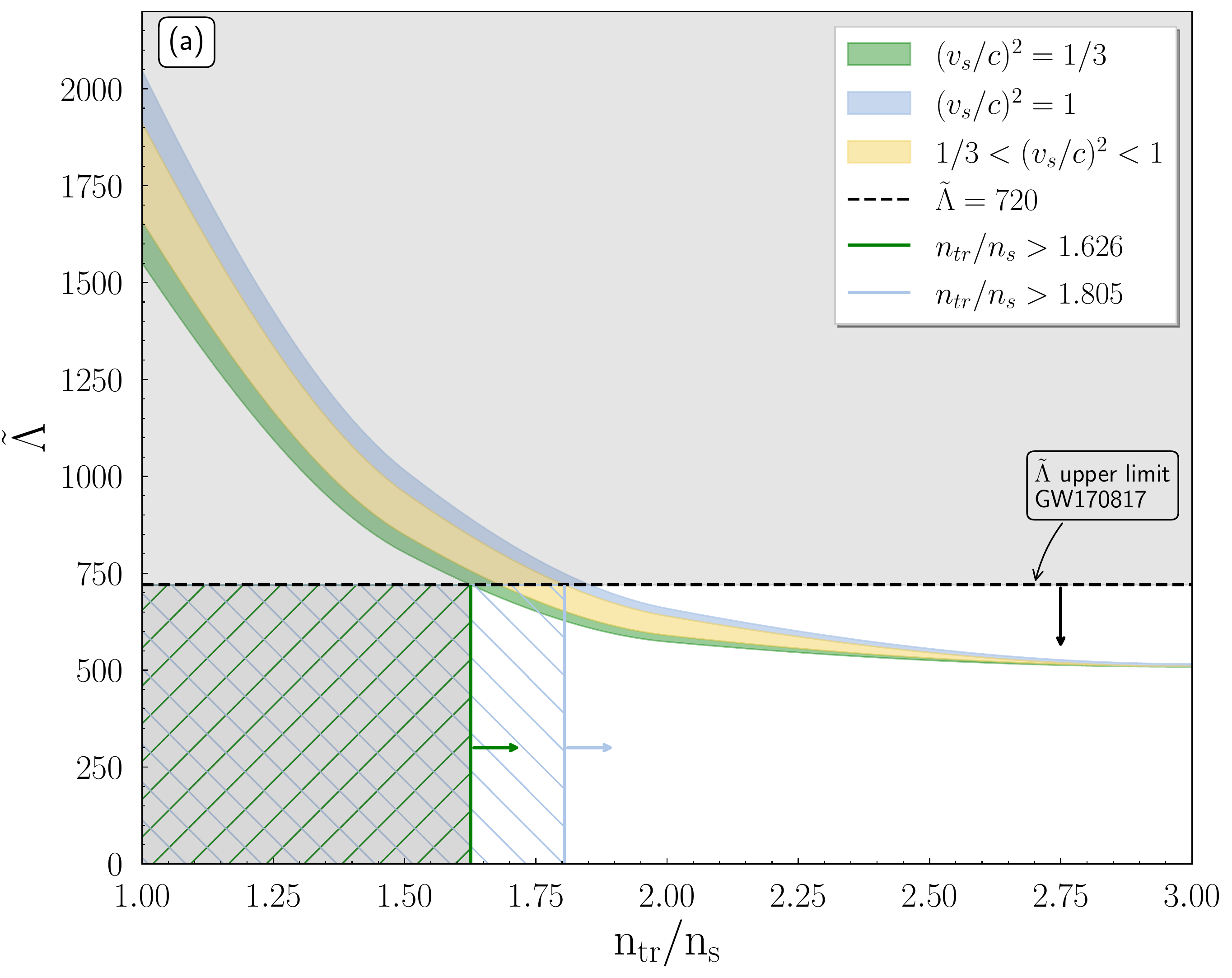}~
	\includegraphics[width=0.5\textwidth]{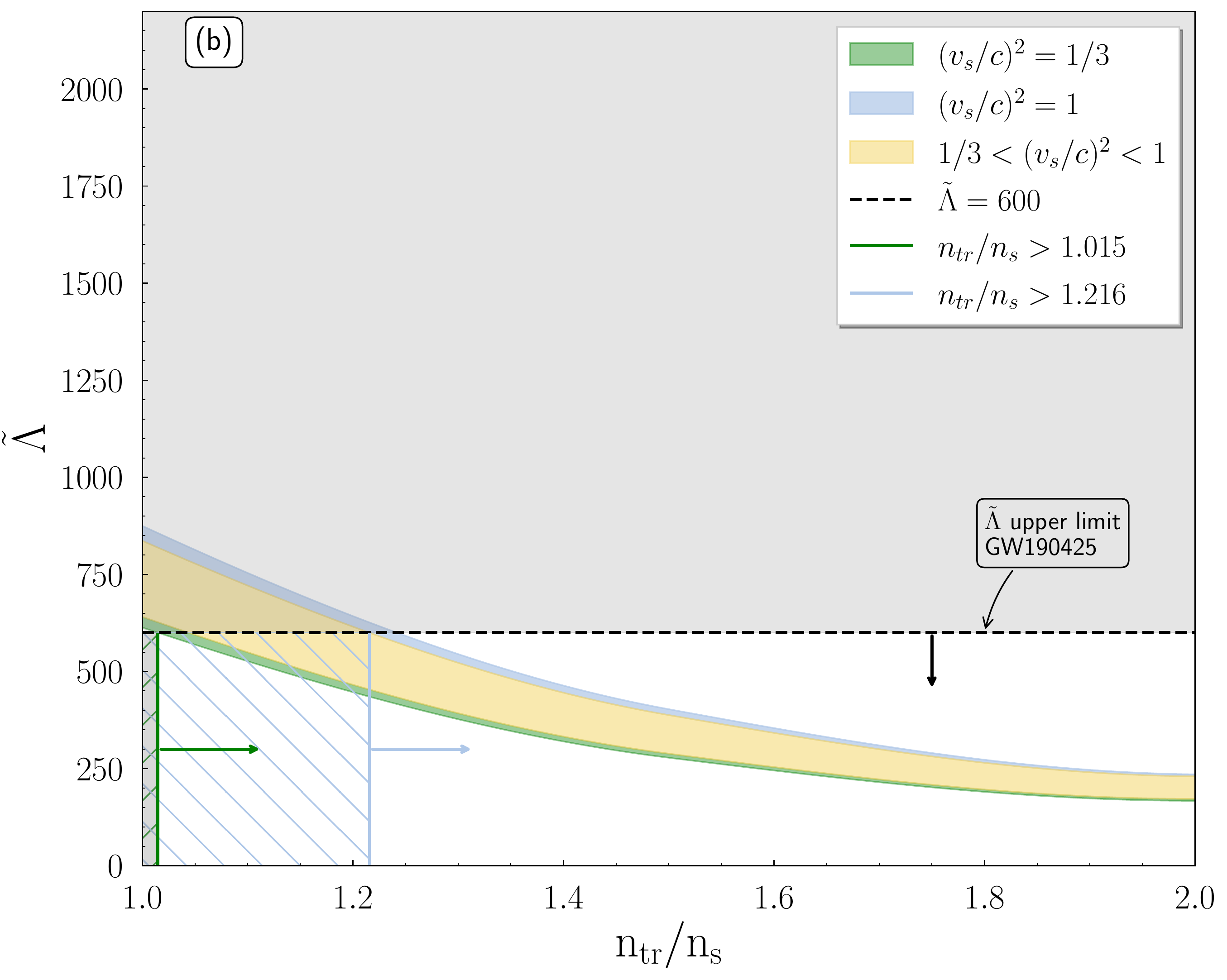}
	\caption{ $\tilde{\Lambda}$  as a function of the transition density $n_{{\rm tr}}$ (in units of saturation density $n_s$) at the maximum mass configuration for the two  speed of sound bounds $v_s=c/\sqrt{3}$ and  $v_s=c$ and for the events (a) GW170817 and (b)~GW190425. The~measured upper limits for $\tilde{\Lambda}$~\cite{Abbott-3,Abbott_2020}  as well as the corresponding  lower values of transition density are also indicated for both events.  The~green (blue) arrow marks the accepted region of transition density for the $v_s=c/\sqrt{3}$ ($v_s=c$) case. The~green lower (blue upper) curved shaded region corresponds to the $v_s=c/\sqrt{3}$ ($v_s=c$) limit. The~yellow shaded area indicates the region between the two cases of bounds of the speed of sound.}
	\label{L-1}
\end{figure*}

In Figure~\ref{Ltildeq1}, we display the effective tidal deformability $\tilde{\Lambda}$ as a function of $q$ for both events. In~Figure~\ref{Ltildeq1}a, we observe that the upper limit on $\tilde{\Lambda}$ (derived from the GW170817 event) leads to the exclusion of both cases of speed of sound with transition density $n_{{\rm tr}}=1,1.5\;n_s$. By~comparing to Figure~\ref{mass-radius2}, the~constraints on the upper limit of $\tilde{\Lambda}$ in Figure \ref{Ltildeq1} make more clear which cases  must be excluded. For~the second event in Figure~\ref{Ltildeq1}b, we observe that all the EoSs are shifted to lower values of $\tilde{\Lambda}$, compared to the GW170817 event. This is because of the higher value of chirp mass in the second event (GW190425). Contrary to the GW170817 event, the~upper limit on $\tilde{\Lambda}$, provided by GW190425 event, excludes only these EoSs with transition density $n_{{\rm tr}}=n_s$, for~both bounds of the speed of sound. In~general, for~both events, the~EoSs corresponding to higher values of transition density $n_{{\rm tr}}$ lead to smaller values of $\tilde{\Lambda}$. Therefore, the~measured upper limits of $\tilde{\Lambda}$ favor softer EoSs. We have to notice that for the GW190425 event, we did not take into consideration the cases with transition density $n_{{\rm tr}}=3\;n_s$ because the EoS with $(v_s/c)^2=1/3$ and $n_{{\rm tr}}=3n_s$ cannot reproduce the masses of this~event. 

Beyond the useful constraints that we obtained so far by the study of the EoSs through the observed upper limit of $\tilde{\Lambda}$, a~more direct connection between this quantity and the speed of sound bounds is still needed. This idea, which lies at the very heart of our study, was the main motive. Such a direct relation between the referred  quantities can be accomplished if we treat the variation of $\tilde{\Lambda}$ in Figure~\ref{Ltildeq1} as a function of the transition density $n_{{\rm tr}}$, i.e.,~the $\tilde{\Lambda}^{(1/3,1)}_{{\rm min}}-n_{{\rm tr}}$ and $\tilde{\Lambda}^{(1/3,1)}_{{\rm max}}-n_{{\rm tr}}$ relations.

In Figure~\ref{L-1}, we display the relation between the effective tidal deformability $\tilde{\Lambda}$ and the transition density $n_{{\rm tr}}$ at the maximum mass configuration for the two bounds of the speed
of sound, $v_s=c/\sqrt{3}$ and  $v_s=c$, and~the two events GW170817 (Figure~\ref{L-1}a)  and GW190425 (Figure~\ref{L-1}b). The~corresponding upper measured limits for $\tilde{\Lambda}$, as~well as the compatible  lower transition density values, are also indicated. The~predictions on the bound of the speed of sound which are considered  between the two referred limits correspond to the middle~region. 

The main remarks from the observation of Figure~\ref{L-1} are the~following
\begin{enumerate} 
	\item The overall thickness decreases as the transition density $n_{{\rm tr}}$ reaches higher values. This behavior can be explained by  the variation of the radius $M(R)$ presented in the M-R diagram (see Figure~\ref{mass-radius2}).
	\item The thickness of each shaded region decreases as the $n_{{\rm tr}}$ reaches higher values.
	\item The shaded areas are shifted downwards in the GW190425 event, compared with the GW170817 event. This behavior is due to the increase in the component masses. A~similar behavior was observed in Figure~\ref{Ltildeq1}b, compared to  Figure~\ref{Ltildeq1}a.
\end{enumerate}

According to our findings, for~the GW170817 event, the lower limit for the transition density is $1.626 n_{s}$  for $v_s=c/\sqrt{3}$   and  $1.805 n_{s}$ for $v_s=c$. In~the case of the second event, GW190425, the~corresponding limits are $1.015 n_{s}$  for $v_s=c/\sqrt{3}$  and  $1.216 n_{s}$ for $v_s=c$. Therefore, the~first event imposes more stringent constraints on the EoS. In~particular, the~value of the speed of sound must be lower than $v_s=c/\sqrt{3}$, at~least  up to  density $1.626 n_{s}$ (so the EoS is soft enough to predict the tidal deformability). Furthermore, the~EoS must remain casual at least up to density  $1.805 n_{s}$. In~addition, according to the Fermi liquid theory (FLT), the speed of sound must be $v_{s,FLT}^2\leq0.163c^2$ for $n=1.5n_s$~\cite{Greif-2019}, meaning that the EoS cannot exceed this value for $n\leq1.5n_s$, which is in agreement with our finding of the lower limit $n_{{\rm tr}}=1.626n_s$ for the case of $v_s=c/\sqrt{3}$.

We notice that so far we used the upper limit on $\tilde{\Lambda}$ to impose stringent constraints on the $n_{{\rm tr}}$. The~existence of a lower limit on $\tilde{\Lambda}$ could provide further information. Indeed, for~the GW170817 event, such a lower limit is provided both by the gravitational wave data~\cite{Abbott-2018,Abbott-3} and the electromagnetic (EM) counterpart of the merger~\cite{Radice-1,Tews-2018b,Radice-2,Kiuchi-2019,Coughlin-2018,Coughlin-2019}. Most~et~al.~\cite{Most-2018} used the bound of Reference~\cite{Radice-1} and demonstrated its significance in order to further constrain  the tidal deformability $\tilde{\Lambda}_{1.4}$ and the radius $R_{1.4}$ of a $M=1.4\;M_\odot$ neutron star. For~our case of interest and especially for the GW170817 event, a~lower limit on $\tilde{\Lambda}$ similar to the proposed values, could not provide any further constraint, even if we consider the more optimistic boundary of $\tilde{\Lambda}\geq400$.

On the contrary, for~the second event GW190425, its higher component masses lead to smaller values of $\tilde{\Lambda}$. Hence, there is an inability for the upper limit of $\tilde{\Lambda}$ to provide further constraints. We speculate that the existence of a lower limit on $\tilde{\Lambda}$ would be able to provide constraints, especially leading to an upper limit for $n_{{\rm tr}}$. Hence, binary neutron stars coalescences with heavy masses would be helpful to constrain the upper limit of $n_{{\rm tr}}$ via the lower limit of $\tilde{\Lambda}$ as provided by the EM counterpart. Unfortunately, an~EM counterpart for the GW190425 event was not detected~\cite{Abbott_2020,Kilpatrick}. 

Furthermore, we provide in Figure~\ref{L-1} an expression for the $\tilde{\Lambda}_{{\rm min}}^{(1/3)}$ and $\tilde{\Lambda}_{{\rm min}}^{(1)}$ boundary curves of the green (lower) and blue (upper) shaded regions, respectively. This expression gives the exact value of the lower limit on $n_{{\rm tr}}^{(1/3)}$  and $n_{{\rm tr}}^{(1)}$, respectively.  The~expression is given by the following equation, and~the coefficients on Table \ref{tab:table1},
\begin{equation}
	\tilde{\Lambda}=c_1\coth\Bigg[c_2\bigg(\frac{n_{{\rm tr}}}{n_s}\bigg)^{2}\Bigg].
	\label{fit1}
\end{equation}

As one can observe from Figure~\ref{mass-radius2}, the~highest mass is provided by the stiffest EoSs, i.e.,~the higher value of speed of sound. Therefore, the~behavior of the maximum mass $M_{{\rm max}}$ and the speed of sound $v_s^2$ has to be studied~further.

\begin{figure*}
	\includegraphics[width=0.5\textwidth]{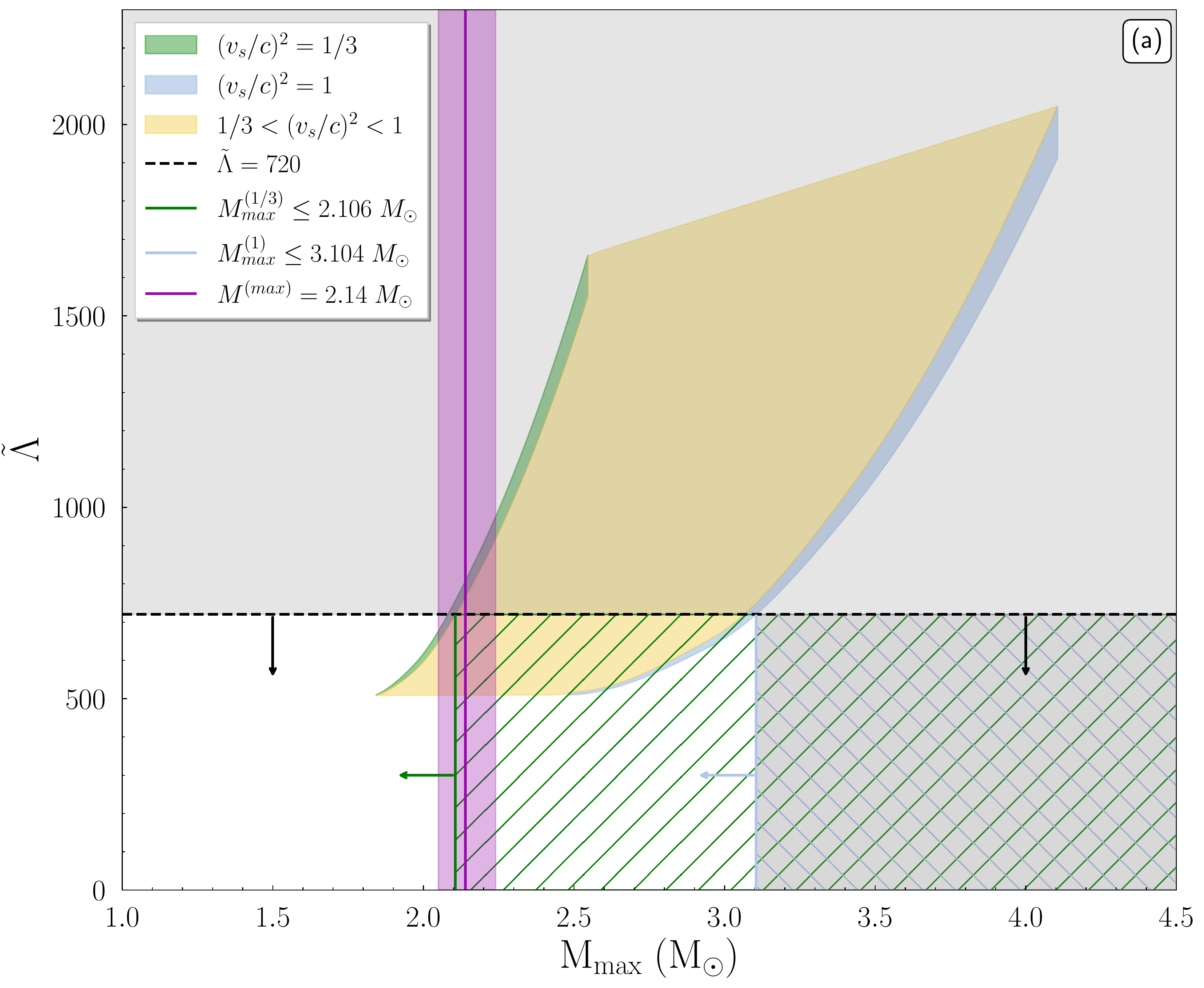}~
	\includegraphics[width=0.5\textwidth]{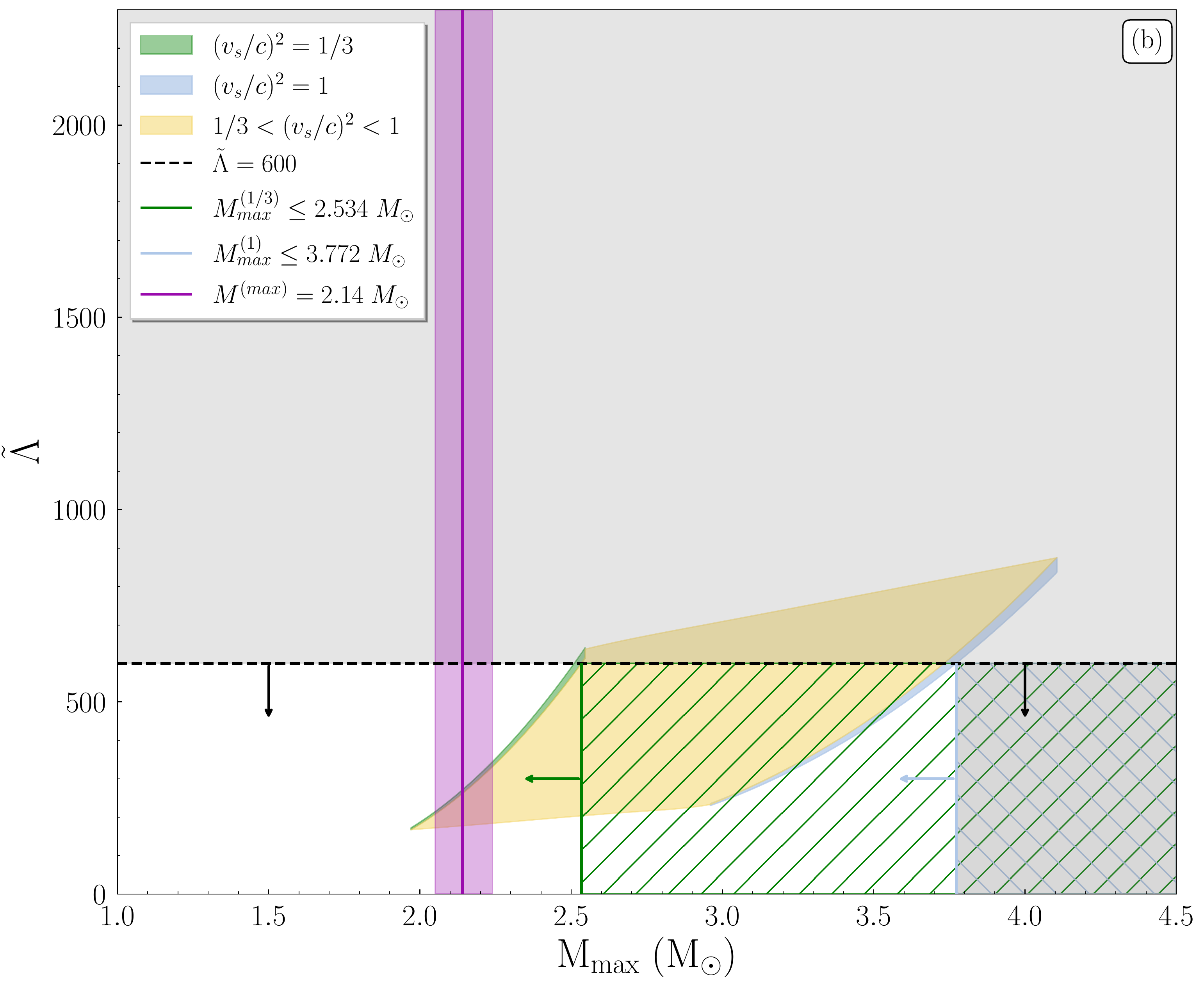}
	\caption{The effective tidal deformability $\tilde{\Lambda}$ as a function of the maximum mass for the two  speed of sound bounds $v_s=c/\sqrt{3}$ and  $v_s=c$ and for the events (a) GW170817 and (b) GW190425. The~measured upper limits for $\tilde{\Lambda}$ (black dashed lines with arrows; see References~\cite{Abbott-3,Abbott_2020}); the~corresponding maximum mass shaded regions, for~the $v_s=c/\sqrt{3}$ (left green) case, for the $v_s=c$ case (right blue), and for the middle cases (yellow); and the current observed maximum neutron star mass  $M=2.14_{-0.09}^{+0.10}\;M_{\odot}$ (purple shaded vertical area; see Reference~\cite{Cromartie-2019}) are also displayed. The~green left (blue right) arrow marks the accepted region of maximum mass $M_{{\rm max}}$ for $v_s=c/\sqrt{3}$ ($v_s=c$) case.}
	\label{L-2}
\end{figure*}

\textls[-15]{In Figure~\ref{L-2}, the behavior of the effective tidal deformability $\tilde{\Lambda}$ as a function of the maximum mass for the two speed of sound bounds and for both events is displayed. The~corresponding upper observational limit for $\tilde{\Lambda}$ (black dashed horizontal line), the~compatible maximum mass in each case (horizontal arrows), and the current observed maximum neutron star mass  $M=2.14_{-0.09}^{+0.10}\ M_{\odot}$ (vertical purple shaded region) are also~indicated.}

\begin{table}
	\caption{Parameters of the Equations~\eqref{fit1} and~\eqref{fit2} for both events and all bounds of the speed of sound.}
	\begin{ruledtabular}
	\begin{tabular}{ccccccccc}
		\multirow{2}{*}{Bounds} & \multicolumn{4}{c}{GW170817} & \multicolumn{4}{c}{GW190425} \\
		& $c_{1}$ & $c_{2}$  & $c_{3}$ & $c_{4}$ & $c_{1}$ & $c_{2}$ & $c_{3}$ & $c_{4}$ \\
		\hline
		$c$ & $500.835$ & $0.258$ & $53.457$ & $0.873$ & $47.821$ & $0.055$ &  $10.651$ & $1.068$ \\
		$c/\sqrt{3}$ & $503.115$ & $0.325$ & $38.991$ & $1.493$  & $43.195$ & $0.069$  & $5.024$ & $1.950$
	\end{tabular}
	\end{ruledtabular}
	\label{tab:table1}
\end{table}

At first glance in Figure~\ref{L-2}, there is a contradiction between the maximum mass and the upper limit of the observed $\tilde{\Lambda}$. For~the first event shown in Figure~\ref{L-2}a, the~upper limit of $\tilde{\Lambda}$ is compatible with a maximum mass value $2.106 M_{\odot}$ for $v_s=c/\sqrt{3}$ and  $3.104 M_{\odot}$ for $v_s=c$. Nonetheless, this bound corresponds to transition density in approximation $1.5 n_{s}$. Experimental evidence disfavors this value. Therefore, the~simultaneous derivation of the maximum mass combined with the experimental knowledge that the EoS cannot take this bound of sound speed for $n_{{\rm tr}}=1.5n_s$ are in contradiction. Furthermore, the~upper limit on $M_{{\rm max}}$ for the case of $(v_s/c)^2=1/3$ lies roughly inside the estimation of the measured maximum mass. In~the general perspective, we notice that two different points of view antagonize each other. The~constraints derived by the upper limit on $\tilde{\Lambda}$ lead to softer EoSs, contrary to the observational estimations of the maximum mass of neutron stars, which lead to stiffer EoSs. As~we move to higher values of the speed of sound, this contrast decreases, with~the causal scenario of $v_s=c$ leading to a very wide area for the maximum~mass.

In the case of the second event GW190425 displayed in Figure~\ref{L-2}b, the~constraints provided by the measured $\tilde{\Lambda}$ are less stringent than the GW170817 event, with~a maximum mass value of $M_{{\rm max}}\leq2.534\;M_\odot$ for $v_s=c/\sqrt{3}$ and $M_{{\rm max}}\leq3.772\;M_\odot$ for $v_s=c$. HOwever, the~presence of a lower limit on $\tilde{\Lambda}$, especially in the case of events with heavy components (such as GW190425), could constrain the lower maximum~mass.

In addition, we provide an expression that describes the $\tilde{\Lambda}$ as a function of the maximum mass $M_{{\rm max}}$. The~expression is given by the following equation and the coefficients in Table \ref{tab:table1},
\begin{equation}
	\tilde{\Lambda}=c_3\bigg(e^{M_{{\rm max}}}-1\bigg)^{c_4}.
	\label{fit2}
\end{equation}

The expression in this form means that $M_{{\rm max}}\to0\Rightarrow\tilde{\Lambda}\to0$. Moreover, the~adoption of an upper limit on the maximum mass $M_{{\rm max}}$ in Figure~\ref{L-2} could provide an additional constraint on the behavior of the speed of sound. Specifically, by~applying the estimated upper limit $M_{{\rm max}}\leq2.33\;M_\odot$~\cite{Rezzolla-2018}, the~case of $(v_s/c)^2=1$ in Figure~\ref{L-2}a for the GW170817 event should be excluded. On~the contrary, the~estimated upper limit $M_{{\rm max}}\leq2.106\;M_\odot$ for the $(v_s/c)^2=1/3$ bound, is a more tight constraint. Additionally, an~upper limit such as $M_{{\rm max}}\leq2.33\;M_\odot$ imposes a general upper bound on the possible intermediate values of speed of sound (intermediate shaded area in the figure). Concerning the second event in Figure~\ref{L-2}b, a~strict upper limit on $M_{{\rm max}}$ could constrain even the $(v_s/c)^2=1/3$ case.

\begin{figure}
	\centering
	\includegraphics[width=0.48\textwidth]{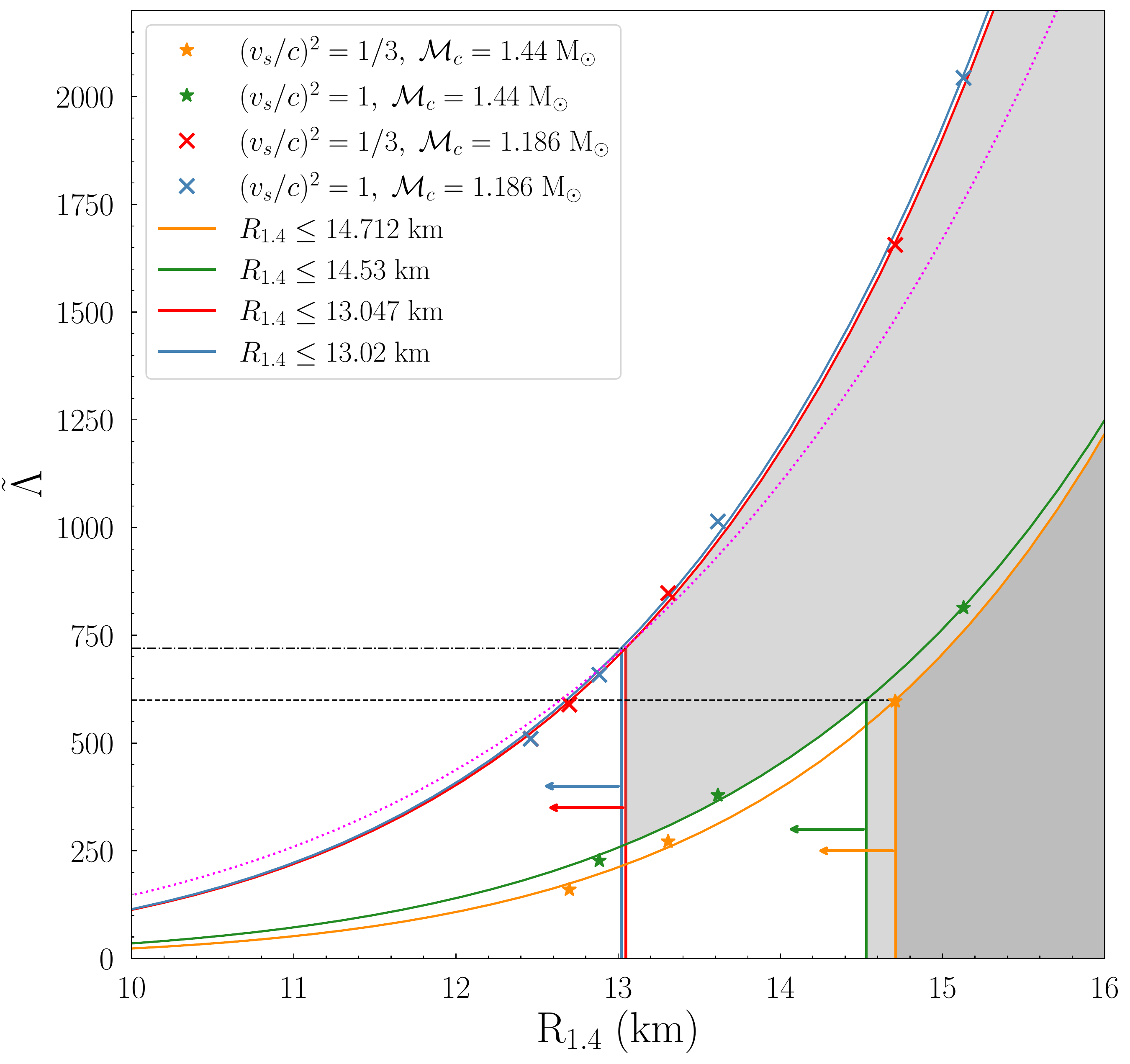}
	\caption{The effective tidal deformability $\tilde{\Lambda}$ as a function of $R_{1.4}$ for both events and all the bounds of the speed of sound. The~dashed  and dash-dotted horizontal black lines correspond to the upper limit on $\tilde{\Lambda}$ for the GW190425  and GW170817 events, respectively, taken from References~\cite{Abbott-3,Abbott_2020}. The~grey shaded regions correspond to the excluded areas. The~horizontal arrows indicate the allowed area for $R_{1.4}$ in each case. The~purple dotted curve demonstrates the proposed expression by Reference~\cite{Zhao,SoumiDe}.}
	\label{LR}
\end{figure}

Another interesting relation is the $\tilde{\Lambda}$ as a function of the radius of a $1.4\;M_\odot$ neutron star, for~both events, which is displayed in Figure~\ref{LR}. First of all, the~upper limit on $\tilde{\Lambda}$ leads to a limitation on the maximum values of the radius, especially in the case of $v_s=c/\sqrt{3}$. Furthermore, there is a trend between $\tilde{\Lambda}$ and $R_{1.4}$, which was also remarked on by Raithel~et~al.~\cite{Raithel-2018}, mentioning that the effective tidal deformability depends strongly on the radii of the stars rather on the component masses. This strong dependence can be observed in Figure~\ref{LR}. 

In particular, for~the GW170817 event, the curves of the two limited cases, red  and blue for the $(v_s/c)^2=1/3$  and $(v_s/c)^2=1$ bounds, respectively, of the speed of sound are almost identical. The~cross marks correspond to the specific values for each case. In our study, we considered four cases of transition density $n_{{\rm tr}}$; therefore, eight marks are expected in total, but~in the diagram, only seven can be seen. This is because of the identical values for the $n_{{\rm tr}}=3n_s$ case that the two bounds provide. This is clear from their behavior in Figure~\ref{mass-radius2} in which, for~the mass range of GW170817 event, their M-R curves are~identical. 

Moreover, as~the effective tidal deformability $\tilde{\Lambda}$  shows higher values, the~distance between them also increases. The~same behavior is present in the curves of  Figure~\ref{Ltildeq1}a, in~which their in-between distance increases for higher values of $\tilde{\Lambda}$. This increment is related to the $n_{{\rm tr}}$, meaning that the differentiation for small values of $n_{{\rm tr}}$ is more obvious. Hence, in~these cases, the effect of each bound of the speed of sound is easier to be~manifested.

The dotted purple line corresponds the approximate relation of References ~\cite{Zhao,SoumiDe}. We notice that this approximate relation is valid only for the first event and for specific assumptions on the components' radii. In particular, the~main assumption of this approximation consists on the $R_1\approx R_2$  relation. From~the comparison of Figure~\ref{LR} with the M-R diagram of Figure~\ref{mass-radius2}, it follows that for smaller values of $n_{{\rm tr}}$ (i.e.,  more stiff EoSs), (a) the inclination of the curves increases and (b) the difference between the M-R curves of boundary cases  also increases. Therefore, these remarks, combined with the strong dependence of $\Lambda_i$ to $R$ (see Equation~(\ref{Lamb-1})), show the~inability of the proposed expression to accurately reproduce  the values of $\tilde{\Lambda}$ in the high-values region can be~explained.

The grey shaded area indicates the excluded area due to the upper limit on $\tilde{\Lambda}$, provided by Reference~\cite{Abbott-3}. The~upper limit of $\tilde{\Lambda}$ leads to constraints on the radius $R_{1.4}$, especially $R_{1.4}\leq 13.047~{\rm km}$ for the $(v_s/c)^2=1/3$ bound and $R_{1.4}\leq 13.02~{\rm km}$ for the $(v_s/c)^2=1$ bound. These upper limits are consistent with other analyses~\cite{Abbott-2018,SoumiDe,Coughlin-2019,Tews-2018b,Most-2018,Raithel-2018,Annala-2018,Fasano-2019,Fattoyev-2018}.

For the second event (GW190425), we notice that the exact range of the component masses is not determinant~\cite{Kilpatrick}. The~orange and green lines and marks correspond to the $(v_s/c)^2=1/3$ and $(v_s/c)^2=1$ bounds, respectively, of the speed of sound. The~shaded grey region indicates the excluded region by the upper limit of $\tilde{\Lambda}$ \cite{Abbott_2020}. The~orange and green arrows indicate the allowed region for each case. For~$(v_s/c)^2=1/3$, the constraint on the radius is $R_{1.4}\leq 14.712~{\rm km}$, while for $(v_s/c)^2=1$ is $R_{1.4}\leq 14.53~{\rm km}$. These are more stringent constraints compared to the $15~{\rm km}$ and $16~{\rm km}$ of Reference~\cite{Abbott_2020}. We notice that recently it was found in Reference~\cite{Landry-2020} that the joint contribution of gravitational waves and NICER data favors the violation of the conformal limit $(v_s/c)^2<1/3$. In particular, this analysis suggests the violation of the conformal limit around  $~4\rho_{{\rm nuc}}$ density, where $\rho_{{\rm nuc}}=2.8\times 10^{14} {\rm g/cm^3}$ is the nuclear saturation~density.

In addition, one can observe the similarity of the curves' behavior between the two events. For~higher values of $n_{{\rm tr}}$, the distance between the points grows. One of the main differences is that for the second event, the~curves and the points are shifted to smaller values of $\tilde{\Lambda}$ because of the higher chirp mass $\mathcal{M}_c$ of the system. Another observation is that the fitting lines are more distinct from each other, contrary to the GW170817 event, in which they were almost identical; nevertheless, there is a common trend (see also Reference~\cite{Raithel-2018}). For~this reason, we applied the following expression
\begin{equation}
	\tilde{\Lambda}=c_1R_{1.4}^{c_2},
	\label{fitR14soundspeed}
\end{equation}
where $R_{1.4}$ is in ${\rm km}$, similar to the proposed relations of References~\cite{Zhao,SoumiDe}. The~coefficients for each case are given in Table~\ref{tab:table3}.

\begin{table}
	\caption{Coefficients of Equation~\eqref{fitR14soundspeed} for the two bounds of the speed of sound.}
	\begin{ruledtabular}
	\begin{tabular}{cccc}
		Event & Bounds & $c_{1}$ & $c_{2}$  \\
		\hline
		\multirow{2}{*}{$\mathrm{GW170817}$} & $c$ & $0.12357\times10^{-4}$ & $6.967$ \\
		& $c/\sqrt{3}$ & $0.12179\times10^{-4}$ & $6.967$ \\
		\multirow{2}{*}{$\mathrm{GW190425}$} & $c$ & $0.870\times10^{-6}$ & $7.605$ \\
		& $c/\sqrt{3}$ & $0.088\times10^{-6}$ & $8.422$ \\
	\end{tabular}
	\end{ruledtabular}
	\label{tab:table3}
\end{table}

\subsection{GW190814: A Postulation of the Most Massive Neutron~Star}
The GW190814 event that arose from the merger of a $\sim 23~M_{\odot}$ black hole with a $\sim 2.6~M_{\odot}$ compact object has provided various scenarios for the nature of the second component. In~particular, the~possibilities for the second merger component are that of (a) the lightest black hole, (b) the most compact neutron star, (c) a rapidly rotating neutron star, and~(d) an exotic compact object. We note that in the present review, we consider only the scenarios where the compact object is a nonrotating (most compact) neutron star or is a rapidly rotating one~\cite{Kanakis-2021}.

\begin{figure}
	\includegraphics[width=\columnwidth]{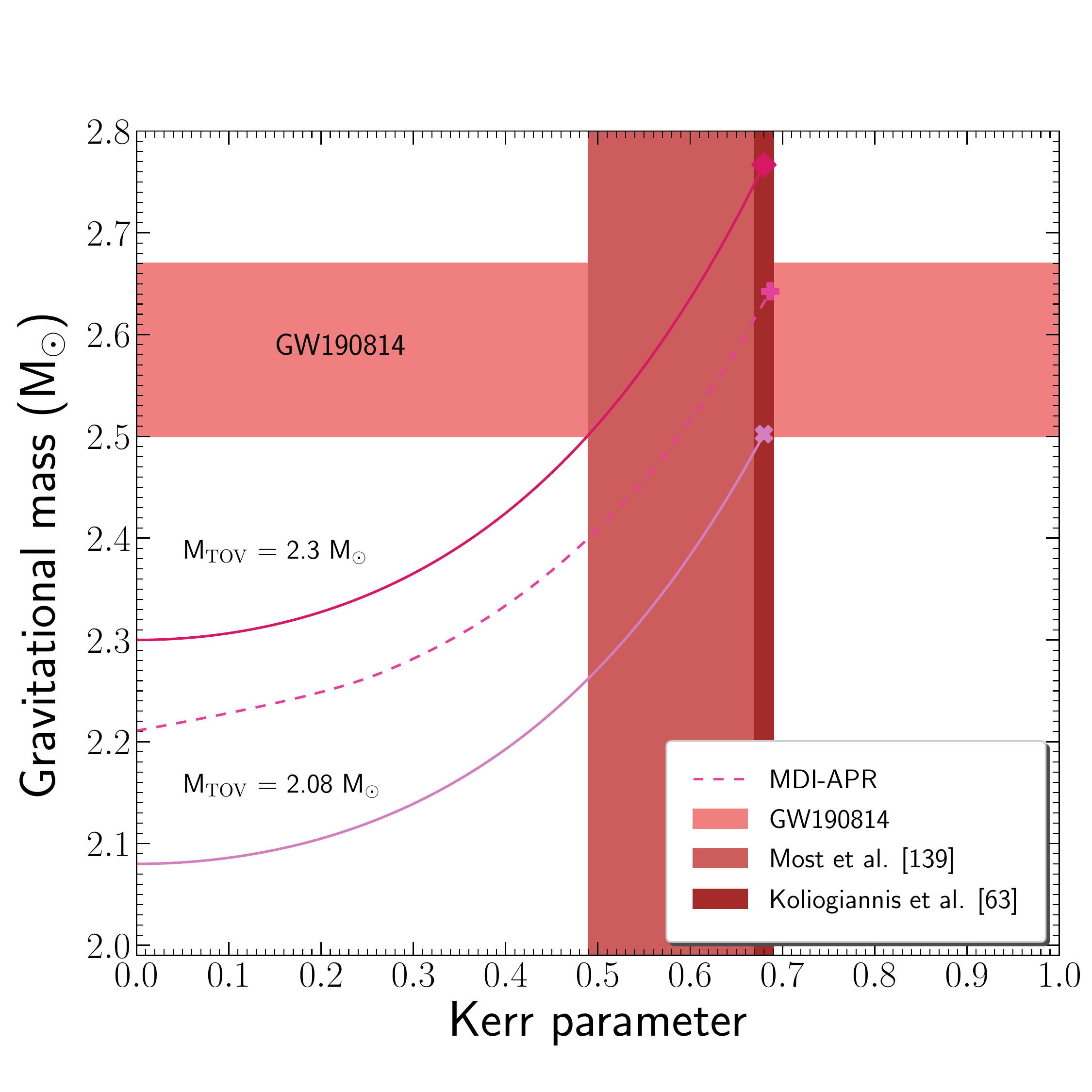}
	\caption{Dependence of the gravitational mass on the Kerr parameter. The~lower solid line represents Equation \eqref{eq:mass_rot_kerr} with $M_{\rm TOV} = 2.08~M_{\odot}$, while the upper solid line represents Equation \eqref{eq:mass_rot_kerr} with $M_{\rm TOV} = 2.3~M_{\odot}$. The~dashed line corresponds to the MDI-APR EoS. In~addition, the~horizontal shaded region notes the mass of the second component of GW190814 event, and the vertical wide shaded region marks the Kerr parameter, $\mathcal{K} = [0.49,0.68]$, according to Reference~\cite{Most-2020}. Furthermore, the~narrow vertical shaded region indicates the Kerr parameter, $\mathcal{K}_{\rm max} = [0.67,0.69]$, extracted from Reference~\cite{Koliogiannis-2020} by assuming that the low mass component was rotating at its mass-shedding limit. The~cross, the~plus sign, and~the diamond show the maximum mass configuration at the mass-shedding~limit.} 
	\label{fig:mass-kerr}
\end{figure}

In Figure~\ref{fig:mass-kerr}, we display the gravitational mass as a function of the Kerr parameter for the pure MDI-APR EoS. In~addition, we note the universal relation 
\begin{equation}
	M_{\rm rot} = M_{\rm TOV} \left[1 + 0.132\left(\frac{\mathcal{K}}{\mathcal{K_{\rm max}}}\right)^{2} + 0.071\left(\frac{\mathcal{K}}{\mathcal{K_{\rm max}}}\right)^{4}\right],
	\label{eq:mass_rot_kerr}
\end{equation}
where $\mathcal{K_{\rm max}}=0.68$ is the Kerr parameter at the mass-shedding limit, for~two limiting cases: \linebreak (a) $M_{\rm TOV} = 2.08~M_{\odot}$ and (b) $M_{\rm TOV} = 2.3~M_{\odot}$~\cite{Most-2020}. The~limiting cases correspond to the minimum and maximum possible mass for a neutron star~\cite{Most-2020}, along with the maximum value of the Kerr parameter (considering the minimum possible mass)~\cite{Most-2020}. The area marked by the intersection of the gravitational mass, $M=2.59^{+0.08}_{-0.09}~M_{\odot}$, with the Kerr parameter, $\mathcal{K}=[0.49,0.68]$~\cite{Most-2020}, notes the area where the compact object can exist. Figures~\ref{fig:pressure_density_sos} and~\ref{fig:mass-kerr} show that the pure MDI-APR EoS, which is well-defined in the above limits, is a suitable hadronic EoS to describe the $\sim$2.6~$M_{\odot}$ compact~object.

Furthermore, by~assuming that the second merger component is rotating at its mass-shedding limit, possible constraints are available through the Kerr parameter, the~equatorial radius, and~the central energy/baryon density. Specifically, by~employing the relation found in Reference~\cite{Koliogiannis-2020}
\vspace{-3pt}
\begin{equation}
	\mathcal{K}_{\rm max} = 0.488 + 0.074 \left(\frac{M_{\rm max}}{M_{\odot}}\right),
\end{equation}
for the observable gravitational mass, the~maximum Kerr parameter is evaluated in the range $\mathcal{K}_{\rm max} = [0.67,0.69]$, a~feature that is also noted in Figure~\ref{fig:mass-kerr}. Moreover, taking into consideration the relation from Reference~\cite{Koliogiannis-2021} that connects the Kerr parameter with the compactness parameter at the mass-shedding limit, namely
\begin{equation}
	\mathcal{K}_{\rm max} = 1.34 \sqrt{\beta_{\rm max}} \quad \text{and} \quad \beta_{\rm max} = \frac{G}{c^{2}}\frac{M_{\rm max}}{R_{\rm max}},
\end{equation} 
the equatorial radius is calculated in the range $R_{\rm max} = [14.77,14.87]~{\rm km}$.

\begin{figure}
	\includegraphics[width=\columnwidth]{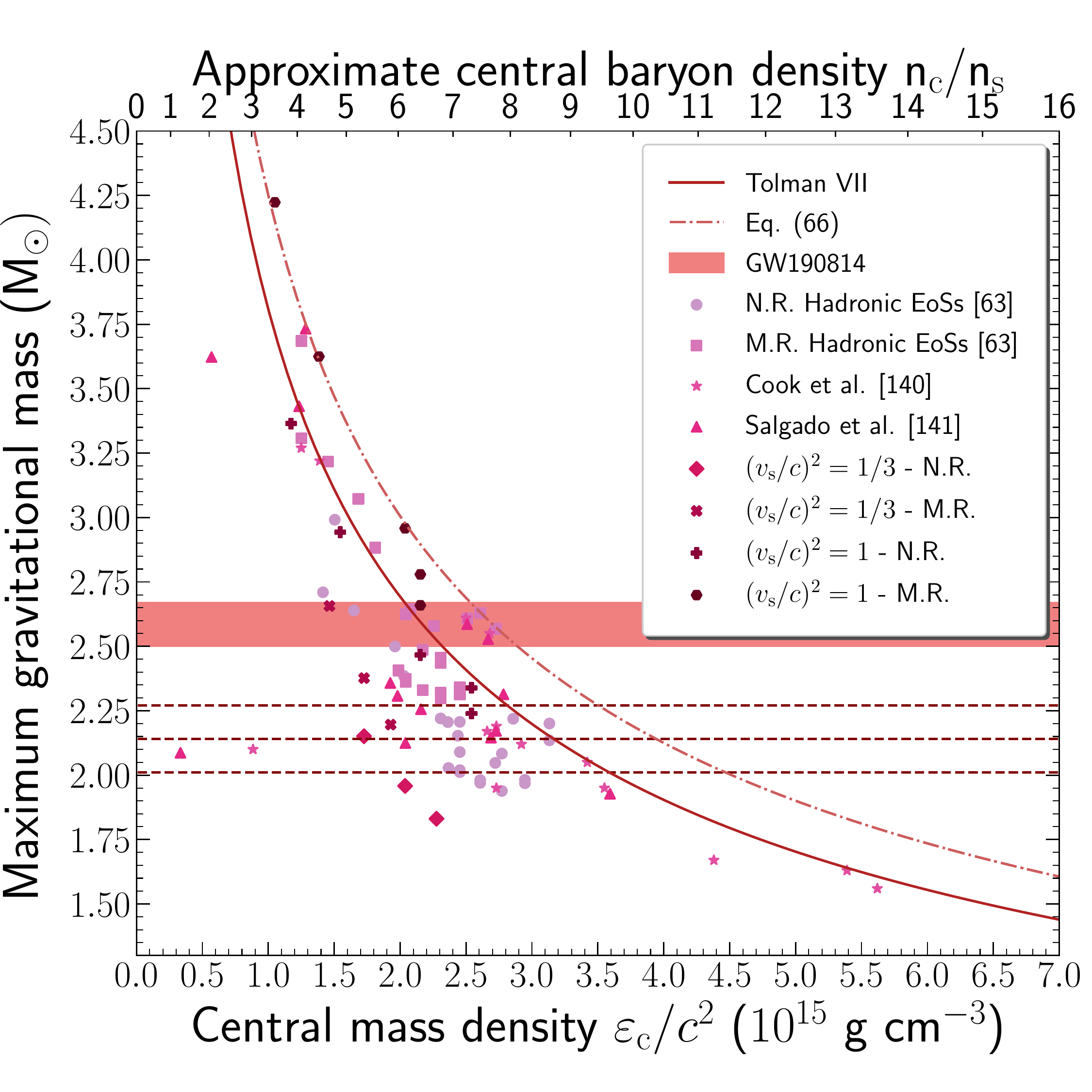}
	\caption{Dependence of the maximum gravitational mass on the central energy/baryon density both at nonrotating and rapidly rotating with the Kepler frequency configurations. Circles and squares correspond to 23 hadronic EoSs~\cite{Koliogiannis-2020} at the nonrotating (N.R.) and maximally-rotating (M.R.) cases, respectively, and~stars and triangles correspond to data of Cook~et~al.~\cite{Cook-1994} and Salgado~et~al.~\cite{Salgado-1994}, respectively. Furthermore, diamonds and plus signs note the nonrotating configuration, while crosses and polygons show the maximally-rotating one, in~the cases of the two limiting values of the sound speed. The~horizontal dashed lines mark the current observed neutron star mass limits ($2.01~M_{\odot}$~\cite{Antoniadis-2013}, $2.14~M_{\odot}$~\cite{Cromartie-2019}, and~$2.27~M_{\odot}$~\cite{Linares-2018}). Equation~\eqref{eq:mass_energy} is noted with the dashed-dotted line, while for comparison, the Tolman VII analytical solution~\cite{Koliogiannis-2020} is added with the solid line. The~horizontal shaded region notes the mass range of the second component of GW190814~event.}
	\label{fig:mass-energy density}
\end{figure}

Finally, we focused on the central energy/baryon density, a~property that is connected to the time evolution of pulsars and the appearance of a possible phase transition. The~above dependence is presented in Figure~\ref{fig:mass-energy density} as~the dependence of the maximum gravitational mass on both the central energy density and the central baryon density. Specifically, Figure~\ref{fig:mass-energy density} contains a wide range of hadronic EoSs (23 EoSs)~\cite{Koliogiannis-2020} both at the nonrotating and maximally rotating configurations, the~analytical solution of Tolman VII, Equation \eqref{eq:mass_energy}, denoted as
\begin{equation}
	\frac{M}{M_{\odot}} = 4.25 \sqrt{\frac{10^{15}~{\rm g}~{\rm cm^{-3}}}{\varepsilon_{\rm c}/c^{2}}},
	\label{eq:mass_energy}
\end{equation}
according to Reference~\cite{Koliogiannis-2020}, the~calculation data from Cook~et~al.~\cite{Cook-1994} and Salgado~et~al.~\cite{Salgado-1994}, and~the recent data for the nonrotating and maximally rotating configurations both in the cases of $(v_{s}/c)^2=1/3$ and $(v_{s}/c)^2=1$ with the corresponding transition~densities.

In addition, via Equation~\eqref{eq:mass_energy}, which is used for the description of the upper bound for the density of cold baryonic matter~\cite{Koliogiannis-2020}, the~central energy density can be constrained in the narrow range $\varepsilon_{\rm c}/c^{2} = [2.53,2.89]~10^{15}~{\rm g}~{\rm cm^{-3}}$. The~latter indicates that neutron stars with higher values of central energy density cannot exist. Furthermore, Figure~\ref{fig:mass-energy density} provides us the tools to extract the corresponding region for the central baryon density, which is $n_{c} = [7.27,8.09]~n_{s}$. It is worth mentioning that the cases in this review meet the limit for the central energy/baryon density as they are included in the region described under Equation~\eqref{eq:mass_energy}.

\begin{figure*}
	\includegraphics[width=\textwidth]{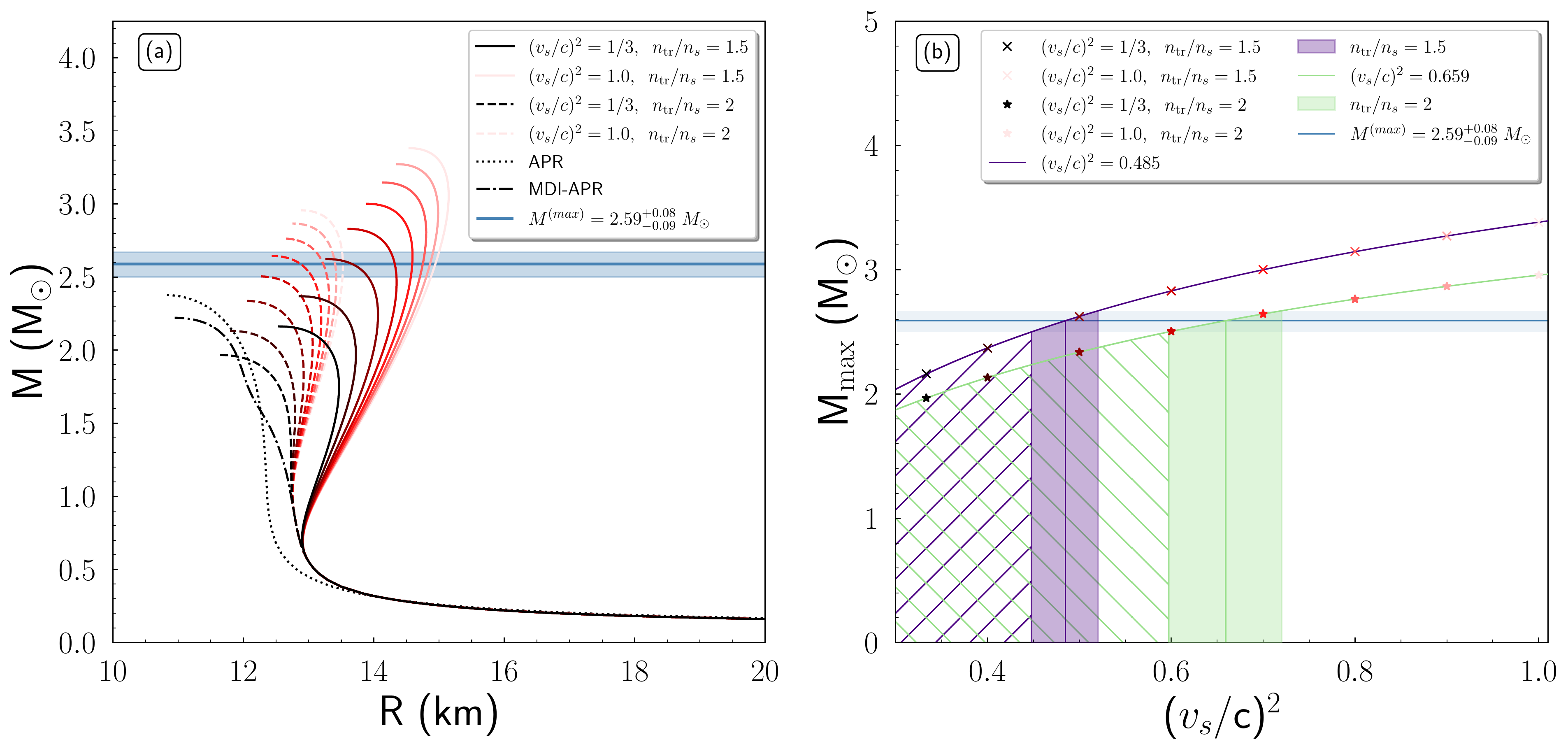}
	\caption{(a) Mass vs. radius for an isolated nonrotating neutron star, for~each transition density $n_{\mathrm{tr}}$ and all speed of sound cases. The~darker curves' color corresponds to the lower values of speed of sound. The~blue horizontal line and region indicate the mass estimation of the massive compact object of Reference~\cite{Abbott_2020_a}. The~dashed-dotted  and dotted curves correspond to the MDI-APR  and APR EoS, respectively. (b) The maximum mass $M_{max}$ of a nonrotating neutron star as a relation to the bounds of the speed of sound  $(v_s/c)^2$ for each transition density $n_{\mathrm{tr}}$ (in units of saturation density $n_s$). The~purple vertical shaded region corresponds to the $n_{\mathrm{tr}}=1.5n_s$ case, while the green one corresponds to the $n_{\mathrm{tr}}=2n_s$ case. The~purple (green) vertical line indicates the corresponding value of the speed of sound for a massive object with $M=2.59\;M_\odot$.} 
	\label{fig:MR}
\end{figure*}

\subsection{The Case of a Very Massive Neutron~Star}
\subsubsection{Isolated Non-Rotating Neutron~Star}
In this study, we extended our previous work in Reference~\cite{Kanakis-2020} for an isolated nonrotating neutron star by~using two transition densities $n_{\mathrm{tr}}=[1.5,2]n_s$ and eight values of speed of sound bounds $(v_s/c)^2=[1/3,0.4,0.5,0.6,0.7,0.8,0.9,1]$~\cite{Kanakis-2021}. The~values of transition density were taken to be close to the constraints of Reference~\cite{Kanakis-2020}. By~solving numerically the system of TOV equations, combined with the previous bounds for the speed of sound, we obtained the M-R diagram, displayed in Figure~\ref{fig:MR}(a).

\begin{figure*}
	\includegraphics[width=\textwidth]{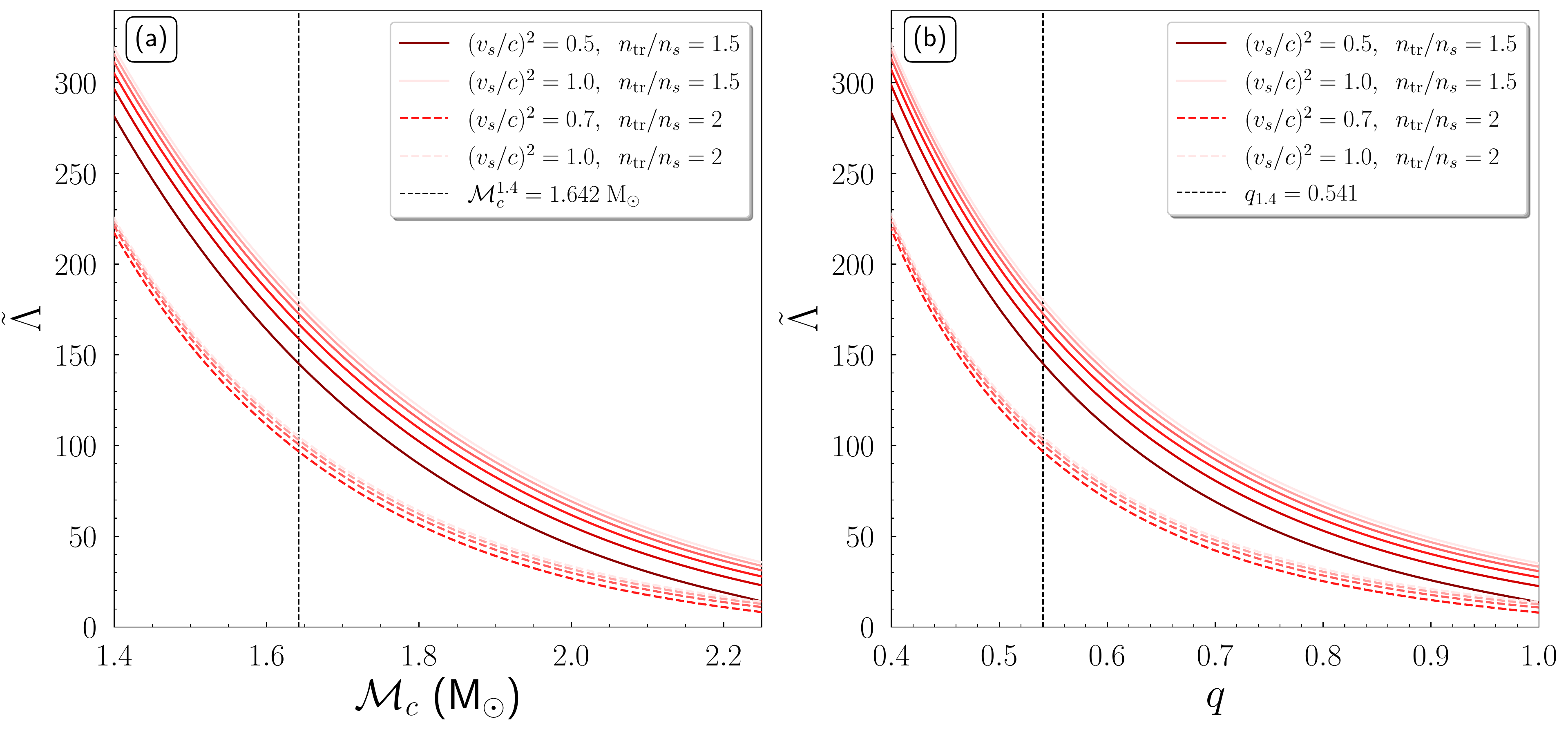}
	\caption{The effective tidal deformability $\tilde{\Lambda}$ as a function of (a) the chirp mass $\mathcal{M}_c$ and (b) binary mass ratio $q$, in~the case of a very massive neutron star component, identical to Reference~\cite{Abbott_2020_a}. The~darker colored curves correspond to lower values of speed of sound. The~black dashed vertical line shows (a) the corresponding chirp mass $\mathcal{M}_c$ and (b) mass ratio $q$, of~a binary neutron star system with $m_1=2.59\;M_\odot$ and $m_2=1.4\;M_\odot$, respectively.}
	\label{fig:chirpmassgw190814}
\end{figure*}

\begin{table*}
	\squeezetable
	\caption{Parameters of Equation \eqref{fiteq1} and bounds of speed of sound value of Figure~\ref{fig:MR}b. The~parameters $c_1$, $c_3$, and~$c_4$ are in solar mass units $M_\odot$.}
	\begin{ruledtabular}
		\begin{tabular}{cccccccc}
			$n_{\mathrm{tr}}$	& $c_{1}$	& $c_{2}$		& $c_{3}$		& $c_{4}$	& $(v_s/c)^2_{min}$		& $(v_s/c)^2$		& $(v_s/c)^2_{max}$\\ 
			\hline
			$1.5n_s$ & $-1.6033\times10^{3}$ &$-7.56\times10^{-4}$ & $-1.64\times10^{-1}$ & $1.6068\times10^{3}$ & 0.448 & 0.485 &  0.52 \\
			$2n_s$ & 5.5754 & 0.2742 & $-$0.6912 & $-$1.9280  & 0.597 &  0.659  &  0.72\\
		\end{tabular}
	\end{ruledtabular}
	\label{tab:tablegw190814}
\end{table*}

At first sight, there are two main branches in Figure~\ref{fig:MR}a related to the transition density. The~solid (dashed) curves correspond to the $n_{\mathrm{tr}}=1.5n_s$ ($n_{\mathrm{tr}}=2n_s$) case. Depending on each value for the bound of the speed of sound, there are bifurcations in the families of EoSs. As~we move to higher values for the speed of sound, the~color of the curves lightens. The~blue solid horizontal line, with~the accompanying shaded region, represents the estimation of the recently observed massive compact object of Reference~\cite{Abbott_2020_a}. In~general, the~branch of EoSs with $n_{\mathrm{tr}}=1.5n_s$ provides stiffer EoSs compared to the $n_{\mathrm{tr}}=2n_s$ branch. Hence, the~EoSs of the $n_{\mathrm{tr}}=1.5n_s$ case are more likely to provide such a massive nonrotating neutron star than~the $n_{\mathrm{tr}}=2n_s$ case in which three EoSs lie outside of the shaded region. Specifically, between~the same kind of transition density $n_{\mathrm{tr}}$ the EoSs with higher bounds of the speed of sound  lead to higher values of neutron star mass and radius. Therefore, a~generally high bound of the speed of sound (as the transition density has higher values, the speed of sound would be closer to the causal scenario) is needed for the description of such a massive compact~object.

As one can observe in Figure~\ref{fig:MR}a, there is a trend across the maximum masses. In~order to study this behavior, we constructed the diagram of Figure~\ref{fig:MR}b. The~cross  and star  marks represent the maximum masses of $n_{\mathrm{tr}}=1.5n_s$ and $n_{\mathrm{tr}}=2n_s$ cases, respectively. The~color of the marks is  lighter for higher values of the speed of sound. The~blue solid horizontal line, with~the accompanying shaded region, indicates the estimation of the recently observed massive compact object of Reference~\cite{Abbott_2020_a}. The~purple and green curves represent the following expression for the $n_{\mathrm{tr}}=1.5n_s$ and $n_{\mathrm{tr}}=2n_s$ cases, respectively, given below
\begin{equation}
	M_{max}=c_1d^{c_2}+c_3d+c_4,
	\label{fiteq1}
\end{equation}
where $d=(v_s/c)^2$. The~coefficients are given in Table~\ref{tab:tablegw190814}.

By applying the formula mentioned above, we were able to obtain estimations of the speed of sound values for each transition density $n_{\mathrm{tr}}$. In~particular, for~a nonrotating massive neutron star with $M=2.59\;M_\odot$ the speed of sound must be (a)  $(v_s/c)^2 = 0.485$ (\mbox{$n_{\mathrm{tr}}=1.5n_s$}), and~(b) $(v_s/c)^2 = 0.659$ ($n_{\mathrm{tr}}=2n_s$). The~exact values' interval is given in Table~\ref{tab:tablegw190814}. We note that in the case of higher values of transition density $n_{\mathrm{tr}}$, the fitted expression and marks are shifted downwards; i.e.,~the higher the point of the transition in density, the~smaller the provided maximum mass. In~addition, as~one can see in Figure~\ref{fig:MR}b, the~ higher values of the speed of sound are better able to predict such massive neutron stars, until~a specific boundary value of transition density $n_{\mathrm{tr}}$ (higher than those we adopted in our study), in which even the causality could not lead to such a massive~neutron star.

Hence, a~very massive nonrotating neutron star favors higher values of the speed of sound than the $v_s=c/\sqrt{3}$ limit. Based on our previous work, the~current observation of neutron star mergers leads to a lower bound on the transition density $n_{\mathrm{tr}}$~\cite{Kanakis-2020}. At~this point, a contradiction  arises; the transition density $n_{\mathrm{tr}}$ must be above a specific lower value but not big enough to predict very massive~masses.

\begin{figure*}
	\includegraphics[width=\textwidth]{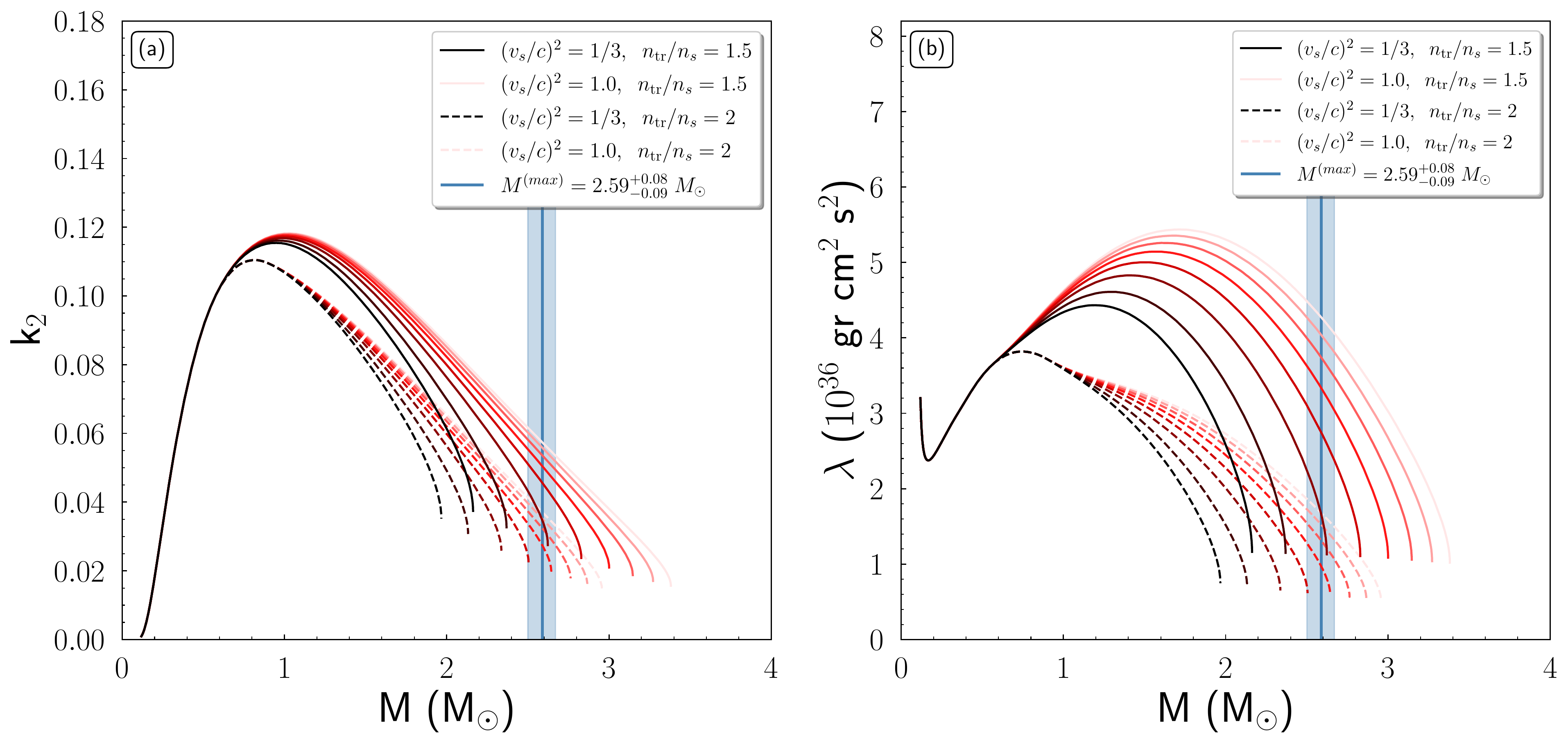}
	\caption{Tidal parameters (a) $k_2$ and (b) $\lambda$ as a function of an neutron star's mass. The~blue vertical line and shaded region indicate the estimation of the recently observed massive compact object of Reference~\cite{Abbott_2020_a}. The~solid (dashed) curves correspond to the $n_{\mathrm{tr}}=1.5n_s$ ($n_{\mathrm{tr}}=2n_s$) case. The~lower values of the speed of sound correspond to the darker-colored~curves.}
	\label{fig:tidalparams}
\end{figure*}

In Figure~\ref{fig:tidalparams}, we display the tidal parameters for the single neutron star case that we study as a function of the mass. The~vertical blue shaded region and line indicate the mass estimation for the second compact object of Reference~\cite{Abbott_2020_a}. One can observe that in both diagrams there are two main families of EoSs, distinguished by the transition density $n_{\mathrm{tr}}$. The~EoSs with higher speed of sound bounds lead to bigger values on both tidal parameters. This means that a neutron star with a higher speed of sound value is more deformable than a more compact one (with a lower speed of sound value). Moreover, the~EoSs with smaller transition density $n_{\mathrm{tr}}$ and higher $(v_s/c)^2$ values are more likely to predict a very massive neutron star of $M=2.59\;M_\odot$.

We notice that taking into consideration cases with higher transition density $n_{\mathrm{tr}}$ could lead to smaller values of tidal parameters, i.e.,~more compact  and less deformable stars. Hence, a~very high value of the speed of sound, close to the causality, would be necessary to provide such a massive nonrotating~neutron star.

\subsubsection{A Very Massive Neutron Star in a Binary Neutron Stars~System}

Beyond the hypothetical scenario of a single nonrotating neutron star, it is of particular interest to consider the binary case of two neutron stars, with~the heavier having a mass of \mbox{$m_1=2.59\; M_\odot$} and letting the secondary lighter neutron star  fluctuate within the range $m_2\in(1,2.59)\;M_\odot$. By~subtracting the component masses $m_1, m_2$ in  Equation~(\ref{chirpmass}), we obtain the corresponding values of $\mathcal{M}_c$. Then, since the masses are defined, from~the  \mbox{Equations~(\ref{L-tild-1})} and~(\ref{Lamb-1}), the~effective tidal deformability $\tilde{\Lambda}$ can be~determined.

In Figure~\ref{fig:chirpmassgw190814}a, we display the effective tidal deformability $\tilde{\Lambda}$ as a function of the chirp mass $\mathcal{M}_c$ of the system, for~all the possible binary neutron star systems with such a massive neutron star component. We have to underline that from all the EoSs that we studied in the single-neutron star case previously, in~the binary case, we used only those that can provide a neutron star with $2.59\;M_\odot$ mass. As~one can see in Figure~\ref{fig:chirpmassgw190814}, there are two families of EoSs, distinguished by the transition density $n_{\mathrm{tr}}$. For~each family of EoSs, the~EoSs with higher values of the speed of sound predict higher values of $\tilde{\Lambda}$. We notice that for a binary system with $m_1=2.59\;M_\odot$ and $m_2=1.4\;M_\odot$, the chirp mass is $\mathcal{M}_c=1.642\;M_\odot$. Another remark is that binary neutron star systems with both heavy components, meaning higher $\mathcal{M}_c$,  lead to smaller values of $\tilde{\Lambda}$. In~this case, a~possible lower limit on $\tilde{\Lambda}$ might provide useful constraints on the~EoS. 

In Figure~\ref{fig:chirpmassgw190814}b, we display the dependence of $\tilde{\Lambda}$ on the corresponding binary mass ratio $q$. We have to underline that this kind of $\tilde{\Lambda}-q$ diagram is different from the usual ones (see in comparison Figure~\ref{Ltildeq1}) because the chirp mass $\mathcal{M}_c$ has no specific value. In particular, in~this work, the $\mathcal{M}_c$ is treated as a variable, and each point of Figure~\ref{fig:chirpmassgw190814}b corresponds to a different binary neutron star system with the heavier component in all cases being a very massive neutron star with mass $2.59\;M_\odot$. Similarly to Figure~\ref{fig:chirpmassgw190814}a, there are two main families for the curves, and the EoSs with higher speed of sound provide higher values of $\tilde{\Lambda}$. The~more symmetric binary neutron star systems ($q\rightarrow1$) lead to smaller values of $\tilde{\Lambda}$. On~the contrary, the~highest values of $\tilde{\Lambda}$ correspond to the most asymmetric binary neutron star systems. We notice that for a binary system with $m_1=2.59\;M_\odot$ and $m_2=1.4\;M_\odot$, the asymmetry ratio is $q=0.541$.

\begin{figure}
	\includegraphics[width=0.48\textwidth]{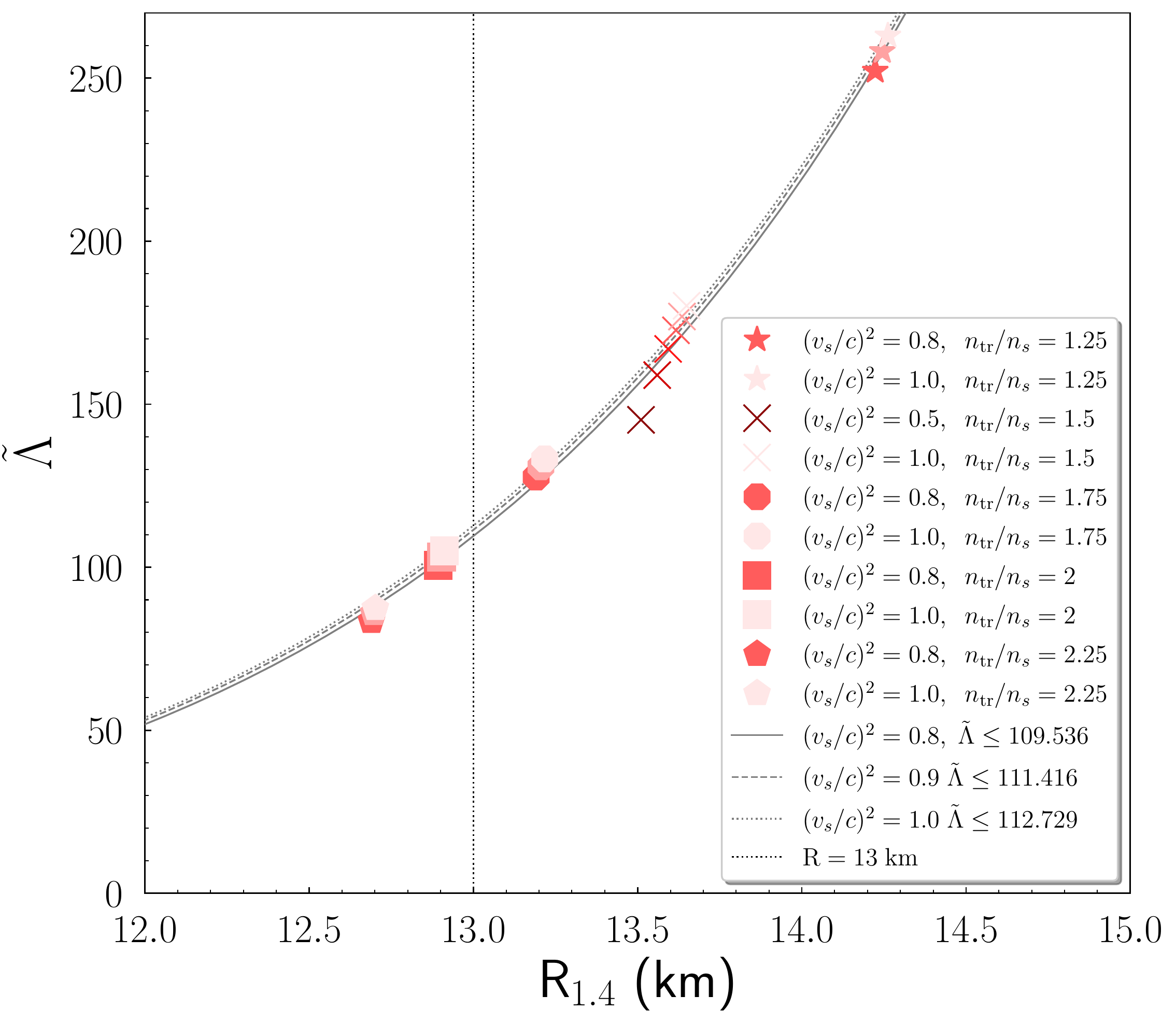}
	\caption{The effective tidal deformability $\tilde{\Lambda}$ as a function of the radius $R_{1.4}$ of an $m_2=1.4\;M_\odot$ neutron star. The~heavier component of the system was taken to be $m_1=2.59\;M_\odot$. The~darker colors correspond to lower values of speed of sound bounds. The~grey lines show the expression of Equation~(\ref{fitR14}). The~black dotted vertical line indicates the proposed upper limit of Reference~\cite{Raithel-2018}.}
	\label{fig:LtildeR14}
\end{figure}

Moreover, we studied the effective tidal deformability $\tilde{\Lambda}$ and the $R_{1.4}$ case of an \mbox{$m_2=1.4\;M_\odot$} secondary component neutron star, with~the heavier component neutron star  taken to be \mbox{$m_1=2.59\;M_\odot$}. In~Figure~\ref{fig:LtildeR14}, we display this dependence. To~be more specific, the~EoSs are in five main groups, characterized by the transition density $n_{\mathrm{tr}}$. Our study has been expanded to transition densities $n_{\mathrm{tr}}=[1.25,1.75,2.25]~n_s$ so that the calculations could be more accurate  The higher speed of sound values correspond to lighter  color. In~analog to the observations of the previous Figure~\ref{fig:chirpmassgw190814}, the~high speed of sound bounds lead to higher $\tilde{\Lambda}$ and $R_{1.4}$. 

In addition, we applied a fitting formula to the $(v_s/c)^2=[0.8,0.9,1]$ cases. The~formula was taken to be in the kind of the proposed relations of References~\cite{Zhao,SoumiDe},
\begin{equation}
	\tilde{\Lambda}=c_5R^{c_6}_{1.4}
	\label{fitR14}
\end{equation}
where the coefficients for each case are given in Table~\ref{tab:table2}. 

\begin{table}
	\caption{Parameters of Equation \eqref{fitR14} and bounds of $\tilde{\Lambda}$ of Figure~\ref{fig:LtildeR14}.}
	\begin{ruledtabular}
		\begin{tabular}{cccc}
			$(v_s/c)^2$	& $c_{5}$ ($\mathrm{km^{-1}}$)	& $c_{6}$	& $\tilde{\Lambda}$	\\
			\hline
			$0.8$ & $4.1897\times10^{-9}$ &$9.3518$  &$109.536$ \\
			$0.9$ & $5.3213\times10^{-9}$ &$9.2652$ &$111.416$  \\
			$1$ & $6.1109\times10^{-9}$ &$9.2159$  &$112.729$ \\
		\end{tabular}
	\end{ruledtabular}
	\label{tab:table2}
\end{table}

According to a recent study, a~similar power-law relation connects the tidal deformability of a single neutron star to the $R_{1.4}$~\cite{Tsang-2020}. The~significance of the tidal deformability $\Lambda_{1.4}$ and $R_{1.4}$ in order to obtain information about microscopic quantities was studied in Reference~\cite{Wei-2020}. By~imposing an upper limit on $R_{1.4}$, one can obtain an upper limit on $\tilde{\Lambda}$. Hence, by~adopting the general limit of Reference~\cite{Raithel-2018}, we obtained the constraints of Table~\ref{tab:table2}.

\begin{figure*}
	\includegraphics[width=\textwidth]{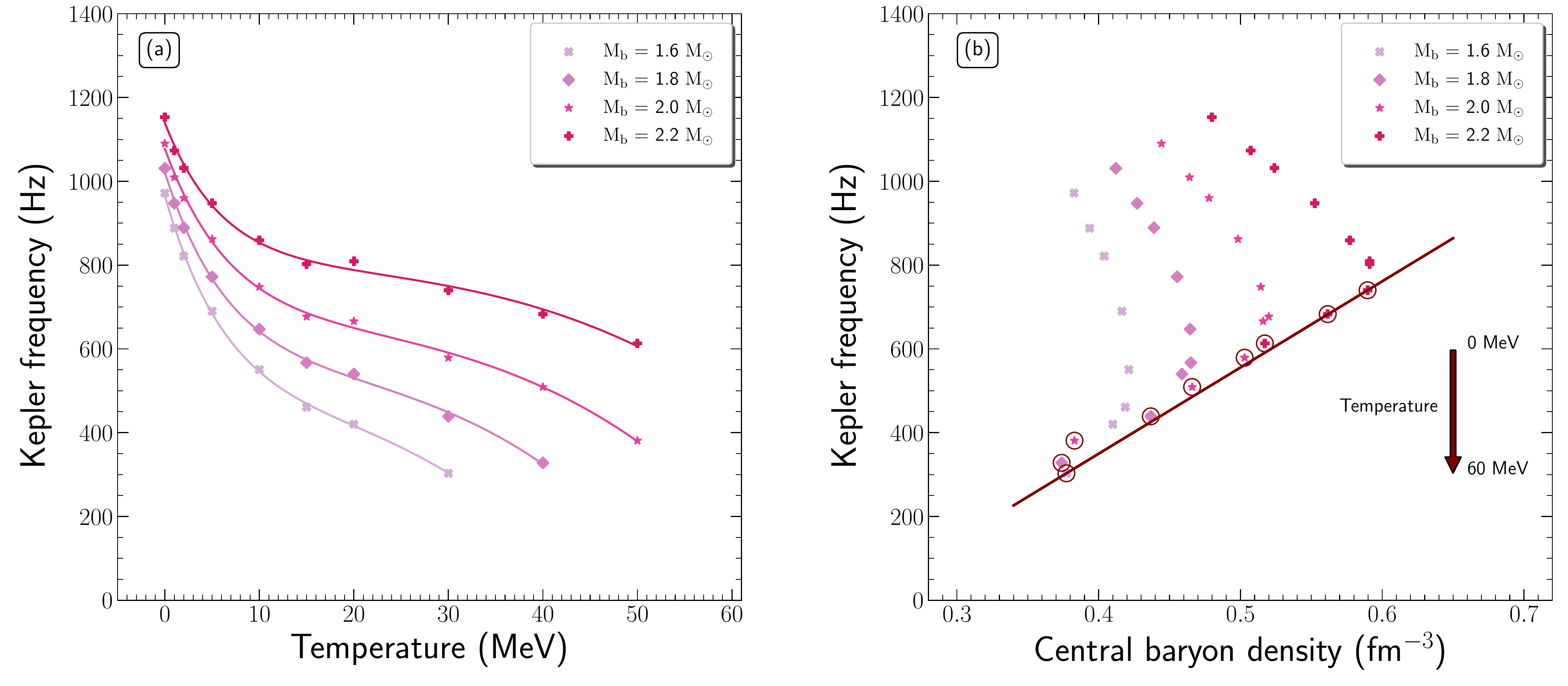}
	\caption{Dependence of the Kepler frequency on (a) the temperature and (b) the central baryon density for baryon masses in the range $[1.6,2.2]~M_{\odot}$. (a) Solid lines represent the ﬁts originated from Equation~\eqref{eq:f_t}. (b) The solid line represents Equation \eqref{eq:f_n}, while open circles note the high-temperature region ($T\ge 30~{\rm MeV}$).}
	\label{fig:constant_baryon_mass}
\end{figure*}

\subsection{Finite Temperature Effects on Rapidly Rotating Neutron~Stars}
\subsubsection{Sequences of Constant Baryon~Mass}
A way to study the effects of finite temperature on the rapidly rotating remnant after a binary neutron star merger is the sequences of constant baryon mass. The~sequences provide us with information about the evolution and instabilities of hot neutron stars. In~particular, in~the case of the isothermal EoSs, we considered the same baryon mass configuration for the EoSs and~constructed a sequence related to the cooling of a hot neutron star~\cite{Koliogiannis-2021}.

Figure~\ref{fig:constant_baryon_mass}a displays the Kepler frequency as a function of the temperature for four baryon masses in the range $[1.6,2.2]~M_{\odot}$. Specifically, in~the range $[0,15]~{\rm MeV}$, the Kepler frequency decreases sharply with the temperature, while for higher temperatures, a smoother behavior is presented. This dependence can be described as
\begin{equation}
	f(T) = \alpha_{0} + \alpha_{1}T^{3} + \alpha_{2}exp[\alpha_{3}T] \quad ({\rm Hz}),
	\label{eq:f_t}
\end{equation}
where $f$ and $T$ are in units of Hz and MeV, respectively, and~the coefficients $\alpha_{i}$ with $i$ = 0--3 are given in Table~\ref{tab:coef_f_t}.

\begin{table}
	\caption{Coefficients $\alpha_{i}$ with $i$ = 0--3 for the empirical relation~\eqref{eq:f_t} and baryon masses in the range [1.6--2.2]~$M_{\odot}$.}
	\begin{ruledtabular}
	\begin{tabular}{lcccc}
		\multirow{2}{*}{Coefficients} & \multicolumn{4}{c}{Baryon Mass} \\
		& $1.6~M_{\odot}$ & $1.8~M_{\odot}$ & $2.0~M_{\odot}$ & $2.2~M_{\odot}$\\
		\hline
		$a_{0}~(\times 10^{2})$ & 4.259 & 5.284 & 6.414 & 7.863 \\
		$a_{1}~(\times 10^{-3})$ & $-$4.787 & $-$3.202 & $-$2.099 & $-$1.443 \\
		$a_{2}~(\times 10^{2})$ & 5.401 & 4.929 & 4.363 & 3.530 \\
		$a_{3}~(\times 10^{-1})$ & $-$1.468 & $-$1.443 & $-$1.424 & $-$1.636 \\
	\end{tabular}
	\end{ruledtabular}
	\label{tab:coef_f_t}
\end{table}

This behavior suggests that the effects of temperature are more pronounced in the range $[0,15]~{\rm MeV}$, leading to significant lower Kepler frequencies, where for higher temperatures than $T>15~{\rm MeV}$, the effects are moderated. 

Afterwards, in~Figure~\ref{fig:constant_baryon_mass}b, the dependence of Kepler frequency on the central baryon density is presented, for~four baryon masses in the range $[1.6,2.2]~M_{\odot}$. While for low values of temperature the central baryon density increases with increasing temperature, for~high values of temperature, ~a reduction in the values of the central baryon density is noted. In addition, for low values of temperature $(T<30~{\rm MeV})$ there is the appearance of a linear relation between the Kepler frequency and the central baryon density, assuming always a constant temperature. The~significance of Figure \ref{fig:constant_baryon_mass}b is focused on temperatures higher than \mbox{$T=30~{\rm MeV}$}, where a linear relation described as
\begin{equation}
	f(n_{b}^{c}) = -473.144 + 2057.271 n_{b}^{c} \quad ({\rm Hz}),
	\label{eq:f_n}
\end{equation}
with $f$ and $n_{b}^{c}$ given in units of Hz and $\rm fm^{-3}$, respectively, interprets the dependence of the Kepler frequency on the central baryon density independent from the specific baryon mass. Henceforth, Equation~\eqref{eq:f_n} defines the allowed region that a neutron star can exist with rotation at its mass-shedding limit for a specific central baryon density and~vice versa.

\subsubsection{Moment of Inertia, Kerr Parameter, and~Ratio $T/W$}
\begin{figure*}
	\includegraphics[width=0.48\textwidth]{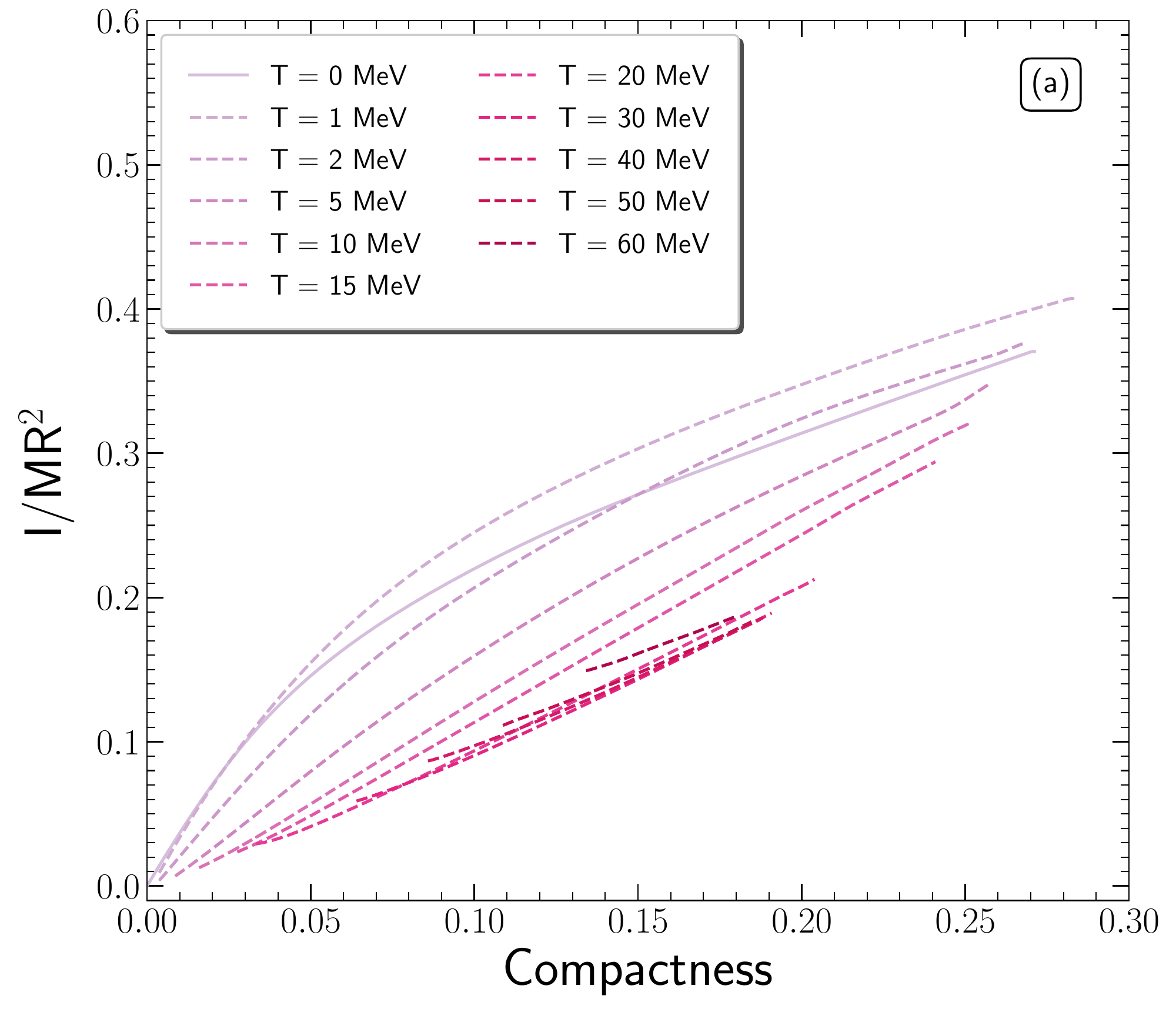}
	\hspace{0.4cm}
	\includegraphics[width=0.48\textwidth]{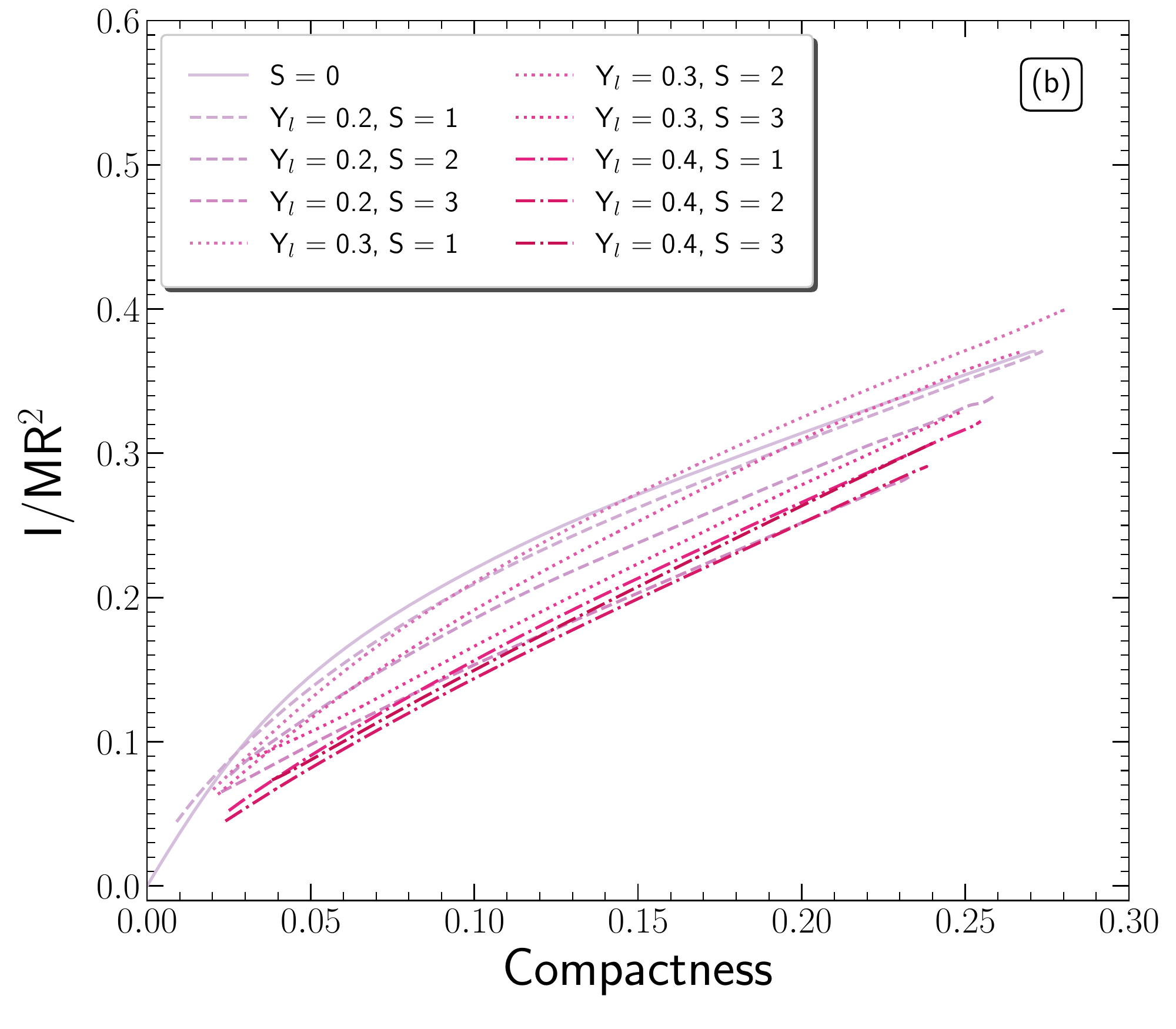}
	\caption{Dependence of the dimensionless moment of inertia on the compactness parameter at the mass-shedding limit in the case of (a) isothermal and (b) isentropic profiles. Dashed lines note the hot configurations, while the solid line notes the cold configuration.}
	\label{fig:moi}
\end{figure*}

\begin{figure*}
	\includegraphics[width=0.3\textwidth]{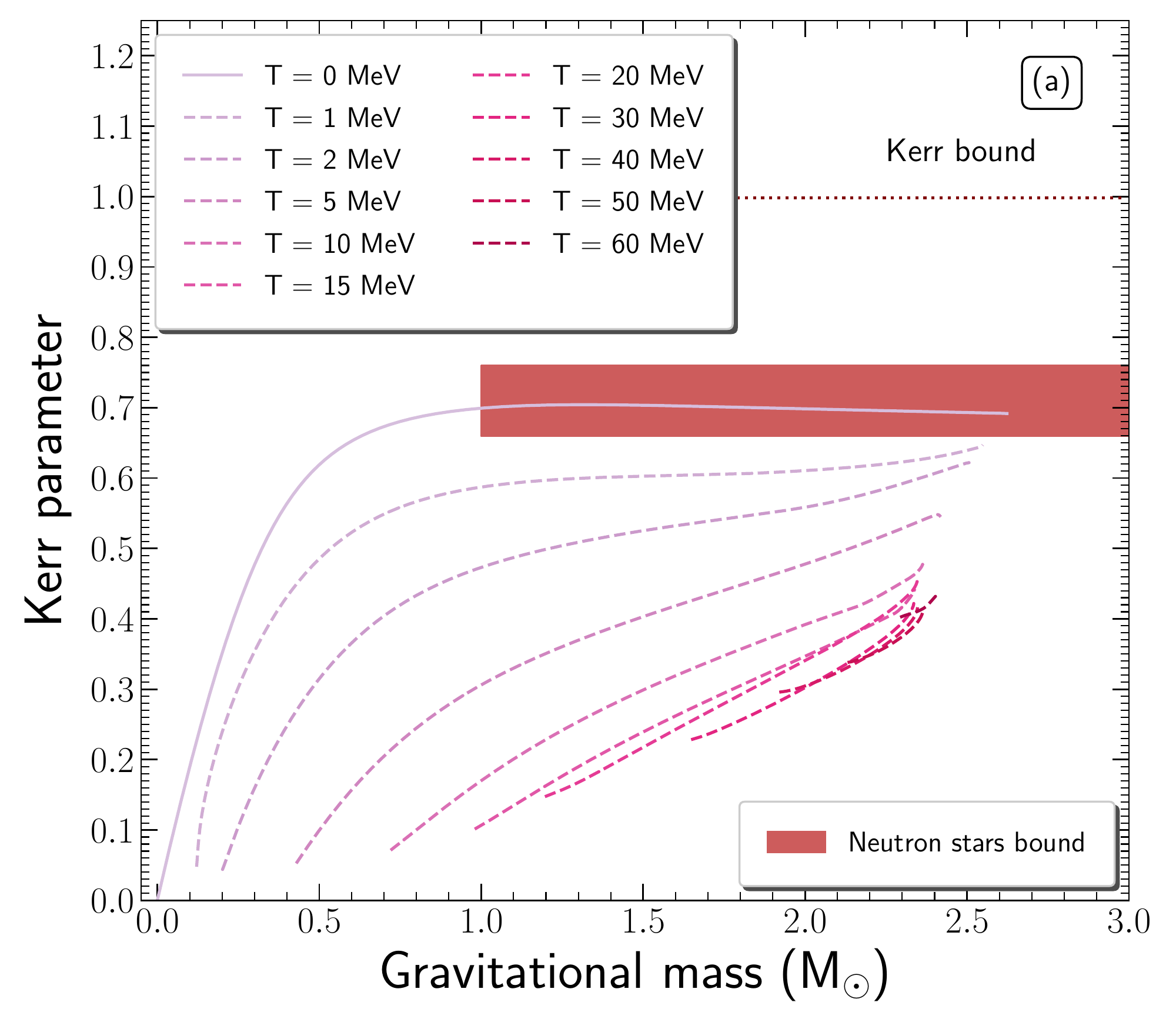}
	\hspace{0.5cm}
	\includegraphics[width=0.3\textwidth]{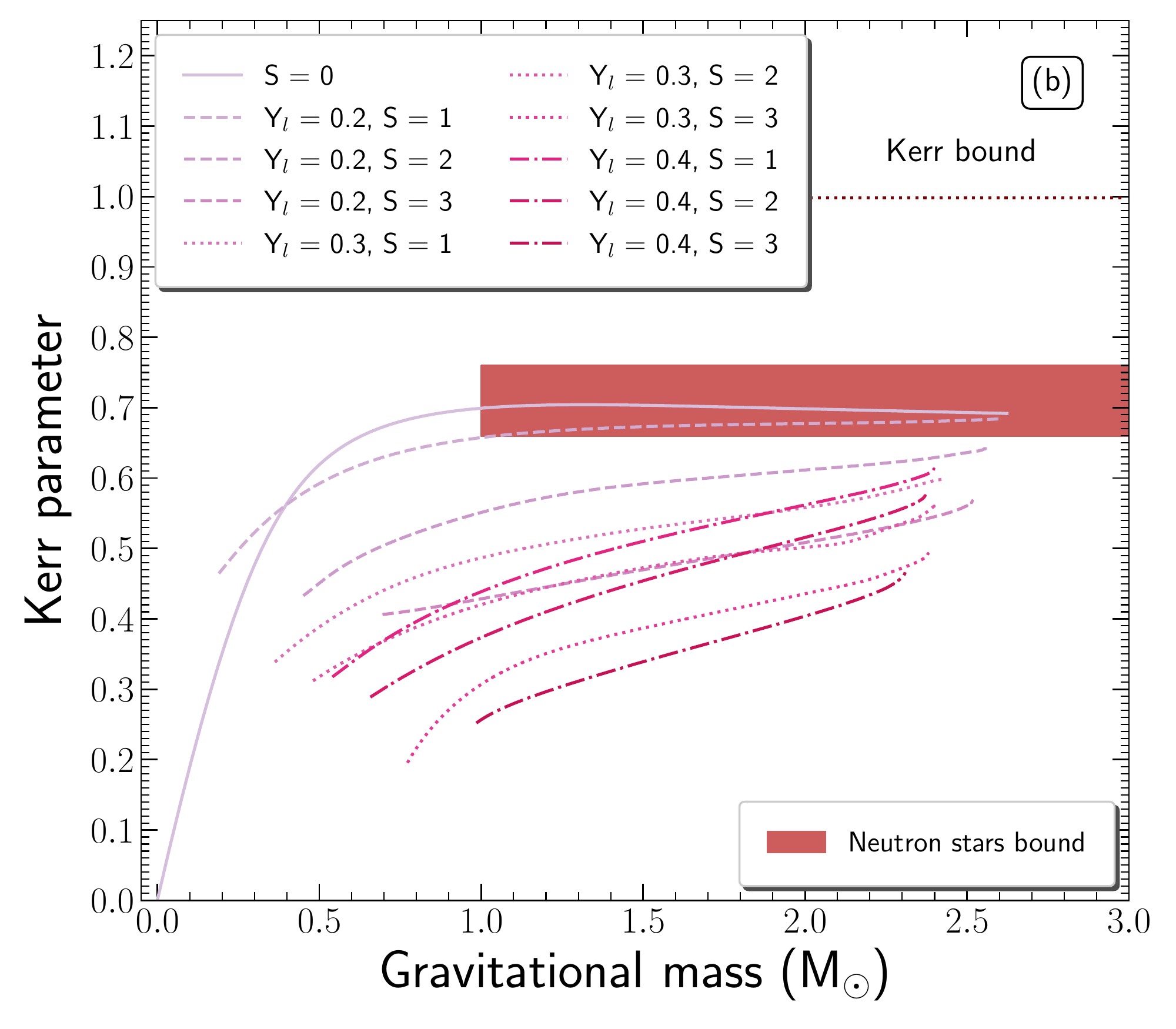}
	\hspace{0.5cm}
	\includegraphics[width=0.3\textwidth]{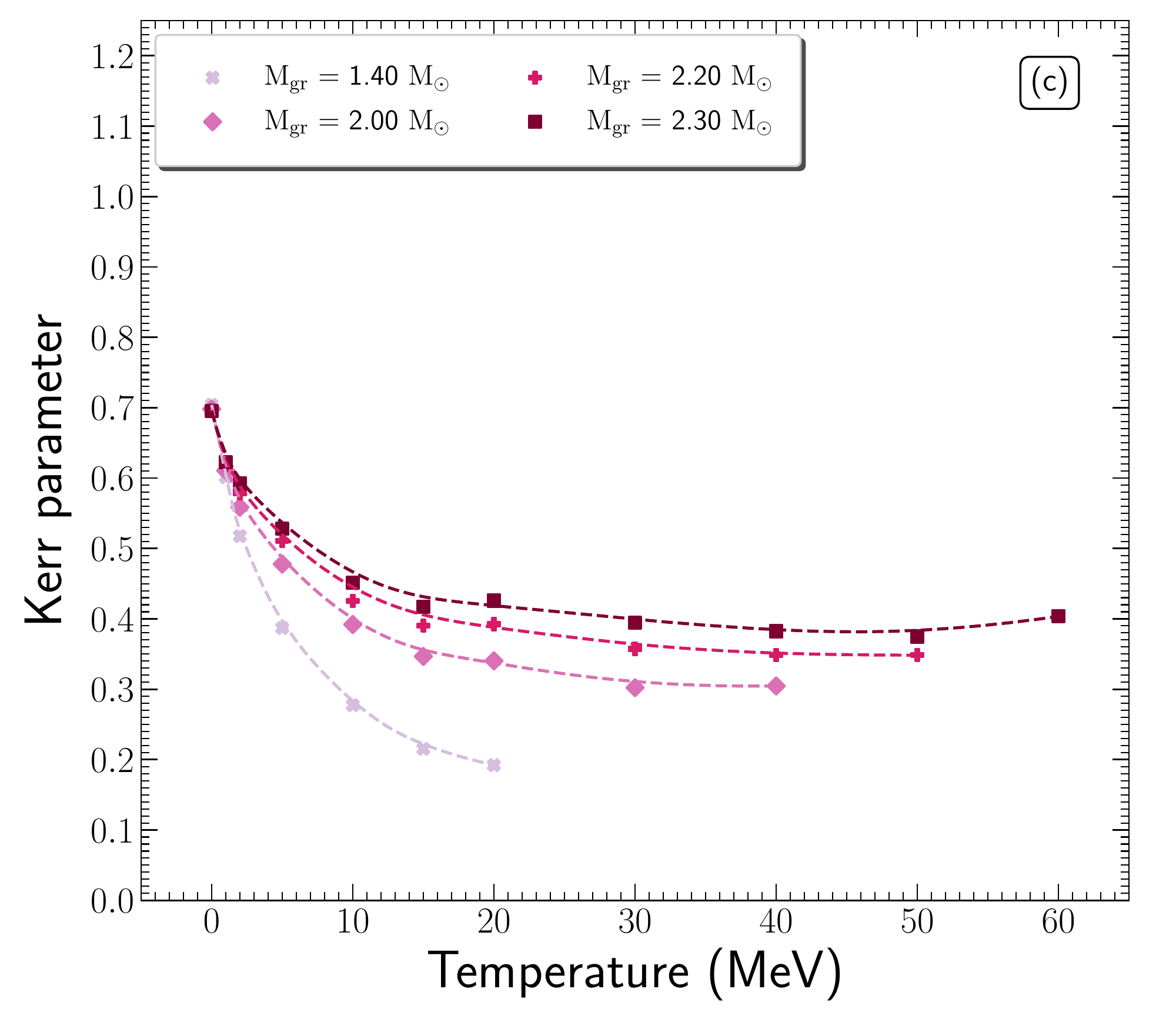}
	\caption{Dependence of the Kerr parameter on the gravitational mass at the mass-shedding limit in the case of (a) isothermal and (b) isentropic profiles. Dashed lines note the hot configurations, while the solid line notes the cold configuration. The~horizontal dotted line corresponds to the Kerr bound for astrophysical Kerr black holes, $\mathcal{K_{\rm B.H.}}=0.998$~\cite{Thorne_1974}, while the shaded region corresponds to the neutron star limits from Reference~\cite{Koliogiannis-2021}. (c) Dependence of the Kerr parameter on the temperature for gravitational masses in the range $[1.4,2.3]~M_{\odot}$ and in the case of isothermal~profile.}
	\label{fig:kerr-1}
\end{figure*}

Figure~\ref{fig:moi} displays the dimensionless moment of inertia as a function of the compactness parameter for neutron stars at the mass-shedding limit in (a) isothermal and (b) isentropic profile. In~both cases, the~increase of the temperature or the entropy per baryon (assuming a constant lepton fraction) leads to lower values of moment of inertia and lesser compact stars than the cold star. However, for~low values of temperature $(T<2~{\rm MeV})$ or low values of entropy per baryon $(S=1$ with $Y_{l}=0.2$ and $0.3)$, the~dimensionless moment of inertia and the compactness parameter exceed the values of the cold neutron~star.

Afterwards, in~Figure~\ref{fig:kerr-1} we present the Kerr parameter as a function of the gravitational mass for neutron stars at the mass-shedding limit in (a) isothermal and (b) isentropic profile. In~addition, we display the constraints for the Kerr parameter of neutron stars with the shaded region~\cite{Koliogiannis-2021} and~the Kerr bound for astrophysical Kerr black holes~\cite{Thorne-1998}. As~the temperature or the entropy per baryon increases in the neutron star, the~Kerr parameter decreases, except~for high values of temperature $(T=60~{\rm MeV})$ where a slightly increase is observed. Furthermore, ~Figure~\ref{fig:kerr-1}c displays the Kerr parameter as a function of the temperature for constant gravitational mass in the isothermal profile. After~$T=30~{\rm MeV}$, it is observed that the Kerr parameter creates a plate, meaning that the increase in temperature does not affect the Kerr~parameter.

The introduction of temperature in neutron stars cannot violate the proposed limit for Kerr black holes~\cite{Thorne-1998} and the one from cold neutron stars~\cite{Koliogiannis-2021}. Therefore, the~gravitational collapse of a hot and uniformly rotating neutron star, constrained to mass--energy and angular momentum conservation, cannot lead to a maximally rotating Kerr black~hole. 

In addition, it is worth mentioning that, while in the cold neutron star, for~$M_{\rm gr} > 1~M_{\odot}$, the~Kerr parameter is almost independent of the gravitational mass, in~hot configurations, Kerr parameter is an increasing function of the gravitational mass. The~latter leads to the conclusion that while the interplay between the angular momentum and the gravitational mass in cold neutron stars is imperceptible, that is not the case in hot neutron stars, where a significant dependence is~suggested.

\begin{figure*}
	\includegraphics[width=0.48\textwidth]{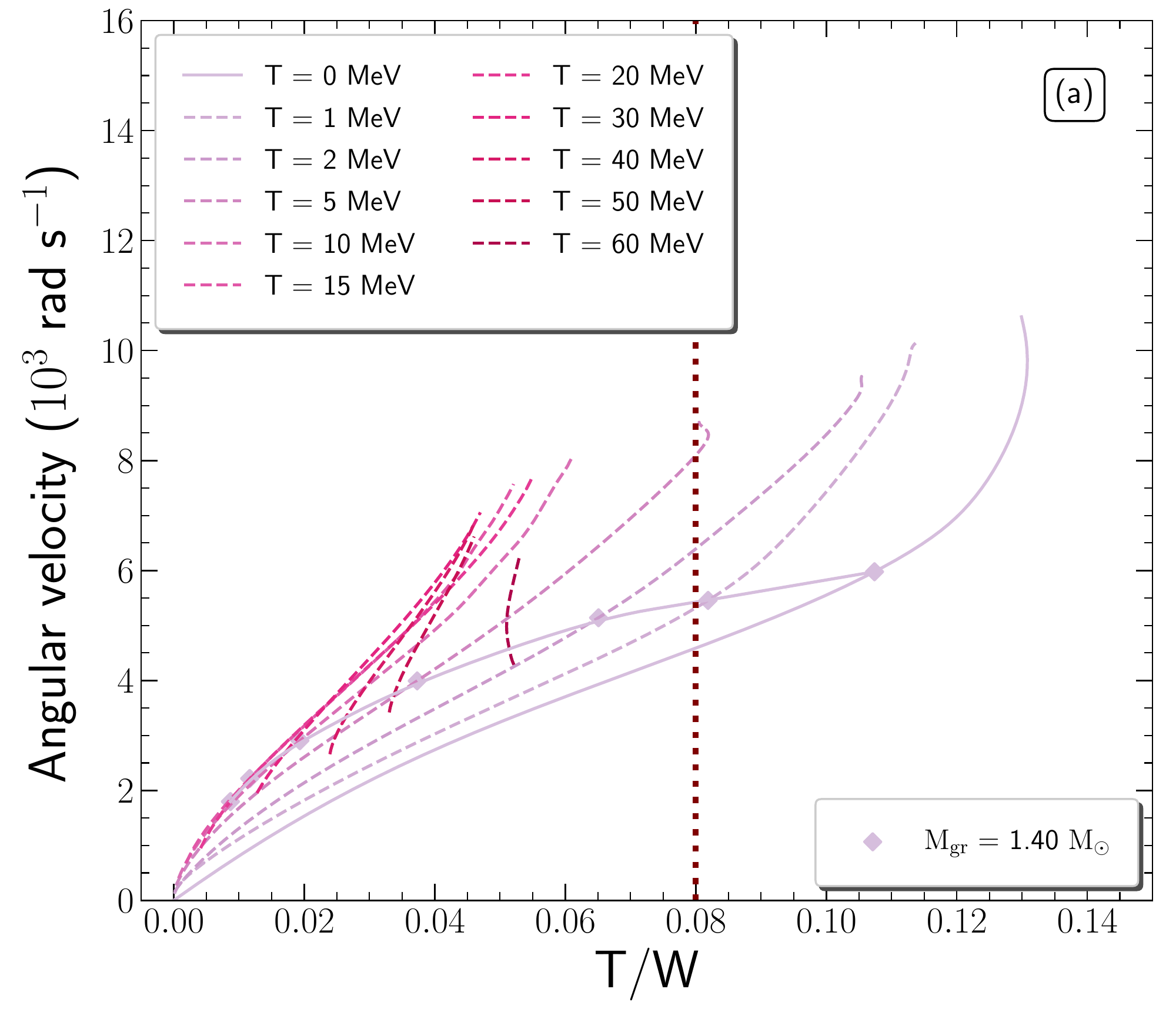}
	\hspace{0.4cm}
	\includegraphics[width=0.48\textwidth]{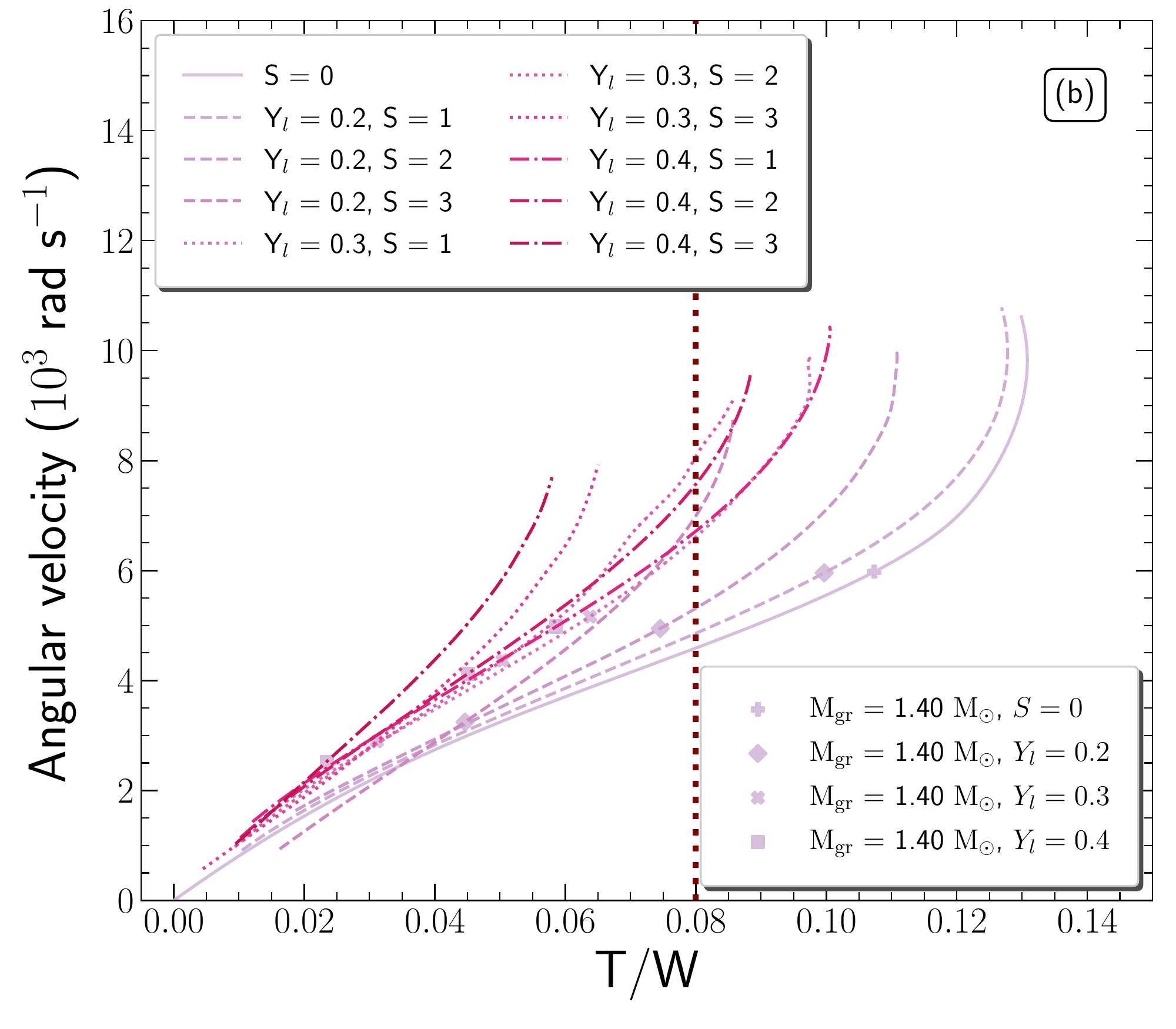}
	\caption{Dependence of the angular velocity on the ratio of rotational kinetic to gravitational binding energy at the mass-shedding limit in the case of (a) isothermal and (b) isentropic profiles. Dashed lines note the hot configurations, while the solid line notes the cold configuration. Markers correspond to the $M_{\rm gr}=1.4~M_{\odot}$ configuration. The~vertical dotted line notes the critical value, $T/W=0.08$, for~gravitational radiation~instabilities.}
	\label{fig:tw}
\end{figure*}

Finally,~Figure~\ref{fig:tw} displays the angular velocity as a function of the ratio of kinetic to gravitational binding energy $T/W$ for neutron stars at the mass-shedding limit in (a) isothermal and (b) isentropic profile. Gravitational waves can be produced from neutron stars through the nonaxisymmetric perturbations. The~vertical dotted line notes the limit for nonaxisymmetric instabilities from gravitational radiation, located at $T/W\sim 0.08$ for models with $M_{\rm gr}=1.4~M_{\odot}$~\cite{Morsink_1999}. The~introduction of temperature leads to the conclusion that nonaxisymmetric instabilities cannot exist in hot neutron stars. However, for~cases with low values of temperature ($T\leq 1~{\rm MeV}$ and $S=1$ with $Y_{l}=0.2$), the~nonaxisymmetric instability would set in before the mass-shedding limit is reached. The~latter indicates that both the maximum gravitational mass and the angular velocity will be~lowered.

The aftermath from the analysis on the compactness parameter, Kerr parameter, and~ratio $T/W$ is the insight for the hot and rapidly rotating remnant after a neutron star merger. Actually, the~remaining object is a compact object consisting of neutron star matter. By~assuming a remnant with at least $T\geq 30~{\rm MeV}$ for isothermal neutron stars and $S=1$ with $Y_{l}=0.2$ for isentropic ones, rotating at its mass-shedding limit, possible constraints are available through the mentioned quantities. In~particular, 
\begin{itemize}
	\item compactness parameter: $\beta_{\rm rem}^{\rm iso}\leq 0.19$ and $\beta_{\rm rem}^{\rm ise}\leq 0.27$,
	\item Kerr parameter: $\mathcal{K}_{\rm rem}^{\rm iso}\leq 0.42$ and $\mathcal{K}_{\rm rem}^{\rm ise}\leq 0.68$,
	\item ratio $T/W$: $(T/W)_{\rm rem}^{\rm iso}\leq 0.05$ and $(T/W)_{\rm rem}^{\rm ise}\leq 0.127$,
\end{itemize}
where the superscripts ``iso'' and ``ise'' correspond to isothermal and isentropic profiles [for more details see Reference~\cite{Koliogiannis-2021}]. Specifically, in~the case that the remnant follows the isothermal profile, the~remaining object is a lesser compact object than the cold neutron star, with~lower values of maximum gravitational mass and frequency, and~stable toward the dynamical instabilities. In~the case that the remnant follows the isentropic profile, the~remaining object is comparable to the cold neutron~star.

\section{Concluding~Remarks} \label{sec:5}
In this review we have presented a suitable EoS  parameterized to reproduce specific values of the speed of sound and gravitational mass of neutron stars. In~addition, we introduced the effects of finite temperature both in isolated neutron stars and in matters of merging. In~particular, we have constructed equilibrium sequences of both nonrotating and rapidly rotating with the Kepler frequency neutron stars and paid special attention to the gravitational and baryon mass, the~radius, the~transition baryon density, the~Kerr parameter, the~moment of inertia, the~ratio $T/W$, and~the tidal deformability. This study is applied in several gravitational wave events, GW170817, GW190425, and~GW190814, and~possible constraints for the EoS are~extracted. 

Firstly, we studied the EoS and~especially imposed constraints on the speed of sound (which affects the stiffness of the EoS) and the transition density by using  recent observations of two binary neutron stars mergers (GW170817 and GW190425 events). The~implemented method that we developed was based on the upper limits of the effective tidal deformability (derived from the mentioned events), combined with measurements of the maximum neutron star mass. As~a base in our study, we used the MDI-APR EoS, for~two cases of speed of sound bounds~\cite{Koliogiannis-2020,Margaritis-2020}; the conformal case $v_s=c/\sqrt{3}$ and the causal one of $v_s=c$. 

The treatment of the effective tidal deformability as a function of the transition density allowed us to extract constraints on the bounds of the speed of sound. Specifically, for~the GW170817 event, we found that the speed of sound must be lower than the value  $v_s=c/\sqrt{3}$, at least up to densities $n_{{\rm tr}}\approx1.6 n_{\rm s}$, and lower than $v_s=c$ up to densities $n_{{\rm tr}}\approx1.8n_s$. For~the GW190425 event, the respective values are $n_{{\rm tr}}\approx n_s$ for the lower speed of sound bound and $n_{{\rm tr}}\approx1.2n_s$ for the upper one. These constraints are less rigorous than those derived from the GW170817~event. 

Moreover, we studied the effective tidal deformability as a function of the maximum mass for both cases of speed of sound bounds. For~the GW170817, we obtained that the maximum mass should be $M_{{\rm max}}\leq2.106\;M_\odot$ for the $v_s=c/\sqrt{3}$ bound and \mbox{$M_{{\rm max}}\leq3.104\;M_\odot$} for the upper bound $v_s=c$. We notice that the limit of $M_{{\rm max}}\approx2.11\;M_\odot$ corresponds to a transition density equal to $n_{{\rm tr}}\approx1.5n_s$. Hence, according to this finding, the~conformal limit $v_s=c/\sqrt{3}$ is in contradiction with the observational estimations of the $M_{{\rm max}}$ of neutron stars. Therefore, it  must be violated in order to be able to simultaneously describe small values of the effective tidal deformability and high values for neutron star mass. The~reason for this contradiction is based on two different points of view that antagonize one another: the upper limit on $\tilde{\Lambda}$ favors softer EoSs (higher values of $n_{{\rm tr}}$), while the maximum mass observations require stiffer EoSs (smaller values of $n_{{\rm tr}}$). For~higher values of the speed of sound, this contradiction becomes less severe (i.e., $M_{{\rm max}}\approx3.1\;M_\odot$ for the case $v_s=c$). We notice that the GW190425 was not able to offer further~information. 

Furthermore, from~the study of the effective tidal deformability and the radius $R_{1.4}$ of a $1.4\;M_\odot$ neutron star, we observed that all the EoSs follow a common trend. This trend is affected mainly by the chirp mass of the binary system. To~be more specific, as~the chirp mass reaches higher values, the trend moves downwards. From~the event GW170817 we obtained an upper limit $R_{1.4}\approx13~{\rm km}$ for both cases, which is consistent with other estimations. The~event GW190425 provided an upper limit $R_{1.4}\approx14.712~{\rm km}$ for the $v_s=c/\sqrt{3}$ bound and $R_{1.4}\approx14.53~{\rm km}$ for the $v_s=c$ bound.

We postulate that the discovery of future events of binary neutron stars mergers will provide rich information and further constraints on the bound of the speed of sound. In particular, the~detection of future events could lead to more stringent constraints on the upper limit of $\tilde{\Lambda}$ and therefore more rigorous constraints on $n_{{\rm tr}}$ and bounds of the sound speed. Based on our approach, the~more useful events for~the lower limit of $n_{{\rm tr}}$ would be those with lighter component masses. Moreover, it would be of great interest to probe the lower limit of $\tilde{\Lambda}$. Such a lower limit might lead to an upper value of the transition density $n_{{\rm tr}}$. We make the assumption that heavier neutron star mergers would be suitable in the direction of a possible upper limit on $n_{{\rm tr}}$. In~any case, further detection of neutron stars mergers will assist in these open~problems.

The baryon mass of the postmerger remnant is considered approximately conserved, a~feature that gives rise to the significance of the temperature. In~particular, in~the case of hot neutron stars, the~baryon mass is lower than the cold ones. As~remnants are considered rapidly rotating, we study them at the mass-shedding limit. Specifically, in~the cold case, the~baryon mass is $3.085\ M_{\odot}$, while for a hot one at $T=30~{\rm MeV}$ is $2.427~M_{\odot}$ and for one at $S=1$ is $3.05~M_{\odot}$. By~considering that the merger components have approximately equal masses, the~above limits correspond to merger components with $\sim$1.5425, $\sim$1.2135, and~$\sim$1.525~$M_{\odot}$ baryon masses, respectively. Furthermore, the~immediate aftermaths of GW170817~\cite{Abbott-2017} and GW190425~\cite{Abbott_2020} events created hot and rapidly rotating remnants probably at the mass-shedding limit. In~the case of GW170817 event, the~remnant with $\sim$2.7~$M_{\odot}$ can be supported under the uniform rotation of cold and isentropic neutron stars with~respect to the baryon mass of MDI+APR1 EoS. In~contrast, isothermal neutron stars cannot support these values of mass. As~far as  the GW190425 event, the assumption of uniform rotation cannot be used to interpret the remnant of $\sim$3.7~$M_{\odot}$. It has to be noted that the postmeger remnant is assumed to rotate differentially. However, uniform rotation is a valid candidate to provide us with useful information about neutron~stars.

In the GW190814 event~\cite{Abbott_2020_a}, a~compact object with a mass of $\sim$2.6~$M_{\odot}$ was observed as~a merger component. It is believed to be either the lightest black hole or the most massive neutron star~\cite{Most-2020}. Nonetheless, Most~et~al.~\cite{Most-2020} suggest that the compact object could be a neutron star rapidly spinning with $\mathcal{K}$ in the range $[0.49,0.68]$. In~this case, the~relevant postulation is in accordance with this study. More specifically, the~values of the gravitational mass and Kerr parameter coincide with the ones from the MDI+APR1 EoS in both cold catalyzed matter and isentropic matter with $S=1$ and $Y_{l}=0.2$. Following the latter conclusion, there is a possibility that the observed star was rotating close to or at its mass-shedding limit and provide us additional constraints on the high density region of the nuclear EoS. In~addition, possible constraints can be extracted for the corresponding equatorial radius. The~Kerr parameter at the mass-shedding limit of the MDI+APR1 EoS lies in the region of $\mathcal{K}_{\rm max} = [0.67,0.69]$. This region also includes the upper limit of the relevant region from Reference~\cite{Most-2020} in a narrow range. Furthermore, by~exploiting the relation between the Kerr parameter and the compactness parameter, a~possible tight region for the equatorial radius of the star is implied as $R_{\rm max} = [14.77,14.87]~{\rm km}$.

The Kerr parameter also has  the role of an indicator of the collapse to a black hole. In~hot neutron stars, the~Kerr parameter   decreases as the temperature inside the neutron star increases and~never exceeds that of  the cold neutron star. In~conclusion, thermal support cannot lead a rapidly rotating star to collapse into a maximally rotating Kerr black hole. In~addition, after~$\sim$1~$M_{\odot}$, while the Kerr parameter is almost constant for the cold neutron stars, for~hot neutron stars, the Kerr parameter is increasing with respect to the gravitational mass. The~latter leads to a specific maximum~value.

The ratio $T/W$ is explicitly linked to the gravitational collapse to a black hole and the existence of stable supramassive neutron stars. We consider in the present study only the first scenario. Instabilities originating from the gravitational radiation, in~which the critical value is at $T/W \sim 0.08$ for the $M_{\rm gr}=1.4~M_{\odot}$ configuration~\cite{Morsink_1999}, do not exist for hot neutron stars. Nonetheless, for~low values of temperatures, as~the ratio $T/W$ exceeds the critical value, this limit sets the upper value for the maximum gravitational mass and angular velocity. It has to be noted that studies concerning the Kerr parameter as~well as the ratio $T/W$ and~the corresponding effect of the temperature are very rare. The~existence of the latter studies may open a new window in neutron star~studies.

A way to manifest the significance of the thermal support in neutron stars is the evolutionary sequences of constant baryon mass. In~this configuration, the~dependence of the Kepler frequency on the central baryon density presents a linear relation for temperatures higher than $T=30~{\rm MeV}$. The~existence of such a relation, independent of the baryon mass, can define the allowed region of the pair of the central baryon density and corresponding Kepler frequency for a rotating hot neutron star at its mass-shedding~limit.

Central baryon/energy density can also be of great interest in cold neutron stars, as~it is connected with the evolution of the neutron star and the possible appearance of a phase transition. The~end point from our study is that the central energy density must be lower than the values in the range $\varepsilon_{\rm c}/c^{2} = [2.53,2.89]~10^{15}~{\rm g}~{\rm cm^{-3}}$, while for the central baryon density, the~corresponding range is $n_{c} = [7.27,8.09]~n_{s}$. The~latter can inform us about the stability of the neutron star, as~a neutron star with higher values of central energy/baryon density cannot exist, as well as the appearance of the back-bending~process.

Moreover, we examined the hypothetical scenario of a very massive neutron star with mass equal to $\sim$$2.59_{-0.09}^{+0.08}~M_{\odot}$, such as the secondary component of the GW190814 system~\cite{Abbott_2020_a}. The study of the maximum mass of each EoS as a function of the speed of sound bounds (for each value of transition density $n_{\mathrm{tr}}$), provide us constraints in the speed of sound. To~be more specific, the~$n_{\mathrm{tr}}=1.5n_s$ case leads to $(v_s/c)^2\in[0.448,0.52]$, while the $n_{\mathrm{tr}}=2n_s$ case leads to $(v_s/c)^2\in[0.597,0.72]$. We postulate that as the transition density $n_{\mathrm{tr}}$ is getting higher values, it is more difficult to achieve such a massive nonrotating neutron star. In particular, above~a specific high-transition-density $n_{\mathrm{tr}}$ value,  the~speed of sound should be close to the causality to provide such a massive~neutron star. 

By studying the tidal parameters for the single case, we observed that the lower transition densities $n_{\mathrm{tr}}$ lead to higher tidal parameters. Hence, the~transition density $n_{\mathrm{tr}}=2n_s$ corresponds to a more compact neutron star (less deformation). Furthermore, between~the same kind of transition density $n_{\mathrm{tr}}$, the~EoSs with higher bounds on the speed of sound predict higher tidal parameters. Therefore, for~the same transition density, the~higher speed of sound bound means that the neutron star is less compact (more deformation).

Moving on to the binary neutron star system case, we adopted the hypothesis of a very massive component with $m_1=2.59\;M_\odot$, allowing us to investigate a variety of hypothetical binary neutron star systems with such a heavy component neutron star. In~the case of heavy components of binary neutron star systems, meaning high value for the system's chirp mass $\mathcal{M}_c$, the~effective tidal deformability $\tilde{\Lambda}$ has smaller values (smaller deformation). This behavior was noticed also in the $\tilde{\Lambda}-q$ diagram, in~which the increasing binary mass symmetric ratio $q$ leads to smaller values of $\tilde{\Lambda}$. For~a binary neutron system with the heavier component equal to $m_1=2.59\;M_\odot$ and the lighter one equal to $m_2=1.4\;M_\odot$, the~chirp mass $\mathcal{M}_c$ and the ratio $q$ were estimated to be $\mathcal{M}_c=1.642\;M_\odot$ and $q=0.541$, respectively.

Lastly, we studied the case of a binary neutron star system with $m_1=2.59\;M_\odot$ with a secondary component $m_2=1.4\;M_\odot$. We especially concentrated on the radius $R_{1.4}$. A~general remark is that the transition density $n_{\mathrm{tr}}=1.5n_s$ provides higher values of $R_{1.4}$ and $\tilde{\Lambda}$ than the $n_{\mathrm{tr}}=2n_s$ case. We extended our study to further transition densities $n_{\mathrm{tr}}$, which confirmed this general behavior. Additionally, the~higher bounds of the speed of sound  provide higher values on both  parameters. By~imposing an upper limit on the radius, we extracted some upper limits on the $\tilde{\Lambda}$ for each value of the speed of sound. In particular, this upper limit on $\tilde{\Lambda}$ shifts to higher values as the bound of the speed of sound increases.

This hypothetical scenario of a very massive neutron star demonstrated the key role of a microscopic quantity of the EoS, the~speed of sound, which dramatically affects  the EoS in combination with the changes in the transition density. We notice that the existence of such a massive nonrotating neutron star would mean a significant difference from all the cases known  so far, which is a challenge for physics.

The underlying physics of neutron star mergers and the hot, rapidly rotating remnant should be investigated by considering differential rotation and cooling mechanisms, as~these are the main features in the early postmerger phase. In~addition, special emphasis should be given in the phase transition region, the~existence of exotic degrees of freedom in the interior of neutron stars, as~well as the accurate measurement of the tidal deformability. Finally, the~observation of binary neutron star mergers and black-hole--neutron-star mergers, combined with the above studies, may provide significant constraints for the construction of the~EoS.

\section*{Acknowledgments}
The authors would like to thank K. Kokkotas for their constructive comments and insight during the preparation of the manuscript and also L. Rezzolla for useful correspondence and clarifications. The~authors also thank  B. Farr and Chatziioannou for their assistance providing computational tools regarding the kernel density estimation of the sample data and D. Radice, N. Stergioulas, and N. Minkov for the useful correspondence. We also acknowledge P. Meszaros and  A. Sedrakian for their useful comments regarding the hot equations of state. In~addition, the~authors wish to thank the Bulgarian Academy of Science and the organizers of the International Workshop ``Shapes and Dynamics of Atomic Nuclei: Contemporary Aspects'' (SDANCA-21) for the their hospitality while this was completed. 
We would like to thank the financial support of the Bulgarian National Science Fund under contracts No. KP-06-N48/1 and  No. KP-06-N28/6.  

\bibliography{koliogiannis}

\end{document}